\documentclass{article}

\usepackage[colorlinks=true,
            citecolor=black,
            linkcolor=black,
            urlcolor=black]{hyperref}
\usepackage{latexsym,amsmath,amsfonts,amssymb,stmaryrd,amsthm}
\usepackage[dvipsnames]{xcolor}
\usepackage{tikz-cd}
\usepackage{url}
\usepackage{proof}
\usepackage[square,numbers]{natbib}
\usepackage[capitalize,noabbrev]{cleveref}
\usepackage[all]{xy} 
\usepackage{manfnt}
\usepackage[backgroundcolor=cyan!50!white!50,linecolor=black]{todonotes}
\usepackage{mathpartir}
\usepackage{xparse}
\usepackage{microtype}
\usepackage{mdframed}
\usepackage{bbm}
\usepackage{fullpage}
\usepackage{enumitem}

\mathchardef\mh="2D
\renewcommand{\_}{\ensuremath{\rule{1ex}{.4pt}}}

\renewcommand{\phi}{\varphi}

\makeatletter
\newtheorem*{rep@theorem}{\rep@title}
\newcommand{\newreptheorem}[2]{%
\newenvironment{rep#1}[1]{%
 \def\rep@title{#2 \ref{##1}}%
 \begin{rep@theorem}}%
 {\end{rep@theorem}}}
\makeatother

\theoremstyle{definition}
\newtheorem{definition}{Definition}[section]
\newreptheorem{definition}{Definition}

\theoremstyle{remark}

\newtheorem{notation}[definition]{Notation}

\theoremstyle{plain}
\newtheorem{lemma}[definition]{Lemma}
\newreptheorem{lemma}{Lemma}
\newtheorem{theorem}[definition]{Theorem}
\newtheorem{proposition}[definition]{Proposition}
\newtheorem{corollary}[definition]{Corollary}
\newreptheorem{corollary}{Corollary}

\newcommand{\eqdef}{\mathrel{:=}}
\newcommand{\iffdef}{\mathrel{:\Leftrightarrow}}

\newcommand{\eq}{\ensuremath{\mathbin{\doteq}}}

\newcommand{\fd}[1]{\ensuremath{\mathsf{FD}(#1)}}
\newcommand{\fb}[1]{\ensuremath{\mathsf{FB}(#1)}}

\newcommand{\ofc}[2]{#1\mathbin{:}#2}
\newcommand{\oft}[2]{#1\mathbin{:}#2}
\newcommand{\cube}{\ensuremath{\text{\mancube}}}

\newcommand{\Label}{\ensuremath{L}}

\newcommand{\Co}{\ensuremath{\formal{\mathcal{C}}}}
\newcommand{\GD}{\ensuremath{\Delta}}
\newcommand{\E}{\ensuremath{\mathcal{E}}}
\newcommand{\F}{\ensuremath{\mathcal{F}}}
\newcommand{\GG}{\ensuremath{\Gamma}}

\newcommand{\K}{\ensuremath{\formal{\mathcal{K}}}}
\newcommand{\N}{\ensuremath{\mathbb{N}}}
\newcommand{\GTh}{\ensuremath{\formal{\Theta}}}
\newcommand{\GF}{\ensuremath{\formal{\Phi}}}
\newcommand{\W}{\ensuremath{\mathsf{W}}}
\newcommand{\Ga}{\ensuremath{\alpha}}
\newcommand{\Gb}{\ensuremath{\beta}}
\newcommand{\Gg}{\ensuremath{\gamma}}
\newcommand{\Gd}{\ensuremath{\delta}}
\newcommand{\Ge}{\ensuremath{\varepsilon}}
\newcommand{\Gf}{\ensuremath{\formal{\phi}}}
\newcommand{\Gh}{\ensuremath{\eta}}
\newcommand{\Gr}{\ensuremath{\rho}}
\newcommand{\Gs}{\ensuremath{\sigma}}
\newcommand{\Gth}{\ensuremath{\formal{\theta}}}

\makeatletter
\def\rightharpoonupfill@{\arrowfill@\relbar\relbar\rightharpoonup}
\newcommand{\overrightharpoonup}{%
\mathpalette{\overarrow@\rightharpoonupfill@}}

\providecommand{\leftsquigarrow}{%
  \mathrel{\mathpalette\reflect@squig\relax}%
}
\newcommand{\reflect@squig}[2]{%
  \reflectbox{$\m@th#1\rightsquigarrow$}%
}
\makeatother

\newcommand{\lst}[1]{\overline{#1}}

\NewDocumentCommand\Infer{o m m}{%
  \IfValueTF{#1}
    {\inferrule*[vcenter,right=#1]{#2}{#3}}
    {\inferrule{#2}{#3}}
  }

\newcommand{\rulename}[1]{(\textsc{#1})}

\usepackage{accsupp} 

\newcommand*{\llbrace}{%
  \BeginAccSupp{method=hex,unicode,ActualText=2983}%
  \textnormal{\usefont{OMS}{lmr}{m}{n}\char102}%
  \mathchoice{\mkern-4.05mu}{\mkern-4.05mu}{\mkern-4.3mu}{\mkern-4.8mu}%
  \textnormal{\usefont{OMS}{lmr}{m}{n}\char106}%
  \EndAccSupp{}%
}
\newcommand*{\rrbrace}{%
  \BeginAccSupp{method=hex,unicode,ActualText=2984}%
  \textnormal{\usefont{OMS}{lmr}{m}{n}\char106}%
  \mathchoice{\mkern-4.05mu}{\mkern-4.05mu}{\mkern-4.3mu}{\mkern-4.8mu}%
  \textnormal{\usefont{OMS}{lmr}{m}{n}\char103}%
  \EndAccSupp{}%
}

\newsavebox{\lXbrace}
\savebox{\lXbrace}{$\llbrace$}
\newsavebox{\rXbrace}
\savebox{\rXbrace}{$\rrbrace$}

\newcommand{\subst}[3]{\ensuremath{#1 [#2 / #3]}}
\newcommand{\dsubst}[3]{\ensuremath{#1 \langle{#2}/{#3}\rangle}}

\newcommand{\arr}[2]{\ensuremath{#1 \to #2}}
\newcommand{\picl}[3]{\ensuremath{({#1}{:}{#2}) \to #3}}
\newcommand{\lam}[2]{\ensuremath{\lambda{#1}.{#2}}}
\newcommand{\app}[2]{\ensuremath{\mathsf{app}({#1},{#2})}}

\newcommand{\sigmacl}[3]{\ensuremath{({#1}{:}{#2}) \times #3}}
\newcommand{\pair}[2]{\ensuremath{\langle #1,#2\rangle}}

\NewDocumentCommand\Eq{g g g}{%
  \ensuremath{\mathsf{Eq}\IfValueT{#1}{_{#1}\IfValueT{#2}{(#2,#3)}}}}

\NewDocumentCommand\Path{g g g}{%
  \ensuremath{\mathsf{Path}\IfValueT{#1}{_{#1}\IfValueT{#2}{(#2,#3)}}}}
\newcommand{\dlam}[2]{\ensuremath{\langle #1 \rangle #2}}
\newcommand{\dapp}[2]{\ensuremath{#1 @ #2}}

\NewDocumentCommand\Id{g g g}{%
  \ensuremath{{\mathsf{Id}\IfValueT{#1}{_{#1}\IfValueT{#2}{(#2,#3)}}}}}
\NewDocumentCommand\Idelim{g g g g g}{%
  \ensuremath{\mathsf{J}\IfValueT{#1}{_{#1}(#2;#3;#4;#5)}}}

\newcommand{\UU}{\mathcal{U}}

\newcommand{\UKan}[1][j]{\ensuremath{\UU^\mathsf{Kan}_{#1}}}

\NewDocumentCommand\formalif{m m}{\IfBooleanTF{#1}{\formal{#2}}{#2}}

\newcommand{\nat}{\ensuremath{\mathsf{nat}}}
\newcommand{\z}{\ensuremath{\mathsf{z}}}
\newcommand{\suc}[1]{\ensuremath{\mathsf{s}(#1)}}
\NewDocumentCommand\natrec{g g g}{%
  \ensuremath{\mathsf{natrec}\IfValueT{#1}{(#1;#2,#3)}}}

\newcommand{\bool}{\ensuremath{\mathsf{bool}}}

\NewDocumentCommand\ifb{G{{}} g g g}{%
  \ensuremath{\mathsf{if}_{#1}
    \IfValueT{#2}{\IfNoValueTF{#3}{({#2})}{({#2};{#3},{#4})}}}}

\newcommand{\C}[1][1]{\ensuremath{{\mathbb{S}^{#1}}}}
\newcommand{\base}{\ensuremath{\mathsf{base}}}
\NewDocumentCommand\lp{g}{\ensuremath{\mathsf{loop}\IfValueT{#1}{^{#1}}}}
\NewDocumentCommand\Celim{G{{}} g g g}{%
  \ensuremath{\C\mh\mathsf{elim}_{#1}
    \IfValueT{#2}{\IfNoValueTF{#3}{({#2})}{({#2};{#3},{#4})}}}}

\NewDocumentCommand\zero{s}{\ensuremath{\mathsf{\formalif{#1}{zero}}}}
\NewDocumentCommand\one{s}{\ensuremath{\mathsf{\formalif{#1}{one}}}}
\NewDocumentCommand\seg{s g}{\ensuremath{\mathsf{\formalif{#1}{seg}}\IfValueT{#2}{^{#2}}}}

\NewDocumentCommand\pushoutcl{g g g g g}{%
  \ensuremath{\amalg\IfValueT{#1}{(#1;#2;#3;#4;#5)}}}
\NewDocumentCommand\pleft{g}{\ensuremath{\mathsf{left}\IfValueT{#1}{(#1)}}}
\NewDocumentCommand\pright{g}{\ensuremath{\mathsf{right}\IfValueT{#1}{(#1)}}}
\NewDocumentCommand\pglue{O{} g g}{%
  \ensuremath{\mathsf{glue}_{#1}\IfValueT{#2}{^{#2}(#3)}}}

\newcommand{\Wcl}[2]{\ensuremath{{\W({#1};{#2})}}}
\NewDocumentCommand\wsup{g g}{%
  \ensuremath{\mathsf{sup}\IfValueT{#1}{(#1;#2)}}}
\NewDocumentCommand\welim{g g g}{%
  \ensuremath{\W\mh\mathsf{elim}\IfValueT{#1}{_{#1}(#2;#3)}}}

\newcommand{\WQ}{\ensuremath{\mathsf{WQ}}}
\newcommand{\WQcl}[5]{\ensuremath{{\WQ({#1};{#2};{#3};{#4};{#5})}}}
\NewDocumentCommand\wqsup{s g g}{%
  \ensuremath{\mathsf{\formalif{#1}{sup}}\IfValueT{#2}{(#2;#3)}}}
\NewDocumentCommand\wqcell{s g g g g}{%
  \ensuremath{\mathsf{\formalif{#1}{cell}}\IfValueT{#2}{^{#2}(#3;#4;#5)}}}
\NewDocumentCommand\wqelim{g g g g}{%
  \ensuremath{\WQ\mh\mathsf{elim}\IfValueT{#1}{_{#1}(#2;#3,#4)}}}

\newcommand{\Trunc}[2][]{\ensuremath{\lVert{#2}\rVert_{#1}}}
\NewDocumentCommand\trpt{s g}{%
  \ensuremath{\mathsf{\formalif{#1}{pt}}\IfValueT{#2}{(#2)}}}
\NewDocumentCommand\trglue{s g g g}{%
  \ensuremath{\mathsf{\formalif{#1}{path}}\IfValueT{#2}{^{#2}(#3;#4)}}}
\NewDocumentCommand\trhub{s g}{%
  \ensuremath{\mathsf{\formalif{#1}{hub}}\IfValueT{#2}{(#2)}}}
\NewDocumentCommand\trspoke{s g g g}{%
  \ensuremath{\mathsf{\formalif{#1}{spoke}}\IfValueT{#2}{^{#2}(#3;#4)}}}
\NewDocumentCommand\trelim{G{{}} g g g}{%
  \ensuremath{\mathsf{trunc}\mh\mathsf{elim}_{#1}
    \IfValueT{#2}{\IfNoValueTF{#3}{({#2})}{({#2};{#3},{#4})}}}}
\NewDocumentCommand\trhselim{G{{}} g g g g}{%
  \ensuremath{\mathsf{trunc}\mh\mathsf{elim}_{#1}
    \IfValueT{#2}{\IfNoValueTF{#3}{({#2})}{({#2};{#3},{#4},{#5})}}}}

\newcommand{\Torus}{\ensuremath{\mathsf{T}}}
\NewDocumentCommand\tbase{s}{\ensuremath{\mathsf{\formalif{#1}{base}}}}
\NewDocumentCommand\tlpa{s g}{\ensuremath{\mathsf{\formalif{#1}{loopa}}\IfValueT{#2}{^{#2}}}}
\NewDocumentCommand\tlpb{s g}{\ensuremath{\mathsf{\formalif{#1}{loopb}}\IfValueT{#2}{^{#2}}}}
\NewDocumentCommand\tsurf{s g g}{\ensuremath{\mathsf{\formalif{#1}{surf}}\IfValueT{#2}{^{#2,#3}}}}
\NewDocumentCommand\telim{G{{}} g g g g g}{%
  \ensuremath{\Torus\mh\mathsf{elim}_{#1}
    \IfValueT{#2}{\IfNoValueTF{#3}{({#2})}{({#2};{#3},{#4},{#5},{#6})}}}}

\NewDocumentCommand\Loc{g g}{\mathcal{L}\IfValueT{#1}{_{#1}(#2)}}
\NewDocumentCommand\loc{s g}{\ensuremath{\mathsf{\formalif{#1}{loc}}\IfValueT{#2}{(#2)}}}
\NewDocumentCommand\ext{s O{} g g g}{%
  \ensuremath{\mathsf{\formalif{#1}{ext}}{#2}\IfValueT{#3}{(#3;#4;#5)}}}
\NewDocumentCommand\rtr{s O{} g g g g}{%
  \ensuremath{\mathsf{\formalif{#1}{rtr}}{#2}\IfValueT{#3}{^{#3}(#4;#5;#6)}}}

\newcommand{\Blass}{\mathsf{F}}

\NewDocumentCommand\indcl{O{\GD} m g}{\mathsf{ind}\IfValueT{#3}{_{#1}}(#2\IfValueT{#3}{;#3})}
\NewDocumentCommand\intro{s O{\K} O{\ell} g g g}{%
  \ensuremath{\mathsf{intro}_{\IfBooleanF{#1}{#2,#3}}\IfValueT{#4}{^{#4}({#5}\IfValueT{#6}{;{#6}})}}}

\NewDocumentCommand\elim{s g g g g}{%
  \ensuremath{\mathsf{elim}\IfValueT{#2}{_{#2;#3}
      \IfBooleanTF{#1}{\left(}{(}
      {#4;#5}
      \IfBooleanTF{#1}{\right)}{)}
  }}
}

\NewDocumentCommand\fibcl{s g g g g}{
  \ensuremath{\mathsf{fib}
  \IfBooleanTF{#1}
    {(#2;#3)}
    {\IfValueT{#2}{(#2;#3;#4;#5)}}}}

\NewDocumentCommand\refl{g}{%
  \ensuremath{\mathsf{refl}\IfValueT{#1}{(#1)}}}
\NewDocumentCommand{\fibelim}{g g g g}{%
  \ensuremath{\mathsf{J}\IfValueT{#1}{_{#1;#2}(#3;#4)}}}

\NewDocumentCommand\func{g g g}{%
  \mathrm{act}\IfValueT{#1}{_{#1}\IfValueT{#2}{(#2;#3)}}}

\NewDocumentCommand\ua{G{{}} g}{%
  \ensuremath{\mathsf{V}_{#1}\IfValueT{#2}{(#2)}}}

\newcommand{\etc}[1]{\ensuremath{\overrightharpoonup{#1}}}
\newcommand{\tube}[2]{\ensuremath{#1\hookrightarrow #2}}
\newcommand{\sys}[2]{\etc{\tube{#1}{#2}}}

\NewDocumentCommand\NewCoercionOperator{m m O{\rightsquigarrow} O{M}}{%
  \NewDocumentCommand#1{s G{{}} g g g}{%
    \ensuremath{\mathsf{#2}_{##2}%
    \IfBooleanTF{##1}
      {^{r #3 r'}(#4)}
      {\IfValueT{##3}{^{##3 #3 ##4}(##5)}}}}
}

\NewCoercionOperator{\coe}{coe}
\NewCoercionOperator{\mcoe}{\mathrm{mcoe}}
\NewCoercionOperator{\fcoe}{fcoe}
\NewCoercionOperator{\tcoe}{tcoe}

\NewDocumentCommand\NewCompositionOperator{s m m O{\rightsquigarrow} O{M} O{y.N_i}}{%
  \IfBooleanTF{#1}
    {\NewDocumentCommand#2{s g g g g}{%
      \IfBooleanTF{##1}
        {\ensuremath{\mathsf{#3}^{r #4 r'}(#5;\sys{##2}{#6})}}
        {\IfNoValueTF{##2}
          {\ensuremath{\mathsf{#3}}}
          {\ensuremath{\mathsf{#3}^{##2 #4 ##3}(##4\IfValueT{##5}{;##5})}}}}}
    {\NewDocumentCommand#2{s G{{}} g g g g}{%
      \IfBooleanTF{##1}
        {\ensuremath{\mathsf{#3}_{##2}^{r #4 r'}(#5;\sys{##3}{#6})}}
        {\IfNoValueTF{##3}
          {\ensuremath{\mathsf{#3}_{##2}}}
          {\ensuremath{\mathsf{#3}_{##2}^{##3 #4 ##4}(##5\IfValueT{##6}{;##6})}}}}}
}
\NewDocumentCommand\NewBigCompositionOperator{s m m O{\rightsquigarrow} O{M} O{y.N_i}}{%
  \IfBooleanTF{#1}
    {\NewDocumentCommand#2{s g g g g}{%
      \IfBooleanTF{##1}
        {\ensuremath{\mathsf{#3}^{r #4 r'}\left(#5;\sys{##2}{#6}\right)}}
        {\IfNoValueTF{##2}
          {\ensuremath{\mathsf{#3}}}
          {\ensuremath{\mathsf{#3}^{##2 #4 ##3}\left(##4\IfValueT{##5}{;##5}\right)}}}}}
    {\NewDocumentCommand#2{s G{{}} g g g g}{%
      \IfBooleanTF{##1}
        {\ensuremath{\mathsf{#3}_{##2}^{r #4 r'}\left(#5;\sys{##3}{#6}\right)}}
        {\IfNoValueTF{##3}
          {\ensuremath{\mathsf{#3}_{##2}}}
          {\ensuremath{\mathsf{#3}_{##2}^{##3 #4 ##4}\left(##5\IfValueT{##6}{;##6}\right)}}}}}
}
\NewCompositionOperator{\hcom}{hcom}
\NewBigCompositionOperator{\bighcom}{hcom}
\NewCompositionOperator{\com}{com}
\NewBigCompositionOperator{\bigcom}{com}
\NewCompositionOperator*{\fhcom}{fhcom}
\NewBigCompositionOperator*{\bigfhcom}{fhcom}
\NewCompositionOperator{\fcom}{fcom}
\NewBigCompositionOperator{\bigfcom}{fcom}
\NewCompositionOperator*{\Kbox}{box}[\rightsquigarrow][M][N_i]
\NewCompositionOperator*{\Kcap}{cap}[\leftsquigarrow][M][y.B_i]

\newcommand{\steps}{\ensuremath{\longmapsto}}
\newcommand{\msteps}{\ensuremath{\longmapsto^*}}
\newcommand{\evals}{\ensuremath{\Downarrow}}

\newcommand{\isval}[1]{\ensuremath{#1\ \mathsf{val}}}

\newcommand{\lift}[1]{#1^\evals}
\NewDocumentCommand\PTy{g}{%
  \ensuremath{\textsc{PTy}\IfValueT{#1}{(#1)}}}

\NewDocumentCommand\Tm{g}{%
  \ensuremath{\textsc{Tm}\IfValueT{#1}{(#1)}}}
\NewDocumentCommand\Vl{g}{%
  \ensuremath{\textsc{Vl}\IfValueT{#1}{(#1)}}}
\NewDocumentCommand\relcts{s m m}{%
  \ensuremath{{#2} \models \IfBooleanTF{#1}{#3}{(#3)}}}
\NewDocumentCommand\Ind{g}{%
  \ensuremath{\textsc{Ind}\IfValueT{#1}{(#1)}}}
\NewDocumentCommand\Fhcom{g}{%
  \ensuremath{\textsc{Fhcom}\IfValueT{#1}{(#1)}}}
\NewDocumentCommand\Fcoe{g}{%
  \ensuremath{\textsc{Fcoe}\IfValueT{#1}{(#1)}}}
\NewDocumentCommand\Fcom{g}{%
  \ensuremath{\textsc{Fcom}\IfValueT{#1}{(#1)}}}
\NewDocumentCommand\Intro{m m g}{%
  \ensuremath{\textsc{Intro}_{#1,#2}\IfValueT{#3}{(#3)}}}
\newcommand{\Pirel}[2]{\textsc{Pi}(#1,#2)}

\NewDocumentCommand\TcoeI{g}{%
  \ensuremath{\textsc{Tcoe}^{-1}\IfValueT{#1}{(#1)}}}
\NewDocumentCommand\opt{m g}{
  \ensuremath{{#1}?\IfValueT{#2}{(#2)}}}
\NewDocumentCommand\multi{m g}{
  \ensuremath{{#1}^*\IfValueT{#2}{(#2)}}}

\newcommand{\dimj}[2][\Psi]{\ensuremath{{#2}\ \mathsf{dim}\ [#1]}}
\newcommand{\tmj}[2][\Psi]{\ensuremath{{#2}\ \mathsf{tm}\ [#1]}}

\newcommand{\bndj}[2][\Psi]{\ensuremath{{#2}\ \mathsf{bnd}\ [#1]}}
\newcommand{\open}[2]{\ensuremath{{#1} \vdash {#2}}}

\newcommand{\vper}[1]{\ensuremath{\llbracket #1 \rrbracket}}

\newcommand{\tds}[3]{\ensuremath{{#2} : {#1} \to {#3}}}
\newcommand{\psitd}[1][]{\tds{\Psi#1'}{\psi#1}{\Psi#1}}
\newcommand{\td}[2]{\ensuremath{{#1}{#2}}}
\newcommand{\id}{\ensuremath{\mathrm{id}}}

\newcommand{\tdss}[3]{\tds{\Psi#1}{\psi#2}{\Psi#3}}

\newcommand{\ix}[2]{{#1}[{#2}]}
\newcommand{\relseq}[2]{{#1} \gg {#2}}

\NewDocumentCommand\judg{O{\Psi} d<> m}
  {#3\ [#1 \IfValueT{#2}{\mid #2}]}

\NewDocumentCommand\cwfctx{m O{\Psi} d<> m}
  {#4\ \mathsf{ctx}_\mathsf{#1}\ [#2 \IfValueT{#3}{\mid #3}]}
\NewDocumentCommand\ceqctx{m O{\Psi} d<> m m}
  {#4 \eq #5\ \mathsf{ctx}_\mathsf{#1}\ [#2 \IfValueT{#3}{\mid #3}]}
\NewDocumentCommand\wfctx{m O{\Psi} d<> m m}
  {#4 \gg #5 \ \mathsf{ctx}_\mathsf{#1}\ [#2 \IfValueT{#3}{\mid #3}]}
\NewDocumentCommand\eqctx{m O{\Psi} d<> m m m}
  {#4 \gg #5 \eq #6\ \mathsf{ctx}_\mathsf{#1}\ [#2 \IfValueT{#3}{\mid #3}]}

\NewDocumentCommand\cwftype{m O{\Psi} d<> m}
  {#4\ \mathsf{type}_\mathsf{#1}\ [#2 \IfValueT{#3}{\mid #3}]}
\NewDocumentCommand\ceqtype{m O{\Psi} d<> m m}
  {#4 \eq #5\ \mathsf{type}_\mathsf{#1}\ [#2 \IfValueT{#3}{\mid #3}]}
\NewDocumentCommand\wftype{m O{\Psi} d<> m m}
  {#4 \gg #5 \ \mathsf{type}_\mathsf{#1}\ [#2 \IfValueT{#3}{\mid #3}]}
\NewDocumentCommand\eqtype{m O{\Psi} d<> m m m}
  {#4 \gg #5 \eq #6\ \mathsf{type}_\mathsf{#1}\ [#2 \IfValueT{#3}{\mid #3}]}

\newcommand{\cwfctxk}{\cwfctx{Kan}}
\newcommand{\ceqctxk}{\ceqctx{Kan}}

\newcommand{\cwftypep}{\cwftype{pre}}
\newcommand{\ceqtypep}{\ceqtype{pre}}

\newcommand{\cwftypek}{\cwftype{Kan}}
\newcommand{\ceqtypek}{\ceqtype{Kan}}
\newcommand{\wftypek}{\wftype{Kan}}
\newcommand{\eqtypek}{\eqtype{Kan}}

\newcommand{\ceqtypex}{\ceqtype{\kappa}}

\newcommand{\eqtypex}{\eqtype{\kappa}}

\NewDocumentCommand\coftype{O{\Psi} d<> m m}
  {#3 \in #4\ [#1 \IfValueT{#2}{\mid #2}]}
\NewDocumentCommand\ceqtm{O{\Psi} d<> m m m}
  {#3 \eq #4 \in #5\ [#1 \IfValueT{#2}{\mid #2}]}
\NewDocumentCommand\oftype{O{\Psi} d<> m m m}
  {#3 \gg #4 \in #5\ [#1 \IfValueT{#2}{\mid #2}]}
\NewDocumentCommand\eqtm{O{\Psi} d<> m m m m}
  {#3 \gg #4 \eq #5 \in #6\ [#1 \IfValueT{#2}{\mid #2}]}

\newcommand{\sch}[1]{\formal{\textsc{#1}}}

\NewDocumentCommand\constr{s m m m m m}{%
  \IfBooleanTF{#1}{\left(}{(}{#3};{#4};{#5};{#2}.{#6}\IfBooleanTF{#1}{\right)}{)}
}
\newcommand{\nilconstrs}{\formal{\bullet}}
\newcommand{\snocconstrs}[2]{\formal{[}{#1}, {#2}\formal{]}}
\newcommand{\listconstrs}[1]{\left[ {#1} \right]}
\newcommand{\constrspre}{\sqsubseteq}

\newcommand{\emp}{\varnothing}

\NewDocumentCommand\tyatrel{s m m}{%
  \IfBooleanTF{#1}
  {\llbrace{#2}\rrbrace_{\mathsf{d}}({#3})}
  {\llbrace{#2}\rrbrace({#3})}}
\NewDocumentCommand\ctxatrel{s m m m}{%
  \IfBooleanTF{#1}
  {\llbrace{#2}\rrbrace^{#3}_{\mathsf{d}}({#4})}
  {\llbrace{#2}\rrbrace^{#3}({#4})}}
\NewDocumentCommand\tyatty{s m m g}{%
  \IfBooleanTF{#1}
  {\llbrace{#2}\rrbrace_{\mathsf{d}}({#3};{#4})}
  {\llbrace{#2}\rrbrace({#3})}}
\NewDocumentCommand\ctxatty{s m m m g}{%
  \IfBooleanTF{#1}
  {\llbrace{#2}\rrbrace^{#3}_{\mathsf{d}}({#4};{#5})}
  {\llbrace{#2}\rrbrace^{#3}({#4})}}
\NewDocumentCommand\insttm{s m m m g g g}{%
  \IfBooleanTF{#1}
  {\llparenthesis{#2}\rrparenthesis^{#3,#4}_{#5}({#6};{#7})}
  {\llparenthesis{#2}\rrparenthesis^{#3}({#4})}}

\definecolor{formal}{rgb}{0,0.2,0.7}
\newcommand{\formal}[1]{{\color{formal}{#1}}}

\NewDocumentCommand\argvar{g}{\formal{\mathsf{X}}\IfValueT{#1}{(#1)}}
\newcommand{\argpi}[3]{({#1}{:}{#2}) \mathbin{\formal{\to}} {#3}}
\newcommand{\argarr}[2]{{#1} \mathbin{\formal{\to}} {#2}}

\newcommand{\ofa}[2]{\oft{#1}{#2}}

\NewDocumentCommand\bndintro{O{\ell} g g g}{%
  \formal{\mathsf{intro}}_{#1}\IfValueT{#2}{^{#2}(#3;#4)}}
\NewCompositionOperator{\bndfhcom}{\formal{fhcom}}[\rightsquigarrow][\sch{m}][y.\sch{n}_i]
\NewBigCompositionOperator{\bigbndfhcom}{\formal{fhcom}}[\rightsquigarrow][\sch{m}][y.\sch{n}_i]
\NewCoercionOperator{\bndfcoe}{\formal{fcoe}}[\rightsquigarrow][\sch{m}]

\newcommand{\bndlam}[2]{\ensuremath{\formal{\lambda}{#1}.{#2}}}
\newcommand{\bndapp}[2]{\ensuremath{\formal{\mathsf{app}}({#1},{#2})}}

\NewDocumentCommand\bndnatrec{g g g}{%
  \ensuremath{\formal{\mathsf{natrec}}\IfValueT{#1}{(#1;#2,#3)}}}

\newcommand{\nilelim}{\bullet}
\newcommand{\snocelim}[2]{[ {#1}, {#2} ]}
\newcommand{\listelim}[1]{\left[ {#1} \right]}
\newcommand{\elimpre}{\sqsubseteq}

\newcommand{\Interp}[4]{\mathrm{Interp}^{#1,#2}_{#3}(#4)}
\newcommand{\ElimBnd}[1]{\mathrm{ElimBnd}(#1)}

\newcommand{\fl}[1]{\mathsf{L}(#1)}
\newcommand{\dom}{\operatorname{dom}}
\newcommand{\height}[2]{\mathrm{ht}_{#1}(#2)}

\newcommand{\lfp}[1]{\mu{#1}}
\newcommand{\gfp}[1]{\nu{#1}}
\newcommand{\indrel}[2][\GD]{\mathbbm{i}_{#1}({#2})}

\NewDocumentCommand\ceqconstrs{O{\Psi} d<> m m m}
  {\judg[#1]<#2>{#3 \rhd #4 \equiv #5\ \mathsf{constrs}}}
\NewDocumentCommand\eqconstrs{O{\Psi} d<> m m m m}
  {\judg[#1]<#2>{#3 \gg #4 \rhd #5 \equiv #6\ \mathsf{constrs}}}
\NewDocumentCommand\cwfconstrs{O{\Psi} d<> m m}
  {\judg[#1]<#2>{#3 \rhd #4\ \mathsf{constrs}}}
\NewDocumentCommand\wfconstrs{O{\Psi} d<> m m m}
  {\judg[#1]<#2>{#3 \gg #4 \rhd #5\ \mathsf{constrs}}}

\NewDocumentCommand\ceqconstr{O{\Psi} d<> m O{\K} m m}
  {\judg[#1]<#2>{#3 \rhd #4 \vdash #5 \equiv #6\ \mathsf{constr}}}
\NewDocumentCommand\eqconstr{O{\Psi} d<> m m O{\K} m m}
  {\judg[#1]<#2>{#3 \gg #4 \rhd #5 \vdash #6 \equiv #7\ \mathsf{constr}}}
\NewDocumentCommand\cwfconstr{O{\Psi} d<> m O{\K} m}
  {\judg[#1]<#2>{#3 \rhd #4 \vdash #5\ \mathsf{constr}}}
\NewDocumentCommand\wfconstr{O{\Psi} d<> m m O{\K} m m}
  {\judg[#1]<#2>{#3 \gg #4 \rhd #5 \vdash #6\ \mathsf{constr}}}

\NewDocumentCommand\ceqargtype{O{\Psi} d<> m m m}
  {\judg[#1]<#2>{#3 \rhd #4 \equiv #5 \ \mathsf{atype}}}
\NewDocumentCommand\eqargtype{O{\Psi} d<> m m m m}
  {\judg[#1]<#2>{#3 \gg #4 \rhd #5 \equiv #6 \ \mathsf{atype}}}
\NewDocumentCommand\cwfargtype{O{\Psi} d<> m m}
  {\judg[#1]<#2>{#3 \rhd #4\ \mathsf{atype}}}
\NewDocumentCommand\wfargtype{O{\Psi} d<> m m m}
  {\judg[#1]<#2>{#3 \gg #4 \rhd #5\ \mathsf{atype}}}

\NewDocumentCommand\ceqargctx{O{\Psi} d<> m m m}
  {#3 \rhd #4 \equiv #5\ \mathsf{actx}\ [#1 \IfValueT{#2}{\mid #2}]}
\NewDocumentCommand\eqargctx{O{\Psi} d<> m m m m}
  {#3 \gg #4 \rhd #5 \equiv #6\ \mathsf{actx}\ [#1 \IfValueT{#2}{\mid #2}]}
\NewDocumentCommand\cwfargctx{O{\Psi} d<> m m}
  {#3 \rhd #4\ \mathsf{actx}\ [#1 \IfValueT{#2}{\mid #2}]}
\NewDocumentCommand\wfargctx{O{\Psi} d<> m m m}
  {#3 \gg #4 \rhd #5\ \mathsf{actx}\ [#1 \IfValueT{#2}{\mid #2}]}

\NewDocumentCommand\ceqargtm{O{\Psi} d<> m O{\K} m m m m}
  {\judg[#1]<#2>{#3 \rhd #4;#5 \vdash #6 \equiv #7 : #8}}
\NewDocumentCommand\eqargtm{O{\Psi} d<> m m O{\K} m m m m}
  {\judg[#1]<#2>{#3 \gg #4 \rhd #5;#6 \vdash #7 \equiv #8 : #9}}
\NewDocumentCommand\cwfargtm{O{\Psi} d<> m O{\K} m m m}
  {\judg[#1]<#2>{#3 \rhd #4;#5 \vdash #6 : #7}}
\NewDocumentCommand\wfargtm{O{\Psi} d<> m m O{\K} m m m}
  {\judg[#1]<#2>{#3 \gg #4 \rhd #5;#6 \vdash #7 : #8}}

\NewDocumentCommand\ceqcasespart{O{\Psi} d<> m m m m m}
  {\judg[#1]<#2>{#3 \rhd #4 \equiv #5 : #6 \rightharpoonup #7}}
\NewDocumentCommand\eqcasespart{O{\Psi} d<> m m m m m m}
  {\judg[#1]<#2>{#3 \rhd #4 \gg #5 \equiv #6 : #7 \rightharpoonup #8}}
\NewDocumentCommand\cwfcasespart{O{\Psi} d<> m m m m}
  {\judg[#1]<#2>{#3 \rhd #4 : #5 \rightharpoonup #6}}
\NewDocumentCommand\wfcasespart{O{\Psi} d<> m m m m m m}
  {\judg[#1]<#2>{#3 \rhd #4 \gg #5 : #6 \rightharpoonup #7}}

\NewDocumentCommand\ceqcases{O{\Psi} d<> m m m m m}
  {\judg[#1]<#2>{#3 \rhd #4 \equiv #5 : #6 \to #7}}
\NewDocumentCommand\eqcases{O{\Psi} d<> m m m m m m}
  {\judg[#1]<#2>{#3 \rhd #4 \gg #5 \equiv #6 : #7 \to #8}}
\NewDocumentCommand\cwfcases{O{\Psi} d<> m m m m}
  {\judg[#1]<#2>{#3 \rhd #4 : #5 \to #6}}
\NewDocumentCommand\wfcases{O{\Psi} d<> m m m m m m}
  {\judg[#1]<#2>{#3 \rhd #4 \gg #5 : #6 \to #7}}

\title{Computational Higher Type Theory IV: \\ Inductive Types}
\author{Evan Cavallo\thanks{\texttt{ecavallo@cs.cmu.edu}}\\Carnegie Mellon University
  \and Robert Harper\thanks{\texttt{rwh@cs.cmu.edu}}\\Carnegie Mellon University}

\date{July, 2018}

\begin{document}

\maketitle

\begin{abstract}
  This is the fourth in a series of papers extending Martin-L\"of's meaning explanation of dependent type
  theory to higher-dimensional types. In this installment, we show how to define cubical type systems
  supporting a general schema of \emph{indexed cubical inductive types} whose constructors may take dimension
  parameters and have a specified boundary. Using this schema, we are able to specify and implement many of
  the higher inductive types which have been postulated in homotopy type theory, including homotopy pushouts,
  the torus, $\W$-quotients, truncations, arbitrary localizations. By including indexed inductive types, we
  enable the definition of identity types.

  The addition of higher inductive types makes computational higher type theory a model of homotopy type
  theory, capable of interpreting almost all of the constructions in the HoTT Book \cite{hott-book} (with the
  exception of inductive-inductive types). This is the first such model with an explicit \emph{canonicity
    theorem}, which specifies the canonical values of higher inductive types and confirms that every term in
  an inductive type evaluates to such a value.
\end{abstract}

\section{Introduction}

Parts I-III of this series \cite{chtt-i,chtt-ii,chtt-iii} introduce \emph{computational higher type theory}
(CHiTT), a relational semantics for higher-dimensional type theory based on the Cartesian cube category (see
e.g., \cite{coquand14,brunerie14,awodey16,buchholtz17}). Cubical type theory centers around the notion of
\emph{path}, a feature of the judgmental apparatus which fills the role played by identity types in
Martin-L\"of's intensional type theory (ITT). This infrastructure is used in Part III to give a computational
interpretation of Voevodsky's \emph{univalence axiom}, which (phrased in cubical terms) asserts an equivalence
between equivalences $E \in A \simeq B$ and paths $P \in \Path{\UU}{A}{B}$ in the universe of types. The
univalence axiom is one component of \emph{homotopy type theory} (HoTT) \cite{hott-book}, an extension of ITT
which enables reasoning about types with higher-dimensional structure. This paper tackles the other component
of HoTT: \emph{higher inductive types}, which provide a way to defined types inductively generated by
higher-dimensional constructors. We develop a schema for \emph{cubical inductive types}, a cubical
reformulation of higher inductive types, which includes cubical equivalents of almost all commonly used higher
inductive types: the circle and torus, pushouts, localizations, and more. In implementing the instances of
this schema in a computational type theory, we specify the \emph{canonical values} of each CIT and obtain a
\emph{canonicity theorem}, which states that every element of a CIT can be evaluated to a canonical value for
that CIT.

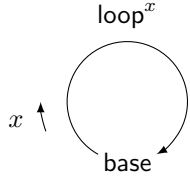
\begin{figure}
  \centering
  \[
    \begin{tikzpicture}
      \def \radius {0.8cm}
      \def \margin {29}
      \node at (-90:\radius) {$\base$};
      \node at (90:\radius+10) {$\lp{x}$};
      \draw[->, >=latex] ({270-\margin}:\radius) 
      arc ({270-\margin}:{-90+\margin}:\radius);
      \node at (190:{\radius+20}) {$x$};
      \draw[->, >=latex] (200:{\radius+10})
      arc (200:180:{\radius+10});
    \end{tikzpicture}
  \]
  \caption{The circle as a cubical inductive type.}
  \label{fig:circle}
\end{figure}

Like an ordinary inductive type, a cubical inductive type is one generated by a list of constructors. The
cubical case introduces two new features to constructors: \emph{dimension parameters} and a \emph{boundary}.
We think of a constructor taking $n$ dimension parameters as constructing an $n$-cube in a type, with its
boundary describing how it is attached to other elements of the type. The classic example, constructed already
in Part I of this series, is the presentation of a \emph{circle} shown in \cref{fig:circle}.

The circle $\C$ is generated by a 0-dimensional ``point'' constructor $\base$ and a 1-dimensional ``path''
constructor $\lp{r}$. The constructor $\lp{r}$ depends on a dimension parameter $r$, which we think of as
ranging over the interval from $0$ to $1$, and has specified boundaries $\lp{0} \steps \base$ and
$\lp{1} \steps \base$. Being inductively generated, the circle must support an eliminator which, given a point
in a type $A$ and a loop at that point, constructs a map from the circle into $A$. However, $\base$ and $\lp$
cannot be the only canonical values of type $\C$: CHiTT requires that types are closed under the \emph{Kan
  operations}, which include such operations as composition and inversion of paths. To give an operational
implementation of CITs, we must first find a set of values which includes such induced terms while maintaining
a reasonable canonicity guarantee.

Going beyond base types like the circle, we can use cubical inductive types to express homotopy-theoretic
constructions on existing types. For example, we can define the homotopy pushout \cite[\S6.8]{hott-book} of a
span of types $A \overset{F}{\leftarrow} C \overset{G}{\rightarrow} B$ as a CIT $\pushoutcl$ generated by
point constructors $\oft{a}{A} \gg \pleft{a} \in \pushoutcl$ and $\oft{b}{B} \gg \pright{b} \in \pushoutcl$
and a path constructor $\oft{c}{C} \gg \pglue{r}{c} \in \pushoutcl$ with boundaries
$\pglue{0}{P} \steps \pleft{F(P)}$ and $\pglue{1}{P} \steps \pright{G(P)}$. Using Part III's universes $\UKan$
of Kan types, we will be able to define a pushout type constructor as a \emph{parameterized CIT}:
\[
  \oft{A}{\UKan}, \oft{B}{\UKan}, \oft{C}{\UKan}, \oft{F}{\arr{C}{A}}, \oft{G}{\arr{C}{B}} \gg
  \pushoutcl{A}{B}{C}{F}{G} \in \UKan.
\]

The most intriguing CITs are those with \emph{recursive constructors}. Here, the traditional example is the
\emph{$(-1)$-truncation of a type} \cite[\S6.9]{hott-book}. Given a type $A$, its $(-1)$-truncation
$\Trunc{A}$ has all of the elements of $A$ and an additional path constructor which connects every pair of
elements of $\Trunc{A}$. Intuitively, $\Trunc{A}$ trivializes the homotopical structure of $A$. We can define
the $(-1)$-truncation $\Trunc{-}$ as a parameterized CIT $\oft{A}{\UKan} \gg \Trunc{A} \in \UKan$ with a point
constructor $\oft{a}{A} \gg \trpt{a} \in \Trunc{A}$ and a recursive path constructor
$\oft{t_0}{\Trunc{A}},\oft{t_0}{\Trunc{A}} \gg \trglue{x}{t_0}{t_1} \in \Trunc{A}$ with boundaries
$\trglue{0}{N_0}{N_1} \steps N_0$ and $\trglue{1}{N_0}{N_1} \steps N_1$. Here, the boundary of $\trglue$ is
given not by previously defined constructors, but by terms which are recursive arguments to $\trglue$. The
$(-1)$-truncation and its higher analogues, the $n$-truncations, are all examples of \emph{localizations}
\cite{shulman11}.

Finally, we can consider \emph{indexed cubical inductive types}. An indexed CIT is a family of types
simultaneously inductively generated by constructors which introduce elements at specified indices. These are
distinguished from the aforementioned parameterized inductive types, which introduce elements uniformly at all
indices. Even without higher-dimensional constructors, constructing indexed inductive types is non-trivial in
the higher setting. The central example is the \emph{identity family}
$\oft{a_0,a_1}{A} \gg \Id{A}{a_0}{a_1} \in \UKan$ generated by the reflexivity constructor
$\oft{a}{A} \gg \refl{a} \in \Id{A}{a}{a}$. The Kan operations ensure that every construction in CHiTT
respects paths, so there must be an element of $\Id{A}{M}{N}$ whenever there is a path from $M$ to $N$ in
$A$. As such, the identity family cannot contain only values of the form $\refl{M}$. As with CITs, indexed
inductive types thus require a re-examination of the canonical values of inductive types.

\paragraph{Outline}

In \cref{sec:chtt}, we review CHiTT following Part III, introducing the notions of cubical programming
language and cubical type system. In \cref{sec:schema}, we define a schema for specifying indexed cubical
inductive types. In \cref{sec:algebra}, we define what it means for a (cubical) relation on values to support
the constructors of an instance of the schema, and in \cref{sec:inductive} we define the inductive type for an
instance as the least relation supporting its constructors. In \cref{sec:rules}, we prove that this definition
supports introduction and elimination rules as well as the Kan operations.

\paragraph{Acknowledgments}

We thank Carlo Angiuli, Steve Awodey, Daniel Gratzer, Kuen-Bang Hou (Favonia), Dan Licata, Ed Morehouse,
Anders M\"ortberg, and Jonathan Sterling for their comments and insights,. Of course, this paper would not
exist without the previous installments \citet{chtt-i}, \citet{chtt-ii}, and \citet{chtt-iii}, and we are
indebted to the other lines of work on cubical type theories, particularly Brunerie and Licata
\cite{brunerie14,licata15} and Coquand et al.\ \cite{bch,cchm,coquand18}.

We gratefully acknowledge the support of the Air Force Office of Scientific Research through MURI grant
FA9550-15-1-0053. Any opinions, findings and conclusions or recommendations expressed in this material are
those of the authors and do not necessarily reflect the views of the AFOSR.


\section{Computational Higher Type Theory}
\label{sec:chtt}

This series studies \emph{cubical type systems}, which are systems for establishing properties of a
\emph{cubical operational semantics}. A cubical operational semantics consists of a grammar of programs,
cubical in the sense that they may contain dimension terms, and a deterministic set of rules explaining how to
execute such programs. These rules are specified by way of two judgments: $\isval{M}$ (``$M$ is a value'') and
$M \steps M'$ (``$M$ steps to $M'$''). A cubical type system, roughly speaking, is a collection of types,
where a type is a named (higher-dimensional) partial equivalence relation on values satisfying certain
conditions. This paper uses the definitions of type and cubical type system given in Part III; in this
section, we recapitulate the definitions necessary for our purposes.

\begin{notation}
  To maximize readability, we use two different notations for lists of terms as the situation demands. When we
  plan to repeatedly refer to indices explicitly, we write $\etc{M_i}$ for a list of terms
  $M_1,\ldots,M_n$. If not, we use the more compact notation $\lst{M}$ and write $\lst{M}[i]$ to select the
  $i$th element.
\end{notation}

\paragraph{Dimensions} A \emph{dimension term} is either 0,1, or one of a fixed set of \emph{dimension
  variables}. We use $r,s,t$ for dimension terms, $x,y,z$ for dimension variables, and $\Ge$ for $0$ or $1$. A
\emph{dimension context} $\Psi = (x_1,\ldots,x_n)$ is a list of dimension variables. We say that $\dimj{r}$
holds when $r \in \{0,1\} \cup \Psi$. We write $\fd{M}$ for the set of dimension variables occurring in a term
$M$, and say that $\tmj{M}$ when $\fd{M} \subseteq \Psi$. A \emph{dimension substitution} $\psitd$ assigns
some $\dimj[\Psi']{\psi(x)}$ to every $x \in \Psi$. Given $\dimj{r}$, we write
$\tds{\Psi}{\dsubst{}{r}{x}}{(\Psi,x)}$ for the substitution which replaces $x$ with $r$. Given $\psitd$ and
$\psitd[']$, we write $\tds{\Psi''}{\psi\psi'}{\Psi}$ for their composition. Given $\psitd$ and $\tmj{M}$, the
substituted term $\tmj[\Psi']{\td{M}{\psi}}$ is obtained by replacing each $x$ occurring in $M$ with
$\psi(x)$. We refer to the terms $\td{M}{\psi}$ for various $\psi$ as the \emph{aspects} of $M$. When
$\td{M}{\psi} \steps \td{M'}{\psi}$ for all $\psi$, we write $M \steps_\cube M'$ and say that $M$ stably steps
to $N$.

\paragraph{$\Psi$-Relations} To capture the denotation of every aspect of a type in context $\Psi$, we
introduce the notion of $\Psi$-relation. A \emph{$\Psi$-relation} $\Ga = (\Ga_\psi)_{\psitd}$ is a collection
of relations indexed by substitutions into $\Psi$, where, for each $\psitd$, the relation $\Ga_\psi$ relates
terms in context $\Psi'$. We abbreviate $\Ga_{\id}(M,M')$ as $\Ga(M,M')$. For $\psitd$, the $\Psi'$-relation
$\td{\Ga}{\psi}$ is defined by $(\td{\Ga}{\psi})_{\psi'} = \Ga_{\psi\psi'}$. A $\Psi$-relation is
\emph{stable} if $\Ga_\psi(M,M')$ implies $\Ga_{\psi\psi'}(\td{M}{\psi'},\td{M'}{\psi'})$ for all
$\psitd[']$. It is a \emph{$\Psi$-PER} when its components are transitive and symmetric. We will write
$\Ga_\psi\{M_1,\ldots,M_n\}$ to mean that $\Ga_\psi(M_i,M_j)$ holds for all $i,j \le n$; for $\Psi$-PERs,
where there is no possibility for confusion, we will simply write $\Ga_\psi(M_1,\ldots,M_n)$. A \emph{value
  $\Psi$-relation} is one which relates only values; these will serve as the denotations of types. We write
$\Vl{\Ga}$ for the restriction of a $\Psi$-relation $\Ga$ to values. $\Psi$-relations ordered by inclusion
form a complete lattice, as do value $\Psi$-relations ordered by inclusion. Given a monotone operator $\F$ on
one of these lattices, we write $\lfp{\F}$ and $\gfp{\F}$ for its least and greatest fixed-points
respectively.

Given a value $\Psi$-relation $\Ga$, we define the $\Psi$-relation $\Tm{\Ga}$, its coherent extension to
terms, by
\[
  \Tm{\Ga}_\psi(M,M') \iffdef \left\{
    \begin{array}{l}
      \forall \tdss{_1}{_1}{}, \tdss{_2}{_2}{_1}.\ \exists M_1,M_1',M_2,M_2',M_{12},M_{12}'. \\
      \td{M}{\psi_1} \evals M_1 \land \td{M_1}{\psi_2} \evals M_2 \land \td{M}{\psi_1\psi_2} \evals M_{12} \land { } \\
      \td{M'}{\psi_1} \evals M_1' \land \td{M_1'}{\psi_2'} \evals M_2' \land \td{M'}{\psi_1\psi_2} \evals M_{12}' \land { } \\
      \Ga_{\psi\psi_1\psi_2}(M_2,M_2') \land \Ga_{\psi\psi_1\psi_2}\{M_2,M_{12}\} \land \Ga_{\psi\psi_1\psi_2}\{M_2',M_{12}'\}
    \end{array}
  \right.
\]
We will interact with $\Tm$ through an interface of lemmas which we prove in \cref{app:lemmas}. The basic
intuition is that $\Tm{\Ga}_\psi(M,M')$ holds when, for any pair of dimension substitutions $\psi_1,\psi_2$,
$M$ and $M'$ compute to values related by $\Ga$ no matter how these substitutions are interleaved with
evaluation.  If $\Ga$ is a $\Psi$-PER, then so is $\Tm{\Ga}$, and $\Tm{\Ga}$ is always stable. A
$\Psi$-relation is \emph{value-coherent} if $\Ga_\psi(V,V')$ implies $\Tm{\Ga}_\psi(V,V')$ for all
$\psi,V,V'$. This condition, which we will impose on all types, implies the following essential property:

\begin{replemma}{lem:value-coherent-evals}
  Let $\Ga$ be a value-coherent $\Psi$-PER. For any $\Tm{\Ga}_\psi(M)$, we have $M \evals V$ and
  $\Tm{\Ga}_\psi(M,V)$.
\end{replemma}
The following lemma is used to prove introduction rules.

\begin{replemma}{lem:introduction}[Introduction]
  Let $\Ga$ be a value $\Psi$-PER. If for all $\psitd$, either $\Ga_\psi(\td{M}{\psi},\td{M'}{\psi})$ or
  $\Tm{\Ga}_\psi(\td{M}{\psi},\td{M'}{\psi})$, then $\Tm{\Ga}(M,M')$.
\end{replemma}

The next, a head expansion lemma, is used to prove computation rules, both for eliminators and for the
boundaries of introduction forms. Roughly, if a term $M'$ is in $\Ga$, and a term $M$ steps to $M'$ at all
aspects modulo equality in $\Ga$, then $M$ and $M'$ are equal in $\Ga$.

\begin{replemma}{lem:expansion}[Coherent expansion]
  Let $\Ga$ be a value $\Psi$-PER and let $\tmj{M,M'}$. If for all $\psitd$, there exists $M''$ such that
  $\td{M}{\psi} \msteps M''$ and $\Tm{\Ga}_\psi(M'',\td{M'}{\psi})$, then $\Tm{\Ga}(M,M')$.
\end{replemma}

A \emph{constraint} $\xi = (r,r')$ specifies an equation on dimension terms; we say that $\models \xi$ holds
for $\xi = (r,r')$ if $r = r'$. Henceforth, We will write constraints as $r = r'$ rather than $(r,r')$. A
\emph{constraint context} $\Xi = (\xi_1,\ldots,\xi_n)$ is an ordered list of constraints; we say that
$\models \Xi$ holds if $\models \xi$ holds for all $\xi \in \Xi$. We say $\Xi$ is \emph{valid} if either
$\models \xi$ for some $\xi \in \Xi$ or there exists some $r$ such that both $(r = 0) \in \Xi$ and
$(r = 1) \in \Xi$. This technical condition was introduced in Part III in order to ensure certain canonicity
properties of zero-dimensional terms. Validity of a constraint context $\Xi$ is a conservative approximation
of the property that for all closing substitutions $\tds{\cdot}{\psi}{\Psi}$ we have $\models \td{\xi}{\psi}$
for some $\xi \in \Xi$.  For a $\Psi$-relation $\Ga$, $\psitd$, and a constraint context $\Xi$ with
$\fd{\Xi} \subseteq \Psi'$, we define
$\Ga_{\psi \mid \Xi}(M,M') \iffdef \forall \psitd[']. (\models \td{\Xi}{\psi'} \implies
\Ga_{\psi\psi'}(M,M'))$. For a $\Psi$-relation $\Ga$ and $\Xi$ with $\fd{\Xi} \subseteq \Psi$, we define a
$\Psi$-relation $(\Ga \mid \Xi)$ by $(\Ga \mid \Xi)_\psi(M,M') \iffdef \Ga_{\psi \mid
  \td{\Xi}{\psi}}(M,M')$. It is convenient to have a variant of the head expansion lemma for restricted
relations.

\begin{repcorollary}{cor:restricted-expansion}[Restricted expansion]
  Let $\Ga$ be a value $\Psi$-PER and $\Xi$ be a constraint context.  Let $\Ga$ be a value $\Psi$-PER and let
  $\tmj{M,M'}$. If for all $\psitd$ with $\models \td{\Xi}{\psi}$, there exists $M''$ such that
  $\td{M}{\psi} \msteps M''$ and $\Tm{\Ga}_\psi(M'',\td{M'}{\psi})$, then $(\Tm{\Ga} \mid \Xi)(M,M')$.
\end{repcorollary}

For elimination rules, we need a notion of a dependent $\Psi$-relation. Given a $\Psi$-relation $\Ga$, we say
that $\Gb = \ix{\Gb_{-}}{-}$ is an \emph{$\Ga$-indexed $\Psi$-relation} when $\ix{\Gb_\psi}{M}$ is a
$\Psi$-relation for all $\psitd$ and $\Ga_\psi\{M\}$, and
\begin{enumerate}
\item $\td{\ix{\Gb_\psi}{M}}{\psi'} = \ix{\Gb_{\psi\psi'}}{\td{M}{\psi'}}$ for all $\psitd$, $\psitd[']$, and
  $\Ga_{\psi}\{M\}$, and
\item $\ix{\Gb_\psi}{M} = \ix{\Gb_\psi}{M'}$ for all $\psitd$ and $\Ga_\psi\{M,M'\}$.
\end{enumerate}
Again, we will abbreviate $\ix{\Gb_{\id}}{M}$ as $\ix{\Gb}{M}$. If $\Ga$ is a $\Psi$-relation, $\Gb$ is an
$\Ga$-indexed $\Psi$-relation, and $\open{a}{\tmj{N,N'}}$, we write $\relseq{a:\Ga}{\ix{\Gb}{a}(N,N')}$ to
mean that $\ix{\Gb_{\psi}}{M}(\subst{\td{N}{\psi}}{M}{a},\subst{\td{N'}{\psi}}{M'}{a})$ holds for all $\psitd$
and $\Ga_\psi(M,M')$.

When we prove elimination rules, we can reduce the problem of proving the eliminator is well-typed on terms to
proving it is well-typed on values if the eliminator is \emph{eager}.

\begin{repdefinition}{def:eager}
  We say that $\open{a}{\tmj{N}}$ is \emph{eager} if for all $\psitd$ and $\tmj[\Psi']{M}$, we have
  $\subst{\td{N}{\psi}}{M}{a} \evals W$ iff there exists $\tmj[\Psi']{V}$ such that $M \evals V$ and
  $\subst{\td{N}{\psi}}{V}{a} \evals W$.
\end{repdefinition}

\begin{replemma}{lem:elimination}[Elimination]
  Let $\Ga$ be a value-coherent $\Psi$-PER and $\Gb$ be a value $\Psi$-PER over $\Tm{\Ga}$. Suppose
  $\relseq{a:\Gg}{\ix{\Tm{\Gb}}{a}(N,N')}$ for some $\Gg \subseteq \Ga$. If $\open{a}{\tmj{N,N'}}$ are eager,
  then $\relseq{a:\Tm{\Gg}}{\ix{\Tm{\Gb}}{a}(N,N')}$.
\end{replemma}

\paragraph{Cubical type systems}

Per Part III, a \emph{candidate cubical type system} $\tau$ is a relation $\tau(\Psi,A_0,A_0',\phi)$ ranging
over dimension contexts $\Psi$, term values $\tmj{A_0,A_0'}$, and relations $\phi$ on values in context
$\Psi$. We think of $\tau(\Psi,A_0,A_0',\phi)$ as saying that the $\Psi$-terms $A_0$ and $A_0'$ are equal
names for the value relation $\phi$. In the same way that $\Tm$ extends value $\Psi$-relations to terms, we
have an operator $\PTy$ extending a candidate cubical type system from values and relations to terms and
$\Psi$-relations. If $\tau$ is a cubical type system, then $\PTy{\tau}(\Psi,A,A',\Ga)$ is a relation ranging
over contexts $\Psi$, terms $\tmj{A,A'}$, and value $\Psi$-relations $\Ga$, defined by
\[
  \PTy{\tau}(\Psi,A,A',\Ga) \iffdef
  \left\{
    \begin{array}{l}
      \forall \tdss{_1}{_1}{}, \tdss{_2}{_2}{_1}.\ \exists A_1,A_1',A_2,A_2',A_{12},A_{12}'. \\
      \td{A}{\psi_1} \evals A_1 \land \td{A_1}{\psi_2} \evals A_2 \land \td{A}{\psi_1\psi_2} \evals A_{12} \land { } \\
      \td{A'}{\psi_1} \evals A_1' \land \td{A_1'}{\psi_2'} \evals A_2' \land \td{A'}{\psi_1\psi_2} \evals A_{12}' \land { } \\
      \tau(\Psi_2,A_2,A_2',\Ga_{\psi_1\psi_2}) \land \tau(\Psi_2,\{A_2,A_{12}\},\Ga_{\psi_1\psi_2}) \land \tau(\Psi_2,\{A_2',A_{12}'\},\Ga_{\psi_1\psi_2})
    \end{array}
  \right.
\]
where $\tau(\Psi,\{A_1,\ldots,A_n\},\phi)$ is defined to hold when $\tau(\Psi,A_i,A_j,\phi)$ holds for all
$i,j \le n$. The intent is for $\PTy{\tau}(\Psi,A,A',\Ga)$ to hold when the aspects of $A$ and $A'$ coherently
name corresponding aspects of the $\Psi$-relation $\Ga$. For our purposes, we will only need the following
fact.

\begin{proposition}
  \label{prop:pty-stable-introduction}
  If $\tau(\Psi',\td{A}{\psi},\td{A'}{\psi},\Ga_\psi)$ holds for all $\psitd$, then $\PTy{\tau}(\Psi,A,A',\Ga)$
  holds.
\end{proposition}

As with $\Tm$, $\PTy{\tau}$ is stable: $\PTy{\tau}(\Psi,A,A',\Ga)$ implies
$\PTy{\tau}(\Psi',\td{A}{\psi},\td{A'}{\psi},\td{\Ga}{\psi})$.

\begin{definition}
  \label{def:cubical type system}
  A \emph{cubical type system} is a candidate
  cubical type system such that
  \begin{enumerate}
  \item if $\tau(\Psi,A,A',\phi)$ and $\tau(\Psi,A,A',\phi')$, then $\phi = \phi'$,
  \item if $\tau(\Psi,A,A',\phi)$, then $\phi$ is a partial equivalence relation,
  \item $\tau(\Psi,-,-,\phi)$ is a partial equivalence relation for all $\Psi$ and $\phi$,
  \item if $\tau(\Psi,A_0,A_0',\phi)$, then $\PTy{\tau}(\Psi,A_0,A_0',\Ga)$ for some $\Ga$.
  \end{enumerate}
\end{definition}
Fixing a candidate cubical type system $\tau$, we can define the central judgments of computational higher
type theory relative to $\tau$. The judgment $\tau \models \ceqtypep{A}{A'}$ is defined to hold when
$\PTy{\tau}(\Psi,A,A',\Ga)$ for a value-coherent $\Psi$-relation $\Ga$. We abbreviate
$\tau \models \ceqtypep{A}{A}$ as $\tau \models \cwftypep{A}$.  We write $\vper{A}$ for the $\Psi$-relation
$\Ga$ above, which is unique when it exists. Presupposing $\tau \models \cwftypep{A}$, the judgment
$\tau \models \ceqtm{M}{M'}{A}$ is defined to hold when $\Tm{\vper{A}}(M,M')$ holds. We abbreviate
$\ceqtm{M}{M}{A}$ as $\tau \models \coftype{M}{A}$. As with relations, we also define restricted versions of
these judgments:
\begin{enumerate}
\item $\tau \models \ceqtypep<\Xi>{A}{A'}$ holds when
  $\tau \models \ceqtypep[\Psi']{\td{A}{\psi}}{\td{A'}{\psi}}$ for all $\psitd$ with $\models \td{\Xi}{\psi}$,
\item Presupposing $\tau \models \cwftypep<\Xi>{A}$, $\tau \models \ceqtm<\Xi>{M}{M'}{A}$ holds when
  $\tau \models \ceqtm[\Psi']{\td{M}{\psi}}{\td{M'}{\psi}}{\td{A}{\psi}}$ for all $\psitd$ with
  $\models \td{\Xi}{\psi}$.
\end{enumerate}
We refer to these $\eq$ judgments as \emph{exact equality} judgments so as to distinguish them from identity
types, which are sometimes called equality types in the literature. When the candidate cubical type system is
understood, we will drop the prefix $\tau \models$ from the judgments above.

A pretype $A$ is a \emph{(Kan) type} when it satisfies the five \emph{Kan conditions}, which require that $A$
supports well-defined composition and coercion operators. The first three Kan conditions concern the
homogeneous composition operator $\hcom$, and the last two concern the coercion operator $\coe$.

A pretype $A$ is $\hcom$-Kan when the operator $\hcom{A}$ implements a homogeneous composition operation for
$A$. Given endpoints $r,r'$ and collection of \emph{tube faces} $\sys{\xi_i}{y.N_i}$ in $A$, homogeneous
composition takes a \emph{cap} $M$ in $A$ which lines up with each term $\dsubst{N_i}{r}{y}$ under the
corresponding constraint $\xi_i$, and constructs a \emph{composite} $\hcom{A}{r}{r'}{M}{\sys{\xi_i}{y.N_i}}$
which lines up with each term $\dsubst{N_i}{r'}{y}$ under $\xi_i$. We imagine $\hcom$ as sliding $M$ from
$y = r$ to $y = r'$ within the ``tube'' created by the terms $\sys{\xi_i}{y.N_i}$. As such, we require that
$\hcom{A}{r}{r'}{M}{\sys{\xi_i}{y.N_i}}$ be equal to $M$ when $r = r'$. As an example, considering the term
$\hcom{A}{0}{y}{M}{\tube{x=0}{y.N_0},\tube{x=1}{y.N_1}}$, we have a diagram
\[
\begin{tikzpicture}
  \draw (0 , 2) [->] to node [above] {\small $x$} (0.5 , 2) ;
  \draw (0 , 2) [->] to node [left] {\small $y$} (0 , 1.5) ;
  \node (tl) at (1.5 , 2) {$\cdot$} ;
  \node (tr) at (5.5 , 2) {$\cdot$} ;
  \node (bl) at (1.5 , 0) {$\cdot$} ;
  \node (br) at (5.5 , 0) {$\cdot$} ;
  \draw (tl) [->] to node [above] {$M$} (tr) ;
  \draw (tl) [->] to node [left] {$N_0$} (bl) ;
  \draw (tr) [->] to node [right] {$N_1$} (br) ;
  \draw (bl) [->,dashed] to node [below] {$\hcom{A}{0}{1}{M}{\cdots}$} (br) ;
  \node at (3.5 , 1) {$\hcom{A}{0}{y}{M}{\cdots}$} ;
\end{tikzpicture}
\]
in $A$. The bottom face of this square can be viewed as the composite path formed by concatenating (a) the
inverse of $N_0$, (b) $M$, and (c) $N_1$.

\begin{definition} Given $\tmj{A,A'}$ and a value $\Psi$-PER $\Ga$, we say that $(A,A',\Ga)$ are \emph{equally
$\hcom$-Kan} if for all $\psitd$, $\dimj[\Psi']{r,r'}$, valid constraint contexts $\Xi = \etc{\xi_i}$, and
  \begin{enumerate}[label=(\alph*)]
  \item $\Tm{\Ga}_\psi(M,M')$,
  \item $\Tm{\Ga}_{\psi;y \mid \xi_i,\xi_j}(N_i,N_j')$ for all $i,j$,
  \item $\Tm{\Ga}_{\psi \mid \xi_i}(\dsubst{N_i}{r}{y},M)$ for all $i$,
  \end{enumerate} we have
  \begin{description}
  \item{K1.} $\Tm{\Ga}_\psi(\hcom*{\td{A}{\psi}}{\xi_i},\hcom{\td{A'}{\psi}}{r}{r'}{M'}{\sys{\xi_i}{y.N'_i}})$,
  \item{K2.} $\Tm{\Ga}_\psi(\hcom{\td{A}{\psi}}{r}{r}{M}{\sys{\xi_i}{y.N_i}},M)$,
  \item{K3.} $\Tm{\Ga}_{\psi \mid \xi_i}(\hcom*{\td{A}{\psi}}{\xi_i},\dsubst{N_i}{r'}{y})$ for all $i$.
  \end{description}
\end{definition}

A pretype $A$ is $\coe$-Kan when the operator $\coe{y.\td{A}{\psi}}$ implements \emph{coercion} for every
$\tds{(\Psi',y)}{\psi}{\Psi}$. Coercion transports elements from one aspect of $A$ to another: if $M$ is an
element of $\dsubst{\td{A}{\psi}}{r}{y}$, then $\coe{y.\td{A}{\psi}}{r}{r'}{M}$ is an element of
$\dsubst{\td{A}{\psi}}{r'}{y}$. Naturally, we require that $\coe{y.\td{A}{\psi}}{r}{r}{M}$ is equal to $M$
itself.

\begin{definition} We say that $(A,A',\Ga)$ are \emph{equally $\coe$-Kan} if for all
$\tds{(\Psi',y)}{\psi}{\Psi}$, $\dimj[\Psi']{r,r'}$, and $\Tm{\Ga}_{\dsubst{\psi}{r}{y}}(M,M')$, we have
  \begin{description}
  \item{K4.} $\Tm{\Ga}_{\dsubst{\psi}{r'}{y}}(\coe{y.A}{r}{r'}{M},\coe{y.A'}{r}{r'}{M'})$,
  \item{K5.} $\Tm{\Ga}_{\dsubst{\psi}{r}{y}}(\coe{y.A}{r}{r}{M},M)$.
  \end{description}
\end{definition}

We say that $(A,A',\Ga)$ are \emph{equally Kan} when they are equally $\hcom$-Kan and equally
$\coe$-Kan. Presupposing $\ceqtypep{A}{A'}$, the judgment $\ceqtypek{A}{A'}$ is defined to hold when
$(A,A',\vper{A})$ are equally Kan.  Using the operators $\hcom$ and $\coe$, we can define a
\emph{heterogeneous composition} operator $\com$, which composes along a type line:
\[ \com{y.A}{r}{r'}{M}{\sys{\xi_i}{y.N_i}} \steps
\hcom{\dsubst{A}{r'}{y}}{r}{r'}{\coe{y.A}{r}{r'}{M}}{\sys{\xi_i}{y.\coe{y.A}{y}{r'}{N_i}}}.
\]

\begin{proposition}
  \label{prop:com}
  If $(A,A',\Ga)$ are equally Kan, then for all $\tds{(\Psi',y)}{\psi}{\Psi}$, $\dimj[\Psi']{r,r'}$, valid
  constraint contexts $\Xi = \etc{\xi_i}$, and
  \begin{enumerate}[label=(\alph*)]
  \item $\Tm{\Ga}_{\dsubst{\psi}{r}{y}}(M,M')$,
  \item $\Tm{\Ga}_{\psi \mid \xi_i,\xi_j}(N_i,N_j')$ for all $i,j$,
  \item $\Tm{\Ga}_{\dsubst{\psi}{r}{y} \mid \xi_i}(\dsubst{N_i}{r}{y},M)$ for all $i$,
  \end{enumerate} we have
  \begin{enumerate}
  \item
$\Tm{\Ga}_{\dsubst{\psi}{r'}{y}}(\com*{y.\td{A}{\psi}}{\xi_i},\com{y.\td{A'}{\psi}}{r}{r'}{M'}{\sys{\xi_i}{y.N'_i}})$,
  \item $\Tm{\Ga}_{\dsubst{\psi}{r}{y}}(\com*{y.\td{A}{\psi}}{r}{r}{M}{\sys{\xi_i}{y.N_i}},M)$,
  \item $\Tm{\Ga}_{\dsubst{\psi}{r'}{y} \mid \xi_i}(\com*{y.\td{A}{\psi}}{\xi_i},\dsubst{N_i}{r'}{y})$ for all
$i$.
  \end{enumerate}
\end{proposition}

Finally, the judgments on closed terms are extended to open term judgments defined by simultaneous induction
on context length:
\begin{itemize}
\item $\tau \models \ceqctx{\kappa}{\GG}{\GG'}$ is defined to hold for $\GG = \etc{\oft{a_i}{A_i}}$ and
  $\GG' = \etc{\oft{a_i}{A'_i}}$ when \\
  $\tau \models \eqtypex{\oft{a_1}{A_1},\ldots,\oft{a_{i-1}}{A_{i-1}}}{A_i}{A'_i}$ holds for all $i$,
\item Presupposing $\tau \models \cwfctx{pre}{\GG}$ with $\GG = \etc{\oft{a_i}{A_i}}$,
  $\tau \models \ceqtm{\etc{M_i}}{\etc{M'_i}}{\GG}$ is defined to hold when \\
  $\tau \models \ceqtm{M_i}{M'_i}{\subst{A_i}{M_1,\ldots,M_{i-1}}{a_1,\ldots,a_{i-1}}}$ holds for all $i$,
\item Presupposing $\tau \models \cwfctx{pre}{\GG}$, $\tau \models \eqtypex{\GG}{A}{A'}$ is defined to hold when \\
  $\tau \models
  \ceqtypex[\Psi']{\subst{\td{A}{\psi}}{\etc{M_i}}{\etc{a_i}}}{\subst{\td{A'}{\psi}}{\etc{M'_i}}{\etc{a_i}}}$
  holds for all $\psitd$ and $\tau \models \ceqtm[\Psi']{\etc{M_i}}{\etc{M'_i}}{\td{\GG}{\psi}}$.
\item Presupposing $\tau \models \cwfctx{pre}{\GG}$, $\tau \models \eqtm{\GG}{N}{N'}{A}$ is defined to hold
  when \\
  $\tau \models
  \ceqtm[\Psi']{\subst{\td{N}{\psi}}{\etc{M_i}}{\etc{a_i}}}{\subst{\td{N'}{\psi}}{\etc{M_i}}{\etc{a_i}}}{\subst{\td{A}{\psi}}{\etc{M_i}}{\etc{a_i}}}$
  for all $\psitd$ and $\tau \models \ceqtm[\Psi']{\etc{M_i}}{\etc{M'_i}}{\td{\GG}{\psi}}$.
\end{itemize}


\section{Schema}
\label{sec:schema}

In this section, we define our schema for specifying indexed cubical inductive types. The schema is defined
by the five judgments shown below.

\begin{center}
  \begin{tabular}{ll}
    $\ceqconstrs{\GD}{\K}{\K'}$ & list of constructors \\
    $\ceqconstr{\GD}{\Co}{\Co'}$ & constructor \\
    $\ceqargtype{\GD}{\sch{a}}{\sch{a}'}$ & argument type \\
    $\ceqargctx{\GD}{\GTh}{\GTh'}$ & argument context \\
    $\ceqargtm{\GD}{\GTh}{\sch{m}}{\sch{m}'}{\sch{a}}$ & boundary term \\
  \end{tabular}
\end{center}
The central judgment, $\ceqconstrs{\GD}{\K}{\K'}$, specifies that $\K$ and $\K'$ are equal specifications for
a cubical inductive type indexed in the context $\GD$. The judgment $\ceqconstr{\GD}{\Co}{\Co'}$ states that
$\Co$ and $\Co'$ are equal constructors in the context of a previously defined list of constructors $\K$. The
judgments $\ceqargtype{\GD}{\sch{a}}{\sch{a}'}$ and $\ceqargtm{\GD}{\GTh}{\sch{m}}{\sch{m}'}{\sch{a}}$
constitute the type theory of \emph{argument types} and \emph{boundary terms}, which are used to specify the
types of recursive arguments to each constructor and the reduction behavior of the constructor when specified
equations hold.

These judgments are extended to the open forms $\eqconstrs{\GG}{\GD}{\K}{\K'}$,
$\eqconstr{\GG}{\GD}{\Co}{\Co'}$, $\eqargtype{\GG}{\GD}{\sch{a}}{\sch{a}'}$,
$\eqargctx{\GG}{\GD}{\GTh}{\GTh'}$, and $\eqargtm{\GG}{\GD}{\GTh}{\sch{m}}{\sch{m}'}{\sch{a}}$ by
functionality in the usual fashion: for example, $\eqconstrs{\ofc{\Gg}{\GG}}{\GD}{\K}{\K'}$ is defined to hold
when
$\ceqconstrs{\subst{\td{\GD}{\psi}}{\lst{M}}{\Gg}}{\subst{\td{\K}{\psi}}{\lst{M}}{\Gg}}{\subst{\td{\K'}{\psi}}{\lst{M}}{\Gg}}$
holds for all $\psitd$ and $\ceqtm[\Psi']{\lst{M}}{\lst{M}'}{\td{\GG}{\psi}}$.

We use $\GG,\GD$ and $\Gg,\Gd,\Gh,\Gr$ for ordinary term contexts and context variables (i.e., lists of term
variables), and $\GTh,\Gf$ and $\Gth,\Gf$ for argument contexts and argument context variables. We reserve
$\formal{p},\formal{q}$ for boundary term variables; other letters denote ordinary term variables.

\begin{definition}
  The grammars of constructor lists, constructors, argument types, argument
  contexts, and boundary terms are given by
  \[
    \begin{array}{rcl}
      \K &::=& \nilconstrs \mid \snocconstrs{\K}{\ell : \Co} \\
      \Co &::=& \constr{\lst{x}}{\GG}{\Gg.\lst{M}}{\Gg.\GTh}{\sys{\xi_k}{\Gg.\Gth.\sch{m}_k}} \; \text{where $|\GG| = |\Gg|$ and $|\GTh| = |\Gth|$} \\
      \sch{a} &::=& \argvar{\lst{M}} \mid \argpi{a}{A}{\sch{a}} \\
      \GTh &::=& \cdot \mid \GTh,\ofa{p}{\sch{a}} \\
      \sch{m} &::=& \bndintro{\lst{r}}{\lst{M}}{\lst{\sch{m}}} \mid \bndfhcom{\lst{M}}{r}{r'}{\sch{m}}{\sys{\xi_i}{y.\sch{m}_i}} \mid \bndfcoe{z.\lst{M}}{r}{r'}{\sch{m}} \mid \bndlam{a}{\sch{m}} \mid \bndapp{\sch{m}}{M}.
    \end{array}
  \]
  Labels $\ell$ are drawn from a fixed set $\Label$. 
\end{definition}

\begin{definition}
  \label{def:schema}
  Fix an index type $\cwfctxk{\GD}$. We define the schema judgments mutually inductively as follows.
  \begin{enumerate}[label=\textbf{\Alph*.}]
  \item The judgment $\ceqconstrs{\GD}{\K}{\K'}$ is defined inductively by the following rules.
    \begin{mathpar}
      \Infer
      { }
      {\ceqconstrs{\GD}{\nilconstrs}{\nilconstrs}}
      \and
      \Infer
      {\ceqconstrs{\GD}{\K}{\K'}
        \\ \ell \not\in \K
        \\ \ceqconstr{\GD}{\Co}{\Co'}}
      {\ceqconstrs{\GD}{\snocconstrs{\K}{\ell:\Co}}{\snocconstrs{\K'}{\ell:\Co'}}}
    \end{mathpar}
    A constructor list is thus, appropriately, a list of constructors, each of which may mention the
    constructors which precede it in its specification. For a constructor list $\K$, we write $\ell \in \K$ to
    mean that $\ell$ labels a constructor in $\K$; we write $\K[\ell]$ for that constructor data and
    $\K_{<\ell}$ for the prefix of $\K$ preceding $\ell.$ We define $\height{\K}{\ell}$ for $\ell \in \K$ to
    be the index at which $\ell$ appears in $\K$, and set
    $\height{\K}{\sch{m}} \eqdef \max(\{-1\} \cup \{ \height{\K}{\ell} : \ell \in \fl{\sch{m}} \})$.

  \item
    Presupposing $\cwfconstrs{\GD}{\K}$, the judgment $\ceqconstr{\GD}{\Co}{\Co'}$ is
    defined to hold when
    $\Co = \constr{\lst{x}}{\GG}{\Gg.\lst{I}}{\Gg.\GTh}{\sys{\xi_k}{\Gg.\Gth.\sch{m}_k}}$ and
    $\Co' = \constr{\lst{x}}{\GG'}{\Gg.\lst{I}'}{\Gg.\GTh'}{\sys{\xi_k}{\Gg.\Gth.\sch{m}_k'}}$ where
    \begin{enumerate}
    \item $\ceqctxk{\GG}{\GG'}$,
    \item $\eqtm{\oft{\Gg}{\GG}}{\lst{I}}{\lst{I}'}{\GD}$,
    \item $\eqargctx{\ofc{\Gg}{\GG}}{\GD}{\GTh}{\GTh'}$,
    \item $\fd{\etc{\xi_k}} \subseteq \{\lst{x}\}$ and $\etc{\xi_k}$ is either empty or valid,
    \item
      $\eqargtm[\Psi,\lst{x}]<\xi_k,\xi_l>{\ofc{\Gg}{\GG}}{\GD}{\ofc{\Gth}{\GTh}}{\sch{m}_k}{\sch{m}'_l}{\argvar{\lst{I}}}$
      for all $k,l$.
    \end{enumerate}
    The list $\lst{x}$ indicates the dimension parameters to the constructor. The context $\GG$ describes the
    types of its non-recursive arguments. The terms $\lst{I}$, which may depend on the variables in $\GG$,
    specify the index in $\GD$ where the constructor lands. The argument context $\GTh$ specifies the
    recursive arguments to the constructor. The constraints $\xi_i$ specify the shape of the constructor's
    boundary, and the corresponding terms $\sch{m}_i$ specify the reduction behavior of the constructor at
    each constraint in terms of its arguments.
  \item The argument type equality judgment $\ceqargtype{\GD}{\sch{a}}{\sch{a}'}$ is inductively defined by
    the following rules.
    \begin{mathpar}
      \Infer
      {\ceqtm{\lst{I}}{\lst{I}'}{\GD}}
      {\ceqargtype{\GD}{\argvar{\lst{I}}}{\argvar{\lst{I}'}}}
      \and
      \Infer
      {\ceqtypek{A}{A'} \\ \eqargtype{\oft{a}{A}}{\GD}{\sch{b}}{\sch{b}'}}
      {\ceqargtype{\GD}{\argpi{a}{A}{\sch{b}}}{\argpi{a}{A'}{\sch{b}'}}}
    \end{mathpar}
    As usual, the judgment $\ceqargtype<\Xi>{\GD}{\sch{a}}{\sch{a}'}$ is defined to hold when
    $\ceqargtype[\Psi']{\td{\GD}{\psi}}{\td{\sch{a}}{\psi}}{\td{\sch{a}'}{\psi}}$ holds for all $\psitd$ such
    that $\models \td{\Xi}{\psi}$. The argument type $\argvar{\lst{I}}$ is the recursive reference to index
    $\lst{I}$ of the inductive type being defined, while the types $\argpi{a}{A}{\sch{b}}$ allow recursive
    arguments parameterized by a (non-recursive) Kan type.
  \item For $\GTh = \etc{\ofa{\formal{p}_i}{\sch{a}_i}}$ and $\GTh' = \etc{\ofa{\formal{p}_i}{\sch{a}'_i}}$,
    the argument context equality judgment $\ceqargctx{\GD}{\GTh}{\GTh'}$ is defined to hold when
    $\ceqargtype{\GD}{\sch{a}_i}{\sch{a}'_i}$ holds for all $i$.
  \item Presupposing $\cwfconstrs{\GD}{\K}$, $\cwfargctx{\GD}{\GTh}$, and $\cwfargtype{\GD}{\sch{a}}$, the
    boundary term equality judgment $\ceqargtm{\GD}{\GTh}{\sch{m}}{\sch{m}'}{\sch{a}}$ is inductively defined
    by the rules shown in \cref{fig:boundary-term-rules}. Elements of the argument type $\argvar{\lst{I}}$ can
    be constructor terms as well as $\bndfhcom$ and $\bndfcoe$ terms, which we will discuss in more detail
    later on. The function type is inhabited by $\formal{\lambda}$-terms, and we can eliminate from it via
    function application.

    Note that this is an \emph{inductive} definition of well-typed terms in argument context $\GTh$; it
    judgment form is \emph{not} defined from the closed judgment form
    $\ceqargtm{\GD}{\cdot}{\sch{m}}{\sch{m}'}{\sch{a}}$ by functionality. On the other hand, the judgment form
    $\eqargtm{\GG}{\GD}{\GTh}{\sch{m}}{\sch{m}'}{\sch{a}}$ \emph{is} defined in terms of the form
    $\ceqargtm{\GD}{\GTh}{\sch{m}}{\sch{m}'}{\sch{a}}$ by functionality; where the ``$\GG$-open'' form occurs
    in \cref{fig:boundary-term-rules}, one should imagine it replaced by its definition.
  \end{enumerate}
\end{definition}

\begin{figure}

\begin{mdframed}
  
  \paragraph{Constructors}
  \begin{mathpar}
    \Infer[$\bndintro$-I]
    {\K[\ell] = \constr{\lst{x}}{\GG}{\Gg.\lst{I}}{\Gg.\GF}{\sys{\xi_k}{\Gg.\Gf.\sch{m}_k}} \\\\
      \dimj{\lst{r}} \\
      \ceqtm{\lst{P}}{\lst{P}'}{\GG} \\
      \ceqargtm{\GD}{\GTh}{\lst{\sch{n}}}{\lst{\sch{n}}'}{\subst{\GF}{\lst{P}}{\Gg}}}
    {\ceqargtm{\GD}{\GTh}{\bndintro{\lst{r}}{\lst{P}}{\lst{\sch{n}}}}{\bndintro{\lst{r}}{\lst{P}'}{\lst{\sch{n}}'}}{\argvar{\subst{\lst{I}}{\lst{P}}{\Gg}}}}
    \and
    \Infer[$\bndintro$-B]
    {\K[\ell] = \constr{\lst{x}}{\GG}{\Gg.\lst{I}}{\Gg.\GF}{\sys{\xi_k}{\Gg.\Gf.\sch{m}_k}} \\\\
      \dimj{\lst{r}} \\
      \models \dsubst{\xi_k}{\lst{r}}{\lst{x}} \\
      \coftype{\lst{P}}{\GG} \\
      \cwfargtm{\GD}{\GTh}{\lst{\sch{n}}}{\subst{\GF}{\lst{P}}{\Gg}}}
    {\ceqargtm{\GD}{\GTh}{\bndintro{\lst{r}}{\lst{P}}{\lst{\sch{n}}}}{\subst{\subst{\dsubst{\sch{m}_k}{\lst{r}}{\lst{x}}}{\lst{P}}{\Gg}}{\lst{\sch{n}}}{\Gf}}{\argvar{\subst{\lst{I}}{\lst{P}}{\Gg}}}}
  \end{mathpar}
  \paragraph{Composition}
  \begin{mathpar}
    \Infer[$\bndfhcom{\lst{I}}$-I]
    {\ceqtm{\lst{I} \eq \lst{J}}{\lst{J}'}{\GD} \\
      \ceqargtm{\GD}{\GTh}{\sch{m}}{\sch{m}'}{\argvar{\lst{I}}} \\
      (\forall i,j)\; \ceqargtm[\Psi\mid\xi_i,\xi_j]{\GD}{\GTh}{\sch{n}_i}{\sch{n}'_j}{\argvar{\lst{I}}} \\
      (\forall i)\; \ceqargtm[\Psi\mid\xi_i]{\GD}{\GTh}{\dsubst{\sch{n}_i}{r}{y}}{\sch{m}}{\argvar{\lst{I}}}}
    {\ceqargtm{\GD}{\GTh}{\bndfhcom*{\lst{J}}{\xi_i}}{\bndfhcom{\lst{J}'}{r}{r'}{\sch{m}'}{\sys{\xi_i}{y.\sch{n}'_i}}}{\argvar{\lst{I}}}}
    \and
    \Infer[$\bndfhcom{\lst{I}}$-C]
    {\ceqtm{\lst{I}}{\lst{J}}{\GD} \\
      \cwfargtm{\GD}{\GTh}{\sch{m}}{\argvar{\lst{I}}} \\
      (\forall i,j)\; \ceqargtm[\Psi\mid\xi_i,\xi_j]{\GD}{\GTh}{\sch{n}_i}{\sch{n}_j}{\argvar{\lst{I}}} \\
      (\forall i)\; \ceqargtm[\Psi\mid\xi_i]{\GD}{\GTh}{\dsubst{\sch{n}_i}{r}{y}}{\sch{m}}{\argvar{\lst{I}}}}
    {\ceqargtm{\GD}{\GTh}{\bndfhcom{\lst{J}}{r}{r}{\sch{m}}{\sys{\xi_i}{y.\sch{n}_i}}}{\sch{m}}{\argvar{\lst{I}}}}
    \and
    \Infer[$\bndfhcom{\lst{I}}$-T]
    {\ceqtm{\lst{I}}{\lst{J}}{\GD} \\
      \cwfargtm{\GD}{\GTh}{\sch{m}}{\argvar{\lst{I}}} \\
      (\forall i,j)\; \ceqargtm[\Psi\mid\xi_i,\xi_j]{\GD}{\GTh}{\sch{n}_i}{\sch{n}_j}{\argvar{\lst{I}}} \\
      (\forall i)\; \ceqargtm[\Psi\mid\xi_i]{\GD}{\GTh}{\dsubst{\sch{n}_i}{r}{y}}{\sch{m}}{\argvar{\lst{I}}} \\
      \models \xi_i}
    {\ceqargtm{\GD}{\GTh}{\bndfhcom{\lst{J}}{r}{r'}{\sch{m}}{\sys{\xi_i}{y.\sch{n}_i}}}{\dsubst{\sch{n}_i}{r'}{y}}{\argvar{\lst{I}}}}
  \end{mathpar}
  \paragraph{Coercion}
  \begin{mathpar}
    \Infer[$\bndfcoe$-I]
    {\ceqtm[\Psi,z]{\lst{I}}{\lst{I}'}{\GD} \\
      \ceqargtm{\GD}{\GTh}{\sch{m}}{\sch{m}'}{\argvar{\dsubst{\lst{I}}{r}{z}}}}
    {\ceqargtm{\GD}{\GTh}{\bndfcoe{z.\lst{I}}{r}{r'}{\sch{m}}}{\bndfcoe{z.\etc{I'_m}}{r}{r'}{\sch{m}'}}{\argvar{\dsubst{\lst{I}}{r'}{z}}}}
    \and
    \Infer[$\bndfcoe$-C]
    {\coftype[\Psi,z]{\lst{I}}{\GD} \\\\
      \cwfargtm{\GD}{\GTh}{\sch{m}}{\argvar{\dsubst{\lst{I}}{r}{z}}}}
    {\ceqargtm{\GD}{\GTh}{\bndfcoe{z.\lst{I}}{r}{r}{\sch{m}}}{\sch{m}}{\argvar{\dsubst{\lst{I}}{r}{z}}}}
    \and
    \Infer[$\bndfcoe$-$\emp$]
    {\cwfargtm{\GD}{\GTh}{\sch{m}}{\argvar{\emp}}}
    {\ceqargtm{\GD}{\GTh}{\bndfcoe{z.\emp}{r}{r'}{\sch{m}}}{\sch{m}}{\argvar{\emp}}}
  \end{mathpar}
  \paragraph{Functions}
  \begin{mathpar}
    \Infer[$\formal{\to}$-I]
    {\cwftypek{A} \\ \eqargtm{\oft{a}{A}}{\GD}{\GTh}{\sch{n}}{\sch{n}'}{\sch{b}}}
    {\ceqargtm{\GD}{\GTh}{\bndlam{a}{\sch{n}}}{\bndlam{a}{\sch{n}'}}{\argpi{a}{A}{\sch{b}}}}
    \and
    \Infer[$\formal{\to}$-E]
    {\ceqargtm{\GD}{\GTh}{\sch{n}}{\sch{n}'}{\argpi{a}{A}{\sch{b}}} \\
      \cwftypek{A} \\
      \ceqtm{M}{M'}{A}}
    {\ceqargtm{\GD}{\GTh}{\bndapp{\sch{n}}{M}}{\bndapp{\sch{n}'}{M'}}{\subst{\sch{b}}{M}{a}}}
    \and
    \Infer[$\formal{\to}$-$\beta$]
    {\wfargtm{\oft{a}{A}}{\GD}{\GTh}{\sch{n}}{\sch{b}} \\
      \cwftypek{A} \\
      \coftype{M}{A}}
    {\ceqargtm{\GD}{\GTh}{\bndapp{\bndlam{a}{\sch{n}}}{M}}{\subst{\sch{n}}{M}{a}}{\sch{b}}}
    \and
    \Infer[$\formal{\to}$-$\eta$]
    {\cwfargtm{\GD}{\GTh}{\sch{m}}{\argpi{a}{A}{\sch{b}}}}
    {\ceqargtm{\GD}{\GTh}{\sch{m}}{\bndlam{a}{(\bndapp{\sch{m}}{a})}}{\argpi{a}{A}{\sch{b}}}}
  \end{mathpar}

\end{mdframed}

\caption{Boundary term typing rules. We omit standard structural rules.}
\label{fig:boundary-term-rules}
\end{figure}



\section{Algebras}
\label{sec:algebra}

We now define what it means for a given family of relations to support the constructors specified by a list
$\K$; in the following section, we will define the denotation of the inductive type generated by $\K$ as the
least such relation. First, we need a notion of $\Psi$-relation family indexed by a Kan context $\GD$.

\begin{definition}
  Let $\cwfctxk{\GD}$. A \emph{$\GD$-indexed $\Psi$-relation} is a family $\Ga$ consisting of a
  $\Psi'$-relation $\ix{\Ga_{\psi}}{\lst{I}}$ for every $\psitd$ and
  $\coftype[\Psi']{\lst{I}}{\td{\GD}{\psi}}$, such that
  \begin{enumerate}
  \item $\td{\ix{\Ga_{\psi}}{\lst{I}}}{\psi'} = \ix{\Ga_{\psi\psi'}}{\td{\lst{I}}{\psi'}}$ for any
    $\psitd[']$,
  \item $\ix{\Ga_{\psi}}{\lst{I}} = \ix{\Ga_{\psi}}{\lst{I}'}$ whenever
    $\ceqtm[\Psi']{\lst{I}}{\lst{I}'}{\td{\GD}{\psi}}$.
  \end{enumerate}
  On account of the first condition, a $\GD$-indexed $\Psi$-relation $\Ga$ is completely determined by the
  relations $\ix{\Ga_\psi}{\lst{I}}_{\id}$ for $\psitd$ and $\coftype[\Psi']{\lst{I}}{\td{\GD}{\psi}}$. As
  such, we will generally give the definition of a $\GD$-indexed $\Psi$-relation by its values at such
  indices. Following our convention for $\Psi$-relations, we abbreviate $\ix{\Ga_\psi}{\lst{I}}_{\id}$ as
  $\ix{\Ga_\psi}{\lst{I}}$ and $\ix{\Ga_{\id}}{\lst{I}}$ as $\ix{\Ga}{\lst{I}}$. We extend $\Tm$ and $\Vl$ to
  indexed relations pointwise: $\ix{\Tm{\Ga}_\psi}{\lst{I}} \eqdef \Tm{\ix{\Ga_\psi}{\lst{I}}}$ and
  $\ix{\Vl{\Ga}_\psi}{\lst{I}} \eqdef \Vl{\ix{\Ga_\psi}{\lst{I}}}$.
\end{definition}

The values of an inductive type come in three forms: $\fhcom$ values (\cref{fig:ind-opsem-fhcom}), $\fcoe$
values (\cref{fig:ind-opsem-fcoe}), and $\intro*$ values (\cref{fig:ind-opsem-intro}). The first two are
``free'' composition and coercion values, necessary in order for the inductive type to be Kan.

The first form, $\fhcom$, will account for composition in the inductive type. The operator $\fhcom$ takes the
same arguments as $\hcom$ (sans type annotation). When a given tube constraint holds, the term
$\fhcom*{\xi_i}$ steps to the corresponding face, while if $r = r'$ it steps to its cap. Otherwise, it is a
value. By adding $\fhcom$ values to any type, we obtain a ``free'' implementation of $\hcom$.

The second form, $\fcoe$, will account for coercion between indices of the inductive type. As we noted in the
case of the identity type, non-constructor values are required to ensure that identity families respects paths
in their indexing context. The term $\fcoe{z.\lst{I}}{r}{r'}{M}$ will be used to coerce a term $M$ in the
inductive type from index $\dsubst{\lst{I}}{r}{z}$ to index $\dsubst{\lst{I}}{r'}{z}$. When $r = r'$, it steps
to $M$; otherwise, it is a value.

Finally, $\intro*$ values are constructor terms. Terms take the form $\intro{\lst{r}}{\lst{P}}{\lst{N}}$ where
$\ell \in \K$ is the label for the constructor in question and $\lst{r}$, $\lst{P}$, and $\lst{N}$ are the
dimension, non-recursive, and recursive arguments respectively. When a boundary constraint for a constructor
holds, it steps to the instantiation (defined below) of the corresponding boundary term. Otherwise, it is a
value.

\begin{figure}
  \begin{mdframed}
    \input{opsem/fhcom}
  \end{mdframed}
  \caption{Operational semantics of $\fhcom$}
  \label{fig:ind-opsem-fhcom}
\end{figure}

\begin{definition}
  For a value $\GD$-indexed $\Psi$-relation $\Ga$, define a value $\GD$-indexed $\Psi$-relation $\Fhcom{\Ga}$
  as generated by
  \begin{enumerate}
  \item
    $\ix{\Fhcom{\Ga}_\psi}{\lst{I}}(\fhcom*{\xi_i},\fhcom{r}{r'}{M'}{\sys{\xi_i}{y.N'_i}})$
    whenever
    \begin{enumerate}
    \item $\etc{\xi_i}$ is valid and $\not\models \xi_i$ for all $i$,
    \item $r \neq r'$,
    \item $\Tm{\ix{\Ga_\psi}{\lst{I}}}(M,M')$,
    \item $\Tm{\ix{\Ga_\psi}{\lst{I}}}_{y\mid \xi_i,\xi_j}(N_i,N'_j)$ for all $i,j$,
    \item $\Tm{\ix{\Ga_\psi}{\lst{I}}}_{\id\mid \xi_i}(\dsubst{N_i}{r}{y},M)$ for all $i$.
    \end{enumerate}
  \end{enumerate}
\end{definition}

\begin{figure}
  \begin{mdframed}
    \input{opsem/fcoe}
  \end{mdframed}
  \caption{Operational semantics of $\fcoe$}
  \label{fig:ind-opsem-fcoe}
\end{figure}

\begin{definition}
  For a value $\GD$-indexed $\Psi$-relation $\Ga$, define a value $\GD$-indexed $\Psi$-relation $\Fcoe{\Ga}$
  as generated by
  \begin{enumerate}
  \item $\ix{\Fcoe{\Ga}_\psi}{\lst{I}}(\fcoe{z.\lst{J}}{r}{r'}{M},\fcoe{z.\lst{J}'}{r}{r'}{M'})$
    whenever
    \begin{enumerate}
    \item $r \neq r'$,
    \item $\ceqtm[\Psi',z]{\lst{J}}{\lst{J}'}{\td{\GD}{\psi}}$,
    \item $\ceqtm[\Psi']{\dsubst{\lst{J}}{r'}{z}}{\lst{I}}{\td{\GD}{\psi}}$,
    \item $\Tm{\ix{\Ga_\psi}{\dsubst{\lst{J}}{r}{z}}}(M,M')$.
    \end{enumerate}
  \end{enumerate}
\end{definition}

\begin{figure}
  \begin{mdframed}
    \input{opsem/intro}
  \end{mdframed}
  \caption{Operational semantics of $\intro*$}
  \label{fig:ind-opsem-intro}
\end{figure}

To define the operational semantics and value relation for $\intro*$ terms, we need to define the
interpretation of argument types and boundary terms as real terms at a given instantiation of the
indeterminant family $\argvar{-}$. We first define these as untyped operations, then establish a typing rule.

\begin{definition}[Syntactic type interpretation]
  For any open term $\open{\Gd}{\tmj{A}}$, we define an open term $\tmj{\tyatty{\sch{b}}{\Gd.A}}$, the
  syntactic interpretation of $\sch{b}$ at $\Gd.A$, by
  \[
    \begin{array}{rcl}
      \tyatty{\argvar{\lst{I}}}{\Gd.A} &\eqdef& \subst{A}{\lst{I}}{\Gd} \\
      \tyatty{\argpi{b}{B}{\sch{c}}}{\Gd.A} &\eqdef& \picl{b}{B}{\tyatty{\sch{c}}{\Gd.A}}.
    \end{array}
  \]
  For a context $\GTh = \etc{\sch{b}_j}$, we will write $\tyatty{\GTh}{\Gd.A}$ for the list
  $\etc{\tyatty{\sch{b}_j}{\Gd.A}}$.
\end{definition}

We also define the semantic instantiation of an argument type with an indexed $\Psi$-relation.

\begin{definition}
  For $\cwftypek{A}$ and a value $A$-indexed $\Psi$-relation $\Gb$, define a value $\Psi$-relation
  $\Pirel{A}{\Gb}$ by
  \[
    \Pirel{A}{\Gb}_\psi \eqdef
    \left\{(\lam{b}{N},\lam{b}{N'}) \middle| \forall \psi',M,M'.
      \begin{array}{l}
        \ceqtm[\Psi'']{M}{M'}{\td{A}{\psi\psi'}} \implies \\
        \Tm{\ix{\Gb_{\psi\psi'}}{M}}(\subst{\td{N}{\psi'}}{M}{b},\subst{\td{N'}{\psi'}}{M'}{b})
      \end{array}\right\}.
  \]
\end{definition}

\begin{definition}[Semantic type interpretation]
  Given $\cwftypek{\GD}$, $\cwfconstrs{\GD}{\K}$, an argument type $\cwfargtype{\GD}{\sch{b}}$, and a value
  $\GD$-indexed $\Psi$-relation $\Ga$, we define a value $\Psi$-relation $\tyatrel{\sch{b}}{\Ga}$, the
  semantic interpretation of $\sch{b}$ at $\Ga$, by recursion on the structure of $\sch{b}$:
  \[
    \begin{array}{rcl}
      \tyatrel{\argvar{\lst{I}}}{\Ga} &\eqdef& \ix{\Ga}{\lst{I}} \\
      \tyatrel{\argpi{b}{B}{\sch{c}}}{\Ga} &\eqdef& \Pirel{B}{(\psi,N) \mapsto \tyatrel{\subst{\td{\sch{c}}{\psi}}{N}{b}}{\td{\Ga}{\psi}}}.
    \end{array}
  \]
  For a context $\GTh = \etc{\sch{b}_j}$, we will write $\tyatrel{\GTh}{\Ga}_\psi(\etc{N_j},\etc{N'_j})$ to
  mean that $\tyatrel{\sch{b}_j}{\Ga}_\psi(N_j,N'_j)$ for each $j$.
\end{definition}

\begin{proposition}
  In a cubical type system with all dependent function types, if
  \begin{enumerate}
  \item $\eqtypek{\oft{\Gd}{\GD}}{A}{A'}$,
  \item $\ceqargtype{\GD}{\sch{b}}{\sch{b}'}$,
  \end{enumerate}
  then $\ceqtypek{\tyatty{\sch{b}}{\Gd.A}}{\tyatty{\sch{b}'}{\Gd.A'}}$.
\end{proposition}

The interpretation function for boundary terms acts on open terms; given an open boundary term
$\open{\Gth}{\bndj{\sch{m}}}$ with $\fb{\sch{m}} \subseteq \Gth$ (but which may contain any ordinary term
variables), we get an ordinary open term $\open{\Gd}{\tmj{\insttm{\Gth.\sch{m}}{\K}{\Gd}}}$ where
$|\Gd| = |\Gth|$. (When we give a typing rule in \cref{lem:insttm-typing}, the types of $\Gd$ will be the
interpretations of the types of $\Gth$.)

\begin{definition}[Boundary term interpretation]
  Given $\K$, an open boundary term $\open{\Gth}{\bndj{\sch{m}}}$ with $\fl{\sch{m}} \subseteq \dom{\K}$ and
  $\fb{\sch{m}} \subseteq \Gth$, and $\tmj{\lst{N}}$ with $|\lst{N}| = |\Gth|$, we define a term
  $\tmj{\insttm{\Gth.\sch{m}}{\K}{\lst{N}}}$, the interpretation of $\Gth.\sch{m}$ at constructor list $\K$
  and term variables $\lst{N}$, by
  \[
    \begin{array}{rcl}
      \insttm{\Gth.\Gth[j]}{\K}{\lst{N}} &\eqdef& \lst{N}[j] \\[.4em]
      \insttm{\Gth.\bndintro{\lst{r}}{\lst{P}}{\lst{\sch{n}}}}{\K}{\lst{N}} &\eqdef& \intro[\K]{\lst{r}}{\lst{P}}{\insttm{\Gth.\lst{\sch{n}}}{\K}{\lst{N}}} \\[.4em]
      \insttm{\Gth.\bndfhcom*{\lst{I}}{\xi_i}}{\K}{\lst{N}} &\eqdef& \fhcom{r}{r'}{\insttm{\Gth.\sch{m}}{\K}{\lst{N}}}{\sys{\xi_i}{y.\insttm{\Gth.\sch{n}_i}{\K}{\lst{N}}}}. \\[.4em]
      \insttm{\Gth.\bndfcoe{z.\lst{I}}{r}{r'}{\sch{m}}}{\K}{\lst{N}} &\eqdef& \fcoe{z.\lst{I}}{r}{r'}{\insttm{\Gth.\sch{m}}{\K}{\lst{N}}} \\[.4em]
      \insttm{\Gth.\bndlam{a}{\sch{n}}}{\K}{\lst{N}} &\eqdef& \lam{a}{(\insttm{\sch{\Gth.n}}{\K}{\lst{N}})} \\[.4em]
      \insttm{\Gth.\bndapp{\sch{n}}{M}}{\K}{\lst{N}} &\eqdef& \app{\insttm{\Gth.\sch{n}}{\K}{\lst{N}}}{M}.
    \end{array}
  \]
  Here and henceforth, we write $\insttm{\Gth.\etc{\sch{m}_n}}{\K}{\lst{N}}$ for a list of terms
  $\open{\Gth}{\etc{\sch{m}_n}}$ to mean $\etc{\insttm{\Gth.\sch{m}_n}{\K}{\lst{N}}}$.
\end{definition}

\begin{proposition}[Basic facts on boundary interpretation]
  \label{prop:insttm-facts}~
  \begin{enumerate}[label=(\alph*)]
  \item $\subst{\insttm{\Gth.\sch{m}}{\K}{\lst{N}}}{P}{a} = \insttm{\Gth.\subst{\sch{m}}{P}{a}}{\subst{\K}{P}{a}}{\subst{\lst{N}}{P}{a}}$.
  \item If $\open{\Gf}{\bndj{\sch{m}}}$ and $\open{\Gth}{\bndj{\lst{\sch{n}}}}$ with $|\Gf| = |\sch{n}|$, then
    $\insttm{\Gth.\subst{\sch{m}}{\lst{\sch{n}}}{\Gf}}{\K}{\lst{N}} =
    \insttm{\Gf.\sch{m}}{\K}{\insttm{\Gth.\lst{\sch{n}}}{\K}{\lst{N}}}$.
  \item If $\K \constrspre \K'$ then $\insttm{\Gth.\sch{m}}{\K}{\lst{N}} = \insttm{\Gth.\sch{m}}{\K'}{\lst{N}}$.
  \end{enumerate}
\end{proposition}

\begin{definition}
  For a value $\GD$-indexed $\Psi$-relation $\Ga$, $\cwfconstrs{\GD}{\K}$ and $\ell \in \K$, define a value
  $\GD$-indexed $\Psi$-relation $\Intro{\K}{\ell}{\Ga}$ as generated by
  \begin{enumerate}
  \item $\ix{\Intro{\K}{\ell}{\Ga}_\psi}{\lst{I}}(\intro[\K']{\lst{r}}{\lst{P}}{\lst{N}},\intro[\K''][\ell]{\lst{r}}{\lst{P}'}{\lst{N}'})$ whenever
    \begin{enumerate}
    \item
      $\td{\K}{\psi}[\ell] =
      \constr{\lst{x}}{\GG}{\Gg.\lst{J}}{\Gg.\GTh}{\sys{\xi_k}{\Gg.\Gth.\sch{m}_k}}$ where
      $\GTh = \etc{\ofa{\formal{p}_j}{\sch{b}_j}}$,
    \item $\ceqconstrs[\Psi']{\td{\GD}{\psi}}{\td{\K}{\psi} \equiv \K'}{\K''}$,
    \item $\not\models \dsubst{\xi_k}{\lst{r}}{\lst{x}}$ for all $k$,
    \item $\ceqtm[\Psi']{\lst{P}}{\lst{P}'}{\GG}$,
    \item $\ceqtm[\Psi']{\subst{\lst{J}}{\lst{P}}{\Gg}}{\lst{I}}{\td{\GD}{\psi}}$,
    \item $\tyatrel{\subst{\GTh}{\lst{P}}{\Gg}}{\td{\Ga}{\psi}}(\lst{N},\lst{N}')$,
    \end{enumerate}
  \end{enumerate}
\end{definition}

\begin{definition}
  For a $\cwfconstrs{\GD}{\K}$, define a monotone operator $\F_\K$ on value $\GD$-indexed $\Psi$-relations:
  \[
    \F_\K(\Ga) \eqdef \Fhcom{\Ga} \cup \Fcoe{\Ga} \cup \bigcup_{\ell \in \K} \Intro{\K}{\ell}{\Ga}
  \]
  We say that $\Ga$ \emph{supports} $\K$ if $\F_\K(\Ga) \subseteq \Ga$. In other words, $\Ga$ supports $\K$ if
  $\Ga$ is an algebra for the functor $\F_\K$ in the category of value $\GD$-indexed $\Psi$-relations and
  inclusions.
\end{definition}

\begin{definition}
  For any operator $\F$ on value $\GD$-indexed $\Psi$-relations, define a monotone operator $\opt{\F}$ by
  $\opt{\F}{\Ga} \eqdef \Ga \cup \F(\Ga)$ and an operator $\multi{\F}$ by
  $\multi{\F}{\Ga} \eqdef \lfp{(\Gb \mapsto \Ga \cup \F(\Gb))}$. Note that
  $\F(\Ga) \subseteq \opt{\F}{\Ga} \subseteq \multi{\F}{\Ga}$.
\end{definition}

We now will now show that the term relations $\Tm{\opt{\Fhcom}{\Ga}}$, $\Tm{\opt{\Fcoe}{\Ga}}$, and
$\Tm{\opt{\Intro{\K}{\ell}}{\Ga}}$ contain $\fhcom$, $\fcoe$, and $\intro$ terms formed from arguments in
$\Ga$ respectively. This implies that if $\Ga$ is a value relation closed under $\fhcom$ values, then
$\Tm{\Ga}$ is closed under all $\fhcom$ terms. Likewise, closure of $\Ga$ under $\fcoe$ or $\intro$ values
implies closure of $\Tm{\Ga}$ under the corresponding terms. In the general case, it is necessary to use
$\Tm{\opt{\Fhcom}{\Ga}}$ rather than $\Tm{\Fhcom{\Ga}}$ because a non-value $\fhcom$ made from terms in
$\Tm{\Ga}$ reduces to a term in $\Tm{\Ga}$; likewise for $\fcoe$ and $\intro$.

\begin{lemma}[$\fhcom$-I]
  \label{lem:supports-I-fhcom}
  For any $\GD$-indexed $\Psi$-PER $\Ga$, $\psitd$, $\coftype[\Psi']{\lst{I}}{\td{\GD}{\psi}}$, and
  $\etc{\xi_i}$ valid, if
  \begin{enumerate}
  \item $\Tm{\ix{\Ga_\psi}{\lst{I}}}(M,M')$,
  \item $\Tm{\ix{\Ga_\psi}{\lst{I}}}_{y\mid \xi_i,\xi_j}(N_i,N'_j)$ for all $i,j$,
  \item $\Tm{\ix{\Ga_\psi}{\lst{I}}}_{\id\mid \xi_i}(\dsubst{N_i}{r}{y},M)$ for all $i$,
  \end{enumerate}
  we have
  \begin{enumerate}[label=(\alph*)]
  \item $\Tm{\ix{\Ga_\psi}{\lst{I}}}_{\id \mid \xi_i}(\fhcom*{\xi_i},\dsubst{N_i}{r'}{y})$ for all $i$,
  \item $\Tm{\ix{\Ga_\psi}{\lst{I}}}_{\id \mid r = r'}(\fhcom*{\xi_i},M)$,
  \item $\Tm{\ix{\opt{\Fhcom}{\Ga}_\psi}{\lst{I}}}(\fhcom*{\xi_i},\fhcom{r}{r'}{M'}{\sys{\xi_i}{y.N'_i}})$.
  \end{enumerate}
\end{lemma}
\begin{proof}
  We prove the three statements in turn.
  \begin{enumerate}[label=(\alph*)]
  \item By \cref{cor:restricted-expansion}. Let $\psitd[']$ be given with $\models \td{\xi_i}{\psi'}$. Take
    $j$ to be least such that $\models \td{\xi_j}{\psi'}$. We have
    $\td{\fhcom*{\xi_i}}{\psi'} \steps \td{\dsubst{N_j}{r'}{y}}{\psi'}$ and
    $\Tm{\ix{\Ga_\psi}{\lst{I}}_{\psi'}}(\td{\dsubst{N_j}{r'}{y}}{\psi'},\td{\dsubst{N_i}{r'}{y}}{\psi'})$ by
    assumption.
  \item By \cref{cor:restricted-expansion}. Let $\psitd[']$ be given with $\td{r}{\psi'} =
    \td{r'}{\psi'}$. Either $\not\models \td{\xi_i}{\psi'}$ for all $i$, or there exists $i$ least such that
    $\models \td{\xi_i}{\psi'}$. In the former case, we have $\td{\fhcom*{\xi_i}}{\psi'} \steps \td{M}{\psi'}$
    and $\Tm{\ix{\Ga_\psi}{\lst{I}}_{\psi'}}(\td{M}{\psi'})$ by assumption. In the latter case, we have
    $\td{\fhcom*{\xi_i}}{\psi'} \steps \td{\dsubst{N_j}{r'}{y}}{\psi'}$ and
    $\Tm{\ix{\Ga_\psi}{\lst{I}}}_{\psi'}(\td{\dsubst{N_j}{r'}{y}}{\psi'},\td{M}{\psi'})$ by assumption.

  \item We go by \cref{lem:introduction}. Let $\psitd[']$ be given. We have three cases
    \begin{itemize}
    \item There exists $i$ such that $\models \td{\xi_i}{\psi'}$.

      Then
      $\Tm{\ix{\Ga_\psi}{\lst{I}}}_{\psi'}(\td{\fhcom*{\xi_i}}{\psi'},\td{\fhcom{r}{r'}{M'}{\sys{\xi_i}{y.N'_i}}}{\psi'})$
      by reducing with (a) on each side and applying assumption 2.
    \item $\td{r}{\psi'} = \td{r'}{\psi'}$ and $\not\models \td{\xi_i}{\psi'}$ for all $i$.

      Then
      $\Tm{\ix{\Ga_\psi}{\lst{I}}}_{\psi'}(\td{\fhcom*{\xi_i}}{\psi'},\td{\fhcom{r}{r'}{M'}{\sys{\xi_i}{y.N'_i}}}{\psi'})$
      by reducing with (b) on each side and applying assumption 1.

    \item $\not\models \td{\xi_i}{\psi'}$ for all $i$ and
      $\td{r}{\psi'} \neq \td{r'}{\psi'}$.

      Then
      $\ix{\Fhcom{\Ga}_\psi}{\lst{I}}_{\psi'}(\td{\fhcom*{\xi_i}}{\psi'},\td{\fhcom{r}{r'}{M'}{\sys{\xi_i}{y.N'_i}}}{\psi'})$
      holds by definition of $\Fhcom{\Ga}$. \qedhere
    \end{itemize}
  \end{enumerate}
\end{proof}

\begin{lemma}[$\fcoe$-I]
  \label{lem:supports-I-fcoe}
  For any $\GD$-indexed $\Psi$-PER $\Ga$, $\psitd$, and $\tmj[\Psi]{r,r'}$, if
  \begin{enumerate}
  \item $\ceqtm[\Psi,z]{\lst{I}}{\lst{I}'}{\td{\GD}{\psi}}$,
  \item $\Tm{\ix{\Ga_\psi}{\dsubst{\lst{I}}{r}{z}}}(M,M')$,
  \end{enumerate}
  then
  \begin{enumerate}[label=(\alph*)]
  \item $\Tm{\ix{\Ga_\psi}{\dsubst{\lst{I}}{r'}{z}}}(\fcoe{z.\lst{I}}{r}{r'}{M},M)$ when $\lst{I} = \emp$,
  \item $\Tm{\ix{\Ga_\psi}{\dsubst{\lst{I}}{r'}{z}}}_{\id\mid r=r'}(\fcoe{z.\lst{I}}{r}{r'}{M},M)$,
  \item $\Tm{\ix{\Ga_\psi}{\dsubst{\lst{I}}{r'}{z}}}(\fcoe{z.\lst{I}}{r}{r'}{M},\fcoe{z.\lst{I}'}{r}{r'}{M'})$.
  \end{enumerate}
\end{lemma}
\begin{proof}
  We prove the three statements in turn.
  \begin{enumerate}[label=(\alph*)]
  \item By \cref{cor:restricted-expansion}. Let $\psitd[']$. We
    have $\td{\fcoe{z.\lst{I}}{r}{r'}{M}}{\psi} \steps \td{M}{\psi}$, and the reduct is in
    $\Tm{\ix{\Ga_\psi}{\dsubst{\lst{I}}{r'}{z}}}_{\psi'}$ by assumption.
  \item By \cref{cor:restricted-expansion}. Let $\psitd[']$ be given with $\td{r}{\psi'} = \td{r'}{\psi'}$. We
    have $\td{\fcoe{z.\lst{I}}{r}{r'}{M}}{\psi} \steps \td{M}{\psi}$, and the reduct is in
    $\Tm{\ix{\Ga_\psi}{\dsubst{\lst{I}}{r'}{z}}}_{\psi'}$ by assumption.
  \item By \cref{lem:introduction}. Let $\psitd[']$ be given. Either $\lst{I} = \emp$,
    $\td{r}{\psi'} = \td{r'}{\psi'}$, or neither is the case. In the former two cases, we have
    $\Tm{\ix{\Ga_\psi}{\dsubst{\lst{I}}{r'}{z}}}_{\psi'}(\td{\fcoe{z.\lst{I}}{r}{r'}{M}}{\psi'},\td{\fcoe{z.\lst{I}'}{r}{r'}{M'}}{\psi'})$
    by reducing with (a) on each side and applying assumption 2. In the latter,
    $\ix{\Fcoe{\Ga}_\psi}{\dsubst{\lst{I}}{r'}{z}}_{\psi'}(\td{\fcoe{z.\lst{I}}{r}{r'}{M}}{\psi'},\td{\fcoe{z.\lst{I}'}{r}{r'}{M'}}{\psi'})$
    holds by definition of $\Fcoe{\Ga}$. \qedhere
  \end{enumerate}
\end{proof}

The proof of the introduction rule for $\intro*$ terms is somewhat more involved than for $\fhcom$ and
$\fcoe$. As the boundary of an $\intro*$ term can step to the interpretation of a boundary term $\sch{m}_k$,
the proof of its coherence relies on the type-correctness of boundary term interpretation. On the other hand,
the correctness of boundary term interpretation appeals to the introduction rule for $\intro*$ terms. We will
therefore prove these two lemmas by a mutual induction.

\begin{definition}
  Let $\ceqconstrs{\GD}{\K}{\K'}$ and a value $\GD$-indexed $\Psi$-PER $\Ga$ be given. We say that the
  property $\Interp{\K}{\K'}{\Ga}{n}$ holds for some $n \le |\K|$ if for every
  \begin{enumerate}
  \item $\ceqargtm{\GD}{\ofc{\Gth}{\GTh}}{\sch{m}}{\sch{m}'}{\sch{b}}$ with
    $\height{\K}{\sch{m},\sch{m}'} = n$,
  \item $\Tm{\tyatrel{\GTh}{\Ga}}(\lst{N},\lst{N}')$,
  \end{enumerate}
  we have
  $\Tm{\tyatrel{\sch{b}}{\Ga}}(\insttm{\Gth.\sch{m}}{\K}{\lst{N}},\insttm{\Gth.\sch{m}'}{\K'}{\lst{N}'})$.
\end{definition}

\begin{lemma}[$\intro*$-I]
  \label{lem:supports-I-intro}
  Let $\cwfconstrs{\GD}{\K}$, $\ell \in \K$, and a value $\GD$-indexed $\Psi$-PER $\Ga$ be given. For all $\psitd$, if
  \begin{enumerate}
  \item
    $\td{\K}{\psi}[\ell] = \constr{\lst{x}}{\GG}{\Gg.\lst{I}}{\Gg.\GTh}{\sys{\xi_k}{\Gg.\Gth.\sch{m}_k}}$,
  \item $\ceqconstrs[\Psi']{\GD}{\td{\K}{\psi} \equiv \K'}{\K''}$,
  \item $\Interp{\K'}{\K''}{\td{\Ga}{\psi}}{n}$ holds for $n < \height{\K}{\ell}$,
  \item $\ceqtm[\Psi']{\lst{P}}{\lst{P}'}{\GG}$,
  \item $\Tm{\tyatrel{\subst{\GTh}{\lst{P}}{\Gg}}{\td{\Ga}{\psi}}}(\lst{N},\lst{N}')$,
  \end{enumerate}
  then
  \begin{enumerate}[label=(\alph*)]
  \item $\Tm{\ix{\Ga_\psi}{\subst{\lst{I}}{\lst{P}}{\Gg}}}_{\id\mid \dsubst{\xi_k}{\lst{r}}{\lst{x}}}(\intro[\K']{\lst{r}}{\lst{P}}{\lst{N}},\insttm{\Gth.\subst{\dsubst{\sch{m}_k}{\lst{r}}{\lst{x}}}{\lst{P}}{\Gg}}{\K'}{\lst{N}})$ for all $k$,
  \item $\Tm{\ix{\opt{\Intro{\K}{\ell}}{\Ga}_\psi}{\subst{\lst{I}}{\lst{P}}{\Gg}}}(\intro[\K']{\lst{r}}{\lst{P}}{\lst{N}},\intro[\K'']{\lst{r}'}{\lst{P}'}{\lst{N}'})$.
  \end{enumerate}
\end{lemma}
\begin{proof}
  We prove the two statements in sequence.
  \begin{enumerate}[label=(\alph*)]
  \item By \cref{cor:restricted-expansion}. Let $\psitd[']$ be given with
    $\models \td{\dsubst{\xi_k}{\lst{r}}{\lst{x}}}{\psi'}$.  Take $l$ to be least such that
    $\models \td{\dsubst{\xi_l}{\lst{r}}{\lst{x}}}{\psi'}$. Then
    $ \td{\intro[\K']{\lst{r}}{\lst{P}}{\lst{N}}}{\psi'} \steps
    \td{\insttm{\Gth.\subst{\dsubst{\sch{m}_l}{\lst{r}}{\lst{x}}}{\lst{P}}{\Gg}}{\K'}{\lst{N}}}{\psi'}$.  We
    have $n \eqdef \height{\K}{\sch{m}_l,\sch{m}_k} < \ell$, so we can apply $\Interp{\K'}{\K''}{\td{\Ga}{\psi}}{n}$ with
    $\eqargtm[\Psi',\lst{x}]<\xi_k,\xi_l>{\Gg:\GG}{\td{\GD}{\psi}}[\K']{\GTh}{\sch{m}_l}{\sch{m}_k}{\argvar{\subst{\lst{I}}{\lst{P}}{\Gg}}}$
    to learn that
    $\Tm{\ix{\Ga_\psi}{\subst{\lst{I}}{\lst{P}}{\Gg}}}_{\psi'}(\insttm{\Gth.\subst{\dsubst{\sch{m}_l}{\lst{r}}{\lst{x}}}{\lst{P}}{\Gg}}{\K'}{\lst{N}},\insttm{\Gth.\subst{\dsubst{\sch{m}_k}{\lst{r}}{\lst{x}}}{\lst{P}}{\Gg}}{\K'}{\lst{N}})$
    holds.
  \item By \cref{lem:introduction}.  Let $\psitd[']$ be given; we have two cases.
    \begin{itemize}
    \item There exists $k$ such that
      $\models \dsubst{\xi_k}{\lst{r}}{\lst{x}}{\psi'}$.

      Then
      $\Tm{\ix{\Ga_\psi}{\subst{\lst{I}}{\lst{P}}{\Gg}}}_{\psi'}(\td{\intro[\K']{\lst{r}}{\lst{P}}{\lst{N}}}{\psi'},\td{\intro[\K'']{\lst{r}}{\lst{P}'}{\lst{N}'}}{\psi'})$
      follows by first reducing with (a) on each side and then applying
      $\Interp{\K'}{\K''}{\td{\Ga}{\psi}}{\height{\K}{\sch{m}_k,\sch{m}'_k}}$ with
      $\eqargtm[\Psi']<\xi_k>{\ofc{\Gg}{\GG}}{\td{\GD}{\psi}}[\K']{\GTh}{\sch{m}_k}{\sch{m}'_k}{\argvar{\lst{I}}}$.
    \item
      $\not\models \td{\dsubst{\xi_k}{\lst{r}}{\lst{x}}}{\psi'}$
      for all $k$.

      Then we have
      $\ix{\Intro{\K}{\ell}{\Ga}_\psi}{\subst{\lst{I}}{\lst{P}}{\Gg}}_{\psi'}(\td{\intro[\K']{\lst{r}}{\lst{P}}{\lst{N}}}{\psi'},\td{\intro[\K'']{\lst{r}}{\lst{P}'}{\lst{N}'}}{\psi'})$
      by our assumptions and the definition of $\Intro{\K}{\ell}{\Ga}$. \qedhere
     \end{itemize}
  \end{enumerate}
\end{proof}

\begin{lemma}[Boundary interpretation typing]
  \label{lem:insttm-typing}
  Let $\ceqconstrs{\GD}{\K}{\K'}$ be given and let $\Ga$ be a $\Psi$-PER which supports $\K$. Then
  $\Interp{\K}{\K'}{\Ga}{n}$ holds for all $n \le |\K|$. That is, for all
$\ceqargtm{\GD}{\ofc{\Gth}{\GTh}}{\sch{m}}{\sch{m}'}{\sch{b}}$ and
 $\Tm{\tyatrel{\GTh}{\Ga}}(\lst{N},\lst{N}')$,
  we have
  $\Tm{\tyatrel{\sch{b}}{\Ga}}(\insttm{\Gth.\sch{m}}{\K}{\lst{N}},\insttm{\Gth.\sch{m}'}{\K'}{\lst{N}'})$.
\end{lemma}
\begin{proof}
  By strong induction on $n$. Suppose that $\Interp{\K}{\K'}{\Ga}{m}$ holds for all $m < n$. We then go by an
  inner induction on the derivation of
  $\ceqargtm{\GD}{\ofc{\Gth}{\GTh}}{\sch{m}}{\sch{m}'}{\sch{b}}$. The proof is entirely
  routine, so we will omit it. For the $\bndfhcom$-, $\bndfcoe$-, and $\bndintro[]$-related cases, we use
  \cref{lem:supports-I-fcom,lem:supports-I-fcoe,lem:supports-I-intro} respectively. In the $\bndintro[]$ case,
  the use of \cref{lem:supports-I-intro} is justified by induction hypothesis. For the cases concerning
  function types, we refer to Part III for proofs of the corresponding rules.
\end{proof}

\begin{figure}
  \begin{mdframed}
    \input{opsem/fcom}
  \end{mdframed}
  \caption{Operational semantics of $\fcom$}
  \label{fig:ind-opsem-fcom}
\end{figure}

Finally, we define a derived operator $\fcom$ in \cref{fig:ind-opsem-fcom}, which combines $\fhcom$ and
$\fcoe$ in the same way that $\com$ combines $\hcom$ and $\coe$. With $\fcom$, we compose along a line
$z.\lst{I}$ of indices in a family.

\begin{definition}
  Define $\Fcom{\Ga} \eqdef \Fhcom{\opt{\Fcoe}{\Ga}}$.
\end{definition}

\begin{lemma}[$\fcom$-I]
  \label{lem:supports-I-fcom}
  For any $\GD$-indexed $\Psi$-PER $\Ga$, $\psitd$, and $\tmj[\Psi]{r,r'}$, if
  \begin{enumerate}
  \item $\ceqtm[\Psi,z]{\lst{I}}{\lst{I}'}{\td{\GD}{\psi}}$,
  \item $\Tm{\ix{\Ga_\psi}{\dsubst{\lst{I}}{r}{z}}}(M,M')$,
  \item $\Tm{\ix{\Ga_\psi}{\lst{I}}}_{y\mid \xi_i,\xi_j}(N_i,N'_j)$ for all $i,j$,
  \item $\Tm{\ix{\Ga_\psi}{\lst{I}}}_{\id\mid \xi_i}(\dsubst{N_i}{r}{y},M)$ for all $i$,
  \end{enumerate}
  we have
  \begin{enumerate}[label=(\alph*)]
  \item $\Tm{\ix{\Ga_\psi}{\lst{I}}}_{\id \mid \xi_i}(\fcom*{z.\lst{I}}{\xi_i},\dsubst{N_i}{r'}{y})$ for all $i$,
  \item $\Tm{\ix{\Ga_\psi}{\lst{I}}}_{\id \mid r = r'}(\fcom*{z.\lst{I}}{\xi_i},M)$,
  \item $\Tm{\ix{\opt{\Fcom}{\Ga}_\psi}{\lst{I}}}(\fcom*{z.\lst{I}}{\xi_i},\fcom{z.\lst{I}'}{r}{r'}{M'}{\sys{\xi_i}{y.N'_i}})$.
  \end{enumerate}
\end{lemma}
\begin{proof}
  By \cref{lem:supports-I-fhcom,lem:supports-I-fcoe}.
\end{proof}


\section{Inductive types}
\label{sec:inductive}

\begin{definition}
  Given $\cwfctxk{\GD}$ and $\cwfconstrs{\GD}{\K}$, define the \emph{inductive $\GD$-indexed $\Psi$-relation
    generated by $\K$} by $\indrel{\K} \eqdef \lfp{F_\K}$. By definition, $\indrel{\K}$ is the least
  $\GD$-indexed $\Psi$-relation which supports $\K$. It is easy to check that each fiber of $\indrel{\K}$ is a
  $\Psi$-PER.
\end{definition}

\begin{definition}
  If $\ceqctxk{\GD}{\GD'}$ and $\ceqconstrs{\GD}{\K}{\K'}$, we say that the candidate cubical type system
  $\tau$ \emph{has their inductive family} if
  $\tau(\Psi',\indcl[\td{\GD}{\psi}]{\td{\K}{\psi}}{\lst{I}},\indcl[\td{\GD'}{\psi}]{\td{\K'}{\psi}}{\lst{I}'},\ix{\indrel{\K}_\psi}{\lst{I}})$
  holds for all $\psitd$ and $\ceqtm[\Psi']{\lst{I}}{\lst{I}'}{\td{\GD}{\psi}}$.
\end{definition}

\begin{proposition}
  There exists a cubical type system $\tau$ which has the inductive family of every
  $\ceqctxk{\GD}{\GD'}$ and $\ceqconstrs{\GD}{\K}{\K'}$. Moreover, there exists such a $\tau$ containing
  a hierarchy of universes closed under inductive type formation.
\end{proposition}
\begin{proof}
  We sketch the construction; for complete details on constructing and establishing basic properties of
  cubical type systems, we refer to Section 3 of Part III. Define an operator $\Ind$ on candidate cubical type
  systems by
  \begin{align*}
    \Ind(\tau)(\Psi,A_0,A_0',\phi)& \iffdef
      \left\{
        \begin{array}{l}
          A_0 = \indcl{\K}{\lst{I}} \land A_0' = \indcl[\GD']{\K'}{\lst{I}'} \\
          \phantom{ } \land \tau \models \ceqctxk{\GD}{\GD'} \\
          \phantom{ } \land \tau \models \ceqconstrs{\GD}{\K}{\K'} \\
          \phantom{ } \land \tau \models \ceqtm{\lst{I}}{\lst{I}'}{\GD} \\
          \phantom{ } \land \phi = \ix{\indrel{\K}}{\lst{I}}_\id
        \end{array}
      \right.
  \end{align*}
  For candidate cubical type systems $\nu, \tau$, define
  \begin{align*}
    K(\nu,\tau) &\eqdef \nu \cup \Ind{\tau} \cup \textsc{Fun}(\tau) \cup {\cdots}
  \end{align*}
  where $\textsc{Fun}(\tau)$ specifies the dependent function types definable from types in $\tau$ and
  $\cdots$ stands for any other type formers from Part III we wish to include. Set
  \begin{align*}
    \nu_0(\Psi,A_0,A'_0,\phi) &\iffdef \bot \\
    \nu_{n+1}(\Psi,A_0,A'_0,\phi) &\iffdef A_0 = A_0' = \UKan \land (j \le n) \land (\phi(B_0,B_0') \iff \tau_n(\Psi,B_0,B_0',\_)) \\
    \tau_n &\eqdef \mu\tau. K(\nu_n, \tau)
  \end{align*}
  Then $\tau_\omega \eqdef \bigcup_{n \in \N} \tau_n$ is a candidate cubical type system with a chain of
  universes $\{ \UKan \mid j \in \N\}$ each of which is closed under inductive type formation. Following Part
  III, it is straightforward to show that $\tau_\omega$ above is in fact a cubical type system.
\end{proof}

For the remainder of \cref{sec:inductive}, we fix $\ceqctxk{\GD}{\GD'}$ and $\ceqconstrs{\GD}{\K}{\K'}$ and
assume we are working in a cubical type system $\tau$ which has their inductive type as well as all dependent
function types. As part of the proof that $\tau$ is a cubical type system, we will have established the
following:

\begin{lemma}
  \label{lem:formation}
  $\PTy{\tau}(\Psi',\indcl[\td{\GD}{\psi}]{\td{\K}{\psi}}{\lst{I}},\indcl[\td{\GD'}{\psi}]{\td{\K'}{\psi}}{\lst{I}'},\ix{\indrel{\K}_\psi}{\lst{I}})$
  holds for every $\psitd$ and $\ceqtm[\Psi']{\lst{I}}{\lst{I}'}{\td{\GD}{\psi}}$.
\end{lemma}
\begin{proof}
  By \cref{prop:pty-stable-introduction}.
\end{proof}

\begin{corollary}
  $\ceqtypep{\indcl[\td{\GD}{\psi}]{\td{\K}{\psi}}{\lst{I}}}{\indcl[\td{\GD'}{\psi}]{\td{\K'}{\psi}}{\lst{I}'}}$ for all
  $\psitd$ and $\ceqtm[\Psi']{\lst{I}}{\lst{I}'}{\td{\GD}{\psi}}$.
\end{corollary}
\begin{proof}
  By definition of the pretype judgment, we must demonstrate in addition to \cref{lem:formation} that
  $\ix{\indrel{\K}_\psi}{\lst{I}}$ is value-coherent for every $\psitd$ and
  $\ceqtm[\Psi']{\lst{I}}{\lst{I}'}{\td{\GD}{\psi}}$. We want to show
  $\indrel{\K} \subseteq \Vl{\Tm{\indrel{\K}}}$. By universal property of $\indrel{\K}$, it is enough to show
  that $\F_\K(\Vl{\Tm{\indrel{\K}}}) \subseteq \Vl{\Tm{\indrel{\K}}}$. As
  $\Vl{\Tm{\indrel{\K}}} \subseteq \indrel{\K}$, it suffices to show
  $\F_\K(\indrel{\K}) \subseteq \Tm{\indrel{\K}}$. This follows from
  \cref{lem:supports-I-fcom,lem:supports-I-fcoe,lem:supports-I-intro}.
\end{proof}

\begin{theorem}[Canonicity]~
  \begin{enumerate}
  \item If $\ceqtm{M}{M'}{\indcl{\K}{\lst{I}}}$, then $M \evals V$ and $M' \evals V'$ for some $V,V'$ with
    $\ix{\indrel{\K}}{\lst{I}}(V,V')$.
  \item If $\ceqtm[\cdot]{M}{M'}{\indcl[\emp]{\K}{\emp}}$, then $M \evals V$ and $M' \evals V'$ for some
    $V,V'$ with $\ix{\Intro{\K}{\ell}}{\emp}(V,V')$ for a constructor $\ell$ which has no specified
    boundaries, i.e., for which $\K[\ell] = \constr{\lst{x}}{\GG}{\Gg.\emp}{\Gg.\GTh}{\emp}$.
  \end{enumerate}
\end{theorem}
\begin{proof}
  The first statement follows immediately from the definition of the term equality judgment. For the second,
  we obtain a stronger result by restricting to the non-indexed case and considering only zero-dimensional
  terms. The case of $\fhcom$ values is excluded because any valid system $\etc{\xi_i}$ containing no
  dimension variables satisfies $\models \xi_i$ for some $i$. The case of $\fcoe$ values is excluded because
  any $\fcoe$ term in a non-indexed inductive type reduces.
\end{proof}

The second part of the canonicity theorem has some interesting consequences. For example, any closed
zero-dimensional term in the $(-1)$-truncation type $\Trunc{A}$ reduces to (and is exactly equal to by
\cref{lem:value-coherent-evals}) a term of the form $\trpt{M}$ for some $M \in A$. We cannot expect the same
for indexed inductive types: while validity still excludes $\fhcom$ values, $\fcoe$ values are essential even
in zero dimensions. This is obvious when we consider the identity type, which cannot respect paths while
containing only $\refl$ even in an empty context.

We will now proceed to prove the introduction, Kan condition, and elimination theorems for $\indcl{\K}$. In
\cref{app:proof-theory}, we list a set of selected rules which could form a proof theory based on these
theorems.


\section{Typing rules}
\label{sec:rules}

\subsection{Introduction}

The introduction rules for $\indcl{\K}{\lst{I}}$ follow immediately from
\cref{lem:supports-I-fhcom,lem:supports-I-fcoe,lem:supports-I-intro} and the fact that $\indrel{\K}$ supports
$\K$.


\subsection{Composition}

\begin{figure}[h!]
  \begin{mdframed}
    \input{opsem/hcom}
  \end{mdframed}
  \caption{Operational semantics of homogeneous composition in $\indcl{\K}{\lst{I}}$.}
  \label{fig:ind-opsem-hcom}
\end{figure}

\begin{lemma}
  \label{lem:ind-hcom-kan}
  $\ceqtypep{\indcl[\td{\GD}{\psi}]{\td{\K}{\psi}}{\lst{I}}}{\indcl[\td{\GD'}{\psi}]{\td{\K'}{\psi}}{\lst{I}'}}$
  are equally $\hcom$-Kan for all $\psitd$ and $\ceqtm[\Psi']{\lst{I}}{\lst{I}'}{\td{\GD}{\psi}}$.
\end{lemma}
\begin{proof}
  We have $\hcom*{\indcl[\td{\GD}{\psi}]{\td{\K}{\psi}}{\lst{I}}}{\xi_i} \steps_\cube \fhcom*{\xi_i}$. By
  \cref{lem:expansion}, it therefore suffices to show that K1-3 hold when
  $\hcom{\indcl[\td{\GD}{\psi}]{\td{\K}{\psi}}{\lst{I}}}$ and $\hcom{\indcl{\td{\K'}{\psi}}{\lst{I}'}}$ are
  replaced with $\fhcom$. This is true by \cref{lem:supports-I-fcom} and the fact that
  $\opt{\Fhcom}{\indrel{\K}} \subseteq \indrel{\K}$.
\end{proof}


\subsection{Coercion}

We decompose coercion in $\indcl{\K}{\lst{I}}$ into two operations: $\fcoe$, which coerces along paths in the
index $\lst{I}$, and $\tcoe$, which coerces along paths in the arguments $\GD$ and $\K$. The operational
semantics of $\coe$ and $\tcoe$ are given in \cref{fig:ind-opsem-coe}. 

To state our target typing rule for $\tcoe$, we first define a meta-operation $\mcoe$ (also in
\cref{fig:ind-opsem-coe}), which implements coercion for lists of terms inhabiting a dependent context. It is
straightforward to derive the following typing rules for $\mcoe$ from the $\coe$-Kan conditions for $\GG$ by
mimicking the proofs of the $\coe$-Kan conditions for dependent product types in Part III.

\begin{proposition}
  \label{prop:mcoe-kan-conditions}
  Let $\ceqctxk{\GG}{\GG'}$. For any $\tds{(\Psi',z)}{\psi}{\Psi}$,
  $\dimj[\Psi']{r,r'}$, and
  $\ceqtm{\lst{M}}{\lst{M}'}{\dsubst{\td{\GG}{\psi}}{r}{z}}$, we have
  \begin{description}
  \item{MK4.} $\ceqtm{\mcoe{z.\GG}{r}{r'}{\lst{M}}}{\mcoe{z.\GG'}{r}{r'}{\lst{M}'}}{\dsubst{\td{\GG}{\psi}}{r'}{z}}$,
  \item{MK5.} $\ceqtm{\mcoe{z.\GG}{r}{r}{\lst{M}}}{\lst{M}}{\dsubst{\td{\GG}{\psi}}{r}{z}}$.
  \end{description}
\end{proposition}
\begin{proof}
  This follows by an argument analogous to the proof of the $\coe$-Kan conditions for dependent pair types;
  see \cite[Rule 12]{chtt-iii}.
\end{proof}

The definition of $\tcoe$ is then intended to satisfy the following typing rules:

\begin{mathpar}
  \Infer
  {\ceqtypek[\Psi,z]{\GD}{\GD'} \\
    \ceqconstrs[\Psi,z]{\GD}{\K}{\K'} \\
    \ceqtm{\lst{I}}{\lst{I}'}{\dsubst{\GD}{r}{z}} \\
    \ceqtm{M}{M'}{\indcl[\dsubst{\GD}{r}{z}]{\dsubst{\K}{r}{z}}{\lst{I}}}}
  {\ceqtm{\tcoe{z.(\GD,\K)}{r}{r'}{M}}{%
      \left\{
        \begin{array}{ll}
          \tcoe{z.(\GD',\K')}{r}{r'}{M'}, \\
          M, &\text{if $r = r'$}
        \end{array}
      \right\}
    }{\indcl[\dsubst{\GD}{r'}{z}]{\dsubst{\K}{r'}{z}}{\mcoe{z.\GD}{r}{r'}{\lst{I}}}}}
\end{mathpar}

Notice that $\tcoe$ transfers terms along paths in $\GD$ and $\K$, carrying along the index $\lst{I}$ via
$\mcoe$. The name $\tcoe$ stands for \emph{total space coercion}, the total space being the pair type
$\sigmacl{\Gd}{\GD}{\indcl{\K}{\Gd}}$: given
$\pair{\lst{I}}{M} \in \dsubst{(\sigmacl{\Gd}{\GD}{\indcl{\K}{\Gd}})}{r}{z}$, we have
\[
  \pair{\mcoe{z.\GD}{r}{r'}{\lst{I}}}{\tcoe{z.(\GD,\K)}{r}{r'}{M}} \in
  \dsubst{(\sigmacl{\Gd}{\GD}{\indcl{\K}{\Gd}})}{r'}{z}.
\]

To implement general coercion, we can use $\tcoe$ to transfer between total spaces and then use $\fcoe$ to
move to the desired fiber:
\[
  \coe{z.\indcl{\K}{\lst{I}}}{r}{r'}{M} \steps \fcoe{z.\mcoe{z.\GD}{z}{r'}{\lst{I}}}{r}{r'}{\tcoe{z.(\GD,\K)}{r}{r'}{M}}.
\]
Here, the result of the $\tcoe$ has type
$\indcl[\dsubst{\GD}{r'}{z}]{\dsubst{\K}{r'}{z}}{\mcoe{z.\GD}{r}{r'}{\dsubst{\lst{I}}{r}{z}}}$. The $\fcoe$
interpolates between the indices $\mcoe{z.\GD}{r}{r'}{\dsubst{\lst{I}}{r}{z}}$ and $\dsubst{\lst{I}}{r'}{z}$,
moving along the path $\lst{I}$ and collapsing the $\mcoe$.

Total space coercion is an eager operator, evaluating its argument to a value and then stepping according to
whether it is an $\fhcom$, $\fcoe$, or $\intro*$ term. In the first two cases, $\tcoe$ pushes inside the
arguments of the value. The third is similar, but an adjustment is necessary to ensure that the result has the
correct boundary and lives in the correct index. The necessity arises from the fact that the index function
$\Gg.\lst{I}$ and boundary functions $\Gg.\Gth.\sch{m}_k$ of a constructor
$\K[\ell] = \constr{\lst{x}}{\GG}{\Gg.\lst{I}}{\Gg.\GTh}{\sys{\xi_k}{\Gg.\Gth.\sch{m}_k}}$ may not commute
with coercion on the nose.

\begin{figure}
  \begin{mdframed}
    \input{opsem/coe}
  \end{mdframed}
  \caption{Operational semantics of coercion in $\indcl{\K}{-}$}
  \label{fig:ind-opsem-coe}
\end{figure}

\begin{definition}
  Let $\ceqctxk{\GD}{\GD'}$, let $\open{\Gd}{A,A'}$ be open terms with $|\Gd| = |\GD|$, and let $\Ga$ be a
  value $\GD$-indexed $\Psi$-relation. We say that $(\Gd.A,\Gd,A',\Ga)$ are equally coe-Kan if
  $(\subst{\td{A}{\psi}}{\lst{I}}{\Gd},\subst{\td{A'}{\psi}}{\lst{I}'}{\Gd},\ix{\Ga_\psi}{\lst{I}})$ are
  equally coe-Kan for every $\psitd$ and $\ceqtm[\Psi']{\lst{I}}{\lst{I}'}{\td{\GD}{\psi}}$.
\end{definition}

\begin{lemma}
  \label{prop:argtype-coe-kan}
  Let $\ceqargtype{\GD}{\sch{b}}{\sch{b}'}$, let $\tmj{A,A'}$, and let $\Ga$ be a $\GD$-indexed $\Psi$-PER. If
  $(A,A',\Ga)$ are equally $\coe$-Kan, then
  $(\tyatty{\sch{b}}{A},\tyatty{\sch{b}'}{A'},\tyatrel{\sch{b}}{\Ga})$ are equally $\coe$-Kan.
\end{lemma}
\begin{proof}
  Per the proof of the $\coe$-Kan conditions for dependent function types; see \cite[Rule 6]{chtt-iii}.
\end{proof}

We will now define a relation $\Gs \subseteq \indrel{\K}$ consisting of values on which $\coe{z.\indcl{\K}}$
and $\coe{z.\indcl{\K'}}$ are well-behaved, then proceed to show that it contains all of $\indrel{\K}$.

\begin{definition}
  Given a value $\GD$-indexed $\Psi$-relation $\Ga$, define a value $\GD$-indexed $\Psi$-relation
  $\TcoeI{\Ga}$ as follows. For any $\psitd$ and $\ceqtm[\Psi']{\lst{I}}{\lst{I}'}{\td{\GD}{\psi}}$,
  $\ix{\TcoeI{\Ga}_\psi}{\lst{I}}(V,V')$ is defined to hold when $\ix{\indrel{\K}_\psi}{\lst{I}}(V,V')$
  and, for all $\tds{(\Psi',z)}{\psi_z}{\Psi}$ and $\dimj[\Psi']{r,r'}$ with $\dsubst{\psi_z}{r}{z} = \psi$,
  we have
  \begin{enumerate}
    \setcounter{enumi}{3}
  \item
    $\Tm{\ix{\Ga_{\dsubst{\psi_z}{r'}{z}}}{\mcoe{z.\td{\GD}{\psi_z}}{r}{r'}{\lst{I}}}}(\tcoe{z.\td{(\GD,\K)}{\psi_z}}{r}{r'}{W},\tcoe{z.\td{(\GD',\K')}{\psi_z}}{r}{r'}{W'})$ for all $W,W' \in \{V,V'\}$,
    and
  \item
    $\Tm{\ix{\Ga_{\dsubst{\psi_z}{r}{z}}}{\lst{I}}}(\tcoe{z.\td{(\GD,\K)}{\psi_z}}{r}{r}{W},W)$ for all $W \in \{V,V'\}$.
  \end{enumerate}
  Define the value $\GD$-indexed $\Psi$-PER $\Gs \eqdef \gfp{(\TcoeI)}$ to be the greatest fixed-point of
  $\TcoeI$.
\end{definition}

We can extend the properties that hold of values in $\TcoeI{\Ga}$ by definition to terms in
$\Tm{\TcoeI{\Ga}}$.

\begin{lemma}[Extension to terms]
  \label{lem:tcoei-term-extension}
  Let $\Ga$ be a value $\GD$-indexed $\Psi$-PER. For any $\tds{(\Psi',z)}{\psi_z}{\Psi}$,
  $\dimj[\Psi']{r,r'}$, and $\ceqtm[\Psi']{\lst{I}}{\lst{I}'}{\dsubst{\td{\GD}{\psi_z}}{r}{z}}$, if
  $\Tm{\ix{\TcoeI{\Ga}_{\dsubst{\psi_z}{r}{z}}}{\lst{I}}}(M,M')$, then
  \begin{enumerate}
    \setcounter{enumi}{3}
  \item
    $\Tm{\ix{\Ga_{\dsubst{\psi_z}{r'}{z}}}{\mcoe{z.\td{\GD}{\psi_z}}{r}{r'}{\lst{I}}}}(\tcoe{z.\td{(\GD,\K)}{\psi_z}}{r}{r'}{M},\tcoe{z.\td{(\GD',\K')}{\psi_z}}{r}{r'}{M'})$ and
  \item
    $\Tm{\ix{\Ga_{\dsubst{\psi_z}{r}{z}}}{\lst{I}}}(\tcoe{z.\td{(\GD,\K)}{\psi_z}}{r}{r}{M},M)$.
  \end{enumerate}
\end{lemma}
\begin{proof}
  To show that this is true, it suffices to show that for every $\tds{(\Psi',z)}{\psi_z}{\psi}$,
  $\dimj[\Psi']{r,r'}$, and $\ceqtm[\Psi']{\lst{I}}{\lst{I}'}{\dsubst{\td{\GD}{\psi}}{r}{z}}$, we have
  \begin{enumerate}
    \setcounter{enumi}{3}
  \item
    $\relseq{\oft{a}{\Tm{\TcoeI{\ix{\Ga_{\dsubst{\psi_z}{r}{z}}}{\lst{I}}}}}}{\Tm{\ix{\Ga_{\dsubst{\psi_z}{r'}{z}}}{\mcoe{z.\td{\GD}{\psi_z}}{r}{r'}{\lst{I}}}}(\tcoe{z.\td{(\GD,\K)}{\psi_z}}{r}{r'}{a},\tcoe{z.\td{(\GD',\K')}{\psi_z}}{r}{r'}{a})}$.
  \item $\relseq{\oft{a}{\Tm{\TcoeI{\ix{\Ga_{\dsubst{\psi_z}{r}{z}}}{\lst{I}}}}}}{\Tm{\ix{\Ga_{\dsubst{\psi_z}{r}{z}}}{\lst{I}}}(\tcoe{z.\td{(\GD,\K)}{\psi_z}}{r}{r}{a},a)}$.
  \end{enumerate}
  Since the coercion operator in the inductive type and the identity operator are eager, we can apply
  \cref{lem:elimination} to reduce these to showing
  \begin{enumerate}
    \setcounter{enumi}{3}
  \item
    $\relseq{\oft{a}{\TcoeI{\ix{\Ga_{\dsubst{\psi_z}{r}{z}}}{\lst{I}}}}}{\Tm{\ix{\Ga_{\dsubst{\psi_z}{r'}{z}}}{\mcoe{z.\td{\GD}{\psi_z}}{r}{r'}{\lst{I}}}}(\tcoe{z.\td{(\GD,\K)}{\psi_z}}{r}{r'}{a},\tcoe{z.\td{(\GD',\K')}{\psi_z}}{r}{r'}{a})}$.
  \item
    $\relseq{\oft{a}{\TcoeI{\ix{\Ga_{\dsubst{\psi_z}{r}{z}}}{\lst{I}}}}}{\Tm{\ix{\Ga_{\dsubst{\psi_z}{r}{z}}}{\lst{I}}}(\tcoe{z.\td{(\GD,\K)}{\psi_z}}{r}{r}{a},a)}$.
  \end{enumerate}
  These follow immediately from the definition of $\TcoeI$.
\end{proof}

We now prove a reduction rule for each of the value forms of the inductive type.

\begin{lemma}[$\tcoe$-$\fhcom$-$\beta$]
  \label{lem:tcoe-beta-fhcom}
  Let $\Ga$ be a value $\GD$-indexed $\Psi$-PER. For any $\tds{(\Psi'',z)}{\psi_z}{\Psi'}$,
  $\dimj[\Psi']{r,r'}$, $\ceqtm[\Psi']{\lst{I}}{\lst{I}'}{\dsubst{\td{\GD}{\psi_z}}{r}{z}}$, and
  $\etc{\xi_i}$ valid, if
  \begin{enumerate}
  \item $\Tm{\ix{\TcoeI{\Ga}_{\dsubst{\psi_z}{r}{z}}}{\lst{I}}}(M)$,
  \item $\Tm{\ix{\TcoeI{\Ga}_{\dsubst{\psi_z}{r}{z}}}{\lst{I}}}_{\id;y \mid \xi_i,\xi_j}(N_i,N'_j)$ for all $i,j$,
  \item $\Tm{\ix{\TcoeI{\Ga}_{\dsubst{\psi_z}{r}{z}}}{\lst{I}}}_{\id; \mid \xi_i}(\dsubst{N_i}{s}{y},M)$ for all $i$,
  \end{enumerate}
  then, abbreviating $\fhcom \eqdef \fhcom{s}{s'}{M}{\sys{\xi_i}{y.N_i}}$, we have that
  $\tcoe{z.\td{(\GD,\K)}{\psi_z}}{r}{r'}{\fhcom}$ is related to \\
  $\fhcom{s}{s'}{\tcoe{z.\td{(\GD,\K)}{\psi_z}}{r}{r'}{M}}{\sys{\xi_i}{y.\tcoe{z.\td{(\GD,\K)}{\psi_z}}{r}{r'}{N_i}}}$
  in $\Tm{\ix{\opt{\Fhcom}{\Ga}_{\dsubst{\psi_z}{r'}{z}}}{\mcoe{z.\td{\GD}{\psi_z}}{r}{r'}{\lst{I}}}}$.
\end{lemma}
\begin{proof}
  By \cref{lem:expansion}. Let $\psitd[']$ be given. We have three cases.
  \begin{enumerate}
  \item There exists a least $i$ such that $\models \td{\xi_i}{\psi'}$.

    Then
    $\td{\tcoe{z.\td{(\GD,\K)}{\psi_z}}{r}{r'}{\fhcom}}{\psi'} \steps
    \td{\tcoe{z.\td{(\GD,\K)}{\psi_z}}{r}{r'}{\dsubst{N_i}{r'}{y}}}{\psi'}$.  By
    \cref{lem:tcoei-term-extension,lem:supports-I-fhcom}, the result of this step is related to 
    $\td{\fhcom{s}{s'}{\tcoe{z.\td{(\GD,\K)}{\psi_z}}{r}{r'}{M}}{\sys{\xi_i}{y.\tcoe{z.\td{(\GD,\K)}{\psi_z}}{r}{r'}{N_i}}}}{\psi'}$
    in \\
    $\Tm{\ix{\opt{\Fhcom}{\Ga}_{\dsubst{\psi_z}{r'}{z}}}{\mcoe{z.\td{\GD}{\psi_z}}{r}{r'}{\lst{I}}}}_{\psi'}$.
    
  \item $\td{s}{\psi'} = \td{s'}{\psi'}$ and $\not\models \td{\xi_i}{\psi'}$ for all $i$.

    Then
    $\td{\tcoe{z.\td{(\GD,\K)}{\psi_z}}{r}{r'}{\fhcom}}{\psi'} \steps
    \td{\tcoe{z.\td{(\GD,\K)}{\psi_z}}{r}{r'}{M}}{\psi'}$.  By
    \cref{lem:tcoei-term-extension,lem:supports-I-fhcom}, the result of this step is related to 
    $\td{\fhcom{s}{s'}{\tcoe{z.\td{(\GD,\K)}{\psi_z}}{r}{r'}{M}}{\sys{\xi_i}{y.\tcoe{z.\td{(\GD,\K)}{\psi_z}}{r}{r'}{N_i}}}}{\psi'}$
    in \\
    $\Tm{\ix{\opt{\Fhcom}{\Ga}_{\dsubst{\psi_z}{r'}{z}}}{\mcoe{z.\td{\GD}{\psi_z}}{r}{r'}{\lst{I}}}}_{\psi'}$.
  \item $\td{s}{\psi'} \neq \td{s'}{\psi'}$ and $\not\models \td{\xi_i}{\psi'}$ for all $i$.

    Then
    $\td{\tcoe{z.\td{(\GD,\K)}{\psi_z}}{r}{r'}{\fhcom}}{\psi'} \steps
    \td{\fhcom{s}{s'}{\tcoe{z.\td{(\GD,\K)}{\psi_z}}{r}{r'}{M}}{\sys{\xi_i}{y.\tcoe{z.\td{(\GD,\K)}{\psi_z}}{r}{r'}{N_i}}}}{\psi'}$. By
    \cref{lem:tcoei-term-extension,lem:supports-I-fhcom}, the result of this step is in
    $\Tm{\ix{\opt{\Fhcom}{\Ga}_{\dsubst{\psi_z}{r'}{z}}}{\mcoe{z.\td{\GD}{\psi_z}}{r}{r'}{\lst{I}}}}_{\psi'}$.
    \qedhere
  \end{enumerate}
\end{proof}

\begin{lemma}[$\tcoe$-$\fcoe$-$\beta$]
  \label{lem:tcoe-beta-fcoe}
  Let $\Ga$ be a value $\GD$-indexed $\Psi$-PER. For any $\tds{(\Psi'',z)}{\psi_z}{\Psi'}$,
  $\dimj[\Psi']{r,r'}$, $\ceqtm[\Psi']{\lst{I}}{\lst{I}'}{\dsubst{\td{\GD}{\psi_z}}{r}{z}}$, and
  $\dimj[\Psi']{s,s'}$, if
  \begin{enumerate}
  \item $\coftype[\Psi',y]{\lst{J}}{\dsubst{\td{\GD}{\psi_z}}{r}{z}}$,
  \item $\ceqtm[\Psi']{\dsubst{\lst{J}}{s'}{y}}{\dsubst{\lst{I}}{r}{z}}{\dsubst{\td{\GD}{\psi_z}}{r}{z}}$,
  \item $\Tm{\ix{\TcoeI{\Ga}_{\dsubst{\psi_z}{r}{z}}}{\dsubst{\lst{J}}{s}{y}}}(M)$,
  \end{enumerate}
  then \\
  $\Tm{\ix{\opt{\Fcoe}{\Ga}_{\dsubst{\psi_z}{r'}{z}}}{\mcoe{z.\td{\GD}{\psi_z}}{r}{r'}{\lst{I}}}}(\tcoe{z.\td{(\GD,\K)}{\psi_z}}{r}{r'}{\fcoe{y.\lst{J}}{s}{s'}{M}},\fcoe{y.\mcoe{z.\td{\GD}{\psi_z}}{r}{r'}{\lst{J}}}{s}{s'}{\tcoe{z.\td{(\GD,\K)}{\psi_z}}{r}{r'}{M}})$.
\end{lemma}
\begin{proof}
  By \cref{lem:expansion}. Let $\psitd[']$. We have two cases.
  \begin{enumerate}
  \item $\lst{J} = \emp$ or $\td{s}{\psi'} = \td{s'}{\psi'}$.

    Then
    $\td{\tcoe{z.\td{(\GD,\K)}{\psi_z}}{r}{r'}{\fcoe{y.\lst{J}}{s}{s'}{M}}}{\psi'} \steps
    \td{\tcoe{z.\td{(\GD,\K)}{\psi_z}}{r}{r'}{M}}{\psi'}$. By
    \cref{lem:tcoei-term-extension,lem:supports-I-fcoe}, the result of this step is related to
    $\fcoe{y.\mcoe{z.\td{\GD}{\psi_z}}{r}{r'}{\lst{J}}}{s}{s'}{\tcoe{z.\td{(\GD,\K)}{\psi_z}}{r}{r'}{M}}$ in
    $\Tm{\ix{\opt{\Fcoe}{\Ga}_{\dsubst{\psi_z}{r'}{z}}}{\mcoe{z.\td{\GD}{\psi_z}}{r}{r'}{\lst{I}}}}_\psi$.
  \item $\td{s}{\psi'} \neq \td{s'}{\psi'}$.

    Then
    $\td{\tcoe{z.\td{(\GD,\K)}{\psi_z}}{r}{r'}{\fcoe{y.\lst{J}}{s}{s'}{M}}}{\psi'} \steps
    \fcoe{y.\mcoe{z.\td{\GD}{\psi_z}}{r}{r'}{\lst{J}}}{s}{s'}{\tcoe{z.\td{(\GD,\K)}{\psi_z}}{r}{r'}{M}}$.
    By \cref{lem:tcoei-term-extension,lem:supports-I-fcoe}, the result of this step is in
    $\Tm{\ix{\opt{\Fcoe}{\Ga}_{\dsubst{\psi_z}{r'}{z}}}{\mcoe{z.\td{\GD}{\psi_z}}{r}{r'}{\lst{I}}}}_\psi$. \qedhere
  \end{enumerate}
\end{proof}

\begin{lemma}
  \label{lem:fhcom-coerceable}
  For any value $\GD$-indexed $\Psi$-PER $\Ga$, we have
  $\Fhcom{\TcoeI{\Ga}} \subseteq \TcoeI{\opt{\Fhcom}{\Ga}}$.
\end{lemma}
\begin{proof}
  Let $\Ga$ be given, and suppose that $\ix{\Fhcom{\TcoeI{\Ga}}_\psi}{\lst{I}}(V,V')$ for some $\psitd$ and
  $\coftype[\Psi']{\lst{I}}{\td{\GD}{\psi}}$. Then $V = \fhcom{s}{s'}{M}{\sys{\xi_i}{y.N_i}}$ and
  $V' = \fhcom{s}{s'}{M'}{\sys{\xi_i}{y.N'_i}}$ where $\not\models \xi_i$ for all $i$, $r \neq r'$,
  and
  \begin{enumerate}
  \item $\ix{\TcoeI{\Ga}_\psi}{\lst{I}}(M,M')$,
  \item $\ix{\TcoeI{\Ga}_\psi}{\lst{I}}_{y \mid \xi_i,\xi_j}(N_i,N'_j)$ for all $i,j$,
  \item $\ix{\TcoeI{\Ga}_\psi}{\lst{I}}_{\id \mid \xi_i}(\dsubst{N_i}{s}{y},M)$ for all $i$.
  \end{enumerate}
  To show $\ix{\TcoeI{\opt{\Fhcom}{\Ga}}_\psi}{\lst{I}}(V,V')$, we need to show that
  $\ix{\indrel{\K}_\psi}{\lst{I}}(V,V')$ holds and that for every $\dimj[\Psi']{r,r'}$,
  $\tds{(\Psi',z)}{\psi_z}{\Psi}$ with $\dsubst{\psi_z}{r}{z} = \psi$, and
  $W,W' \in \{V,V'\}$, we have
  \begin{enumerate}
    \setcounter{enumi}{3}
  \item
    $\Tm{\ix{\opt{\Fhcom}{\Ga}_{\dsubst{\psi_z}{r'}{z}}}{\mcoe{z.\td{\GD}{\psi_z}}{r}{r'}{\lst{I}}}}(\tcoe{z.\td{(\GD,\K)}{\psi_z}}{r}{r'}{W},\tcoe{z.\td{(\GD',\K')}{\psi_z}}{r}{r'}{W'})$ for all $W,W' \in \{V,V'\}$,
    and
  \item
    $\Tm{\ix{\opt{\Fhcom}{\Ga}_{\dsubst{\psi_z}{r}{z}}}{\lst{I}}}(\tcoe{z.\td{(\GD,\K)}{\psi_z}}{r}{r}{W},W)$ for all $W \in \{V,V'\}$.
  \end{enumerate}
  We know that $\indrel{\K}_\psi(V,V')$ holds because $\indrel{\K}$ supports $\K$.  To prove the two remaining
  conditions, we apply \cref{lem:tcoe-beta-fhcom} to reduce each occurrence of $\tcoe$ and then apply
  \cref{lem:tcoei-term-extension,lem:supports-I-fhcom} to equate the resulting $\fhcom$ terms.
\end{proof}

\begin{corollary}
  \label{cor:tcoe-sigma-supports-fhcom}
  $\opt{\Fhcom}{\Gs} \subseteq \Gs$.
\end{corollary}
\begin{proof}
  By definition of $\Gs$, it suffices to show that $\opt{\Fhcom}{\Gs}$ is a post-fixed-point of $\TcoeI$,
  i.e., that $\TcoeI{\opt{\Fhcom}{\Gs}} \subseteq \opt{\Fhcom}{\Gs}$. This follows from
  \cref{lem:fhcom-coerceable} and the fact that $\TcoeI{\Gs} = \Gs$.
\end{proof}

\begin{lemma}
  \label{lem:fcoe-coerceable}
  For any value $\GD$-indexed $\Psi$-PER $\Ga$, we have
  $\Fcoe{\TcoeI{\Ga}} \subseteq \TcoeI{\opt{\Fcoe}{\Ga}}$.
\end{lemma}
\begin{proof}
  The proof is directly analogous to that of \cref{lem:fhcom-coerceable}: given an $\fcoe$ term applied to an
  element of $\TcoeI{\Ga}$, we apply \cref{lem:tcoe-beta-fcoe} to reduce it to a $\tcoe$ applied to an
  $\fcoe$, then use \cref{lem:tcoei-term-extension,lem:supports-I-fcoe} (with \cref{prop:mcoe-kan-conditions})
  to show the reduct is well-typed.
\end{proof}

\begin{corollary}
  \label{cor:tcoe-sigma-supports-fcoe}
  $\opt{\Fcoe}{\Gs} \subseteq \Gs$.
\end{corollary}
\begin{proof}
  As in \cref{cor:tcoe-sigma-supports-fhcom}.
\end{proof}

\begin{corollary}
  \label{cor:fcom-coerceable}
  For any value $\GD$-indexed $\Psi$-PER $\Ga$, we have
  $\Fcom{\TcoeI{\Ga}} \subseteq \TcoeI{\opt{\Fcom}{\Ga}}$.
\end{corollary}

\begin{corollary}
  \label{cor:tcoe-sigma-supports-fcom}
  $\opt{\Fcom}{\Gs} \subseteq \Gs$.
\end{corollary}

\begin{lemma}
  \label{lem:tcoe-sigma-coe-kan}
  $(\Gd.\indcl{\K}{\Gd},\Gd.\indcl[\GD']{\K'}{\Gd},\Gs)$ are equally $\coe$-Kan.
\end{lemma}
\begin{proof}
  By consolidating quantifications over dimension substitutions, it suffices to show that for every
  $\tds{(\Psi',z)}{\psi_z}{\Psi}$, $\ceqtm[\Psi,z]{\lst{I}}{\lst{I}'}{\td{\GD}{\psi_z}}$,
  $\dimj[\Psi']{r,r'}$ and $\ix{\Gs_{\psi_z}}{\lst{I}}_{\dsubst{}{r}{z}}(M,M')$, we have
  \begin{enumerate}
    \setcounter{enumi}{3}
  \item $\Tm{\ix{\Gs_{\psi_z}}{\lst{I}}}_{\dsubst{}{r'}{z}}(\coe{z.\td{\indcl{\K}{\lst{I}}}{\psi_z}}{r}{r'}{M},\coe{z.\td{\indcl[\GD']{\K'}{\lst{I}'}}{\psi_z}}{r}{r'}{M'})$,
  \item $\Tm{\ix{\Gs_{\psi_z}}{\lst{I}}}_{\dsubst{}{r}{z}}(\coe{z.\td{\indcl{\K}{\lst{I}}}{\psi_z}}{r}{r}{M},M)$.
  \end{enumerate}
  We have
  $\coe{z.\td{\indcl{\K}{\lst{I}}}{\psi_z}}{r}{r'}{M} \steps_\cube
  \fcoe{z.\mcoe{z.\td{\GD}{\psi_z}}{z}{r'}{\td{\lst{I}}}{\psi_z}}{r}{r'}{\tcoe{z.\td{(\GD,\K)}{\psi_z}}{r}{r'}{M}}$
  and likewise for $\GD',\K',M'$, so it suffices by \cref{lem:expansion} to prove these equations where each
  coercion is replaced by its reduct. By applying \cref{lem:supports-I-fcoe} with
  \cref{cor:tcoe-sigma-supports-fcoe}, we reduce the problem to showing that
  \begin{enumerate}
    \setcounter{enumi}{3}
  \item
    $\Tm{\ix{\Gs_{\dsubst{\psi_z}{r'}{z}}}{\mcoe{z.\td{\GD}{\psi_z}}{r}{r'}{\lst{I}}}}(\tcoe{z.\td{(\GD,\K)}{\psi_z}}{r}{r'}{M},\tcoe{z.\td{(\GD',\K')}{\psi_z}}{r}{r'}{M'})$ and
  \item
    $\Tm{\ix{\Gs_{\dsubst{\psi_z}{r}{z}}}{\lst{I}}}(\tcoe{z.\td{(\GD,\K)}{\psi_z}}{r}{r}{M},M)$.
  \end{enumerate}
  These follow directly from \cref{lem:tcoei-term-extension} and the fact that $\TcoeI{\Gs} = \Gs$.
\end{proof}

For $\intro*$ terms, $\coe$ has two separate reduction rules dealing with constructors without and with
boundary respectively. In each of these, an outer $\fcoe$ is necessary to ensure that the result lives in
desired index. In the second case, we also need an $\fhcom$ to ensure that the result has the right boundary;
we combine the $\fcoe$ and $\fhcom$ into an $\fcom$.

\begin{lemma}[$\tcoe$-$\intro*$-$\beta_{0}$]
  \label{lem:tcoe-beta-intro0}
  For any $\tds{(\Psi',z)}{\psi_z}{\Psi}$, $\dimj[\Psi']{r,r'}$,
  $\ceqtm[\Psi']{\lst{I}}{\lst{I}'}{\dsubst{\td{\GD}{\psi_z}}{r}{z}}$, and $\ell \in \K$, if
  \begin{enumerate}
  \item $\ceqconstrs[\Psi']{\GD}{\K_1}{\dsubst{\td{\K}{\psi_z}}{r}{z}}$,
  \item $\td{\K}{\psi_z}[\ell] = \constr{\lst{x}}{\GG}{\Gg.\lst{J}}{\Gg.\GTh}{\emp}$ where
    $\GTh = \etc{\ofa{\formal{p}_j}{\sch{b}_j}}$,
  \item $\coftype[\Psi']{\lst{P}}{\dsubst{\GG}{r}{z}}$,
  \item $\ceqtm[\Psi']{\subst{\dsubst{\lst{J}}{r}{z}}{\lst{P}}{\Gg}}{\lst{I}}{\dsubst{\td{\GD}{\psi_z}}{r}{z}}$,
  \item
    $\Tm{\tyatrel{\subst{\dsubst{\GTh}{r}{z}}{\lst{P}}{\Gg}}{\dsubst{\td{\Gs}{\psi_z}}{r}{z}}}(\etc{N_j})$,
  \end{enumerate}
  then, abbreviating
  \begin{align*}
    \lst{P}^s &\eqdef \mcoe{z.\GG}{r}{s}{\lst{P}} \\
    (\forall j)\; N_j^s &\eqdef \coe{z.\tyatty{\subst{\sch{b}_j}{\lst{P}^z}{\Gg}}{\Gd.\td{\indcl{\K}{\Gd}}{\psi_z}}}{r}{s}{N_j},
  \end{align*}
  we have that $\tcoe{z.\td{(\GD,\K)}{\psi_z}}{r}{r'}{\intro[\K_1]{\lst{r}}{\lst{P}}{\etc{N_j}}}$ is related
  to
  $\fcoe{z.\mcoe{z.\td{\GD}{\psi_z}}{z}{r'}{\subst{\lst{J}}{\lst{P}^z}{\Gg}}}{r'}{r}{\intro[\dsubst{\td{\K}{\psi_z}}{r'}{z}]{\lst{r}}{\lst{P}^{r'}}{\etc{N_j^{r'}}}}$
  in
  $\Tm{\ix{\opt{\Intro{\K}{\ell}}{\Gs}_{\dsubst{\psi_z}{r'}{z}}}{\mcoe{z.\td{\GD}{\psi_z}}{r}{r'}{\lst{I}}}}$.
\end{lemma}
\begin{proof}
  First, observe the following:
  \begin{enumerate}
  \item By \cref{prop:mcoe-kan-conditions}, we have (a)
    $\coftype[\Psi',z]{\lst{P}^z}{\GG}$ and (b)
    $\ceqtm[\Psi']{\lst{P}^r}{\lst{P}}{\dsubst{\GG}{r}{z}}$.
  \item By \cref{lem:tcoe-sigma-coe-kan,prop:argtype-coe-kan}, we know that \\
    $(\tyatty{\subst{\sch{b}_j}{\lst{P}^z}{\Gg}}{\Gd.\indcl[\GD']{\td{\K}{\psi_z}}{\Gd}},\tyatty{\subst{\sch{b}_j}{\lst{P}^z}{\Gg}}{\Gd.\indcl{\td{\K'}{\psi_z}}{\Gd}},\tyatty{\subst{\sch{b}_j}{\lst{P}^z}{\Gg}}{\td{\Gs}{\psi_z}})$
    are equally $\coe$-Kan, so we have
    \begin{enumerate}
    \item
      $\Tm{\tyatrel{\subst{\sch{b}_j}{\lst{P}^z}{\Gg}}{\td{\Gs}{\psi_z}}}(N_j^z)$
      for all $j$,
    \item
      $\Tm{\tyatrel{\subst{\dsubst{\sch{b}_j}{r}{z}}{\lst{P}^r}{\Gg}}{\dsubst{\td{\Gs}{\psi_z}}{r}{z}}}(N_j^r,N_j)$
      for all $j$.
    \end{enumerate}
  \end{enumerate}
  We obtain
  $\Tm{\ix{\opt{\Intro{\K}{\ell}}{\Gs}_{\dsubst{\psi_z}{r'}{z}}}{\subst{\dsubst{\lst{J}}{r'}{z}}{\lst{P}^{r'}}{\Gg}}}(\intro[\dsubst{\td{\K}{\psi_z}}{r'}{z}]{\lst{r}}{\lst{P}^{r'}}{\etc{N_j^{r'}}})$
  by supplying 1(a) and 2(a) to \cref{lem:supports-I-intro}. 
  \begin{enumerate}
    \setcounter{enumi}{2}
  \item By \cref{prop:mcoe-kan-conditions}, fact 1 above,
    and the assumption
    $\ceqtm[\Psi']{\subst{\dsubst{\lst{J}}{r}{z}}{\lst{P}}{\Gg}}{\lst{I}}{\dsubst{\td{\GD}{\psi_z}}{r}{z}}$, we
    have
    \begin{enumerate}
    \item $\coftype[\Psi',z]{\mcoe{z.\td{\GD}{\psi_z}}{z}{r'}{\subst{\lst{J}}{\lst{P}^z}{\Gg}}}{\dsubst{\td{\GD}{\psi_z}}{r'}{z}}$.
    \item $\ceqtm[\Psi',z]<z=r>{\mcoe{z.\td{\GD}{\psi_z}}{z}{r'}{\subst{\lst{J}}{\lst{P}^z}{\Gg}}}{\mcoe{z.\td{\GD}{\psi_z}}{z}{r'}{\lst{I}}}{\dsubst{\td{\GD}{\psi_z}}{r'}{z}}$,
    \item $\ceqtm[\Psi',z]<z=r'>{\mcoe{z.\td{\GD}{\psi_z}}{z}{r'}{\subst{\lst{J}}{\lst{P}^z}{\Gg}}}{\subst{\dsubst{\lst{J}}{r'}{z}}{\lst{P}^{r'}}{\Gg}}{\dsubst{\td{\GD}{\psi_z}}{r'}{z}}$,
    \end{enumerate}
  \end{enumerate}
  By \cref{lem:supports-I-fcoe}, we thus have
  \[\Tm{\ix{\opt{\Intro{\K}{\ell}}{\Gs}_{\dsubst{\psi_z}{r'}{z}}}{\mcoe{z.\td{\GD}{\psi_z}}{r}{r'}{\lst{I}}}}(\fcoe{z.\mcoe{z.\td{\GD}{\psi_z}}{z}{r'}{\subst{\lst{J}}{\lst{P}^z}{\Gg}}}{r'}{r}{\intro[\dsubst{\td{\K}{\psi_z}}{r'}{z}]{\lst{r}}{\lst{P}^{r'}}{\etc{N_j^{r'}}}}).\]
  The desired equality then follows by \cref{lem:expansion}, as
  $\tcoe{z.\td{(\GD,\K)}{\psi_z}}{r}{r'}{\intro[\K_1]{\lst{r}}{\lst{P}}{\etc{N_j}}}$ stably steps to the term
  above.
\end{proof}

\begin{lemma}[$\tcoe$-$\intro*$-$\beta_{>0}$]
  \label{lem:tcoe-beta-intron}
  For any $\tds{(\Psi',z)}{\psi_z}{\Psi}$, $\dimj[\Psi']{r,r'}$,
  $\ceqtm[\Psi']{\lst{I}}{\lst{I}'}{\dsubst{\td{\GD}{\psi_z}}{r}{z}}$, and $\ell \in \K$ such that $\Gs$
  supports $\K_{<\ell}$, if
  \begin{enumerate}
  \item $\ceqconstrs[\Psi']{\GD}{\K_1}{\dsubst{\td{\K}{\psi_z}}{r}{z}}$,
  \item $\td{\K}{\psi_z}[\ell] = \constr{\lst{x}}{\GG}{\Gg.\lst{I}}{\Gg.\GTh}{\sys{\xi_k}{\Gg.\Gth.\sch{m}_k}}$ where
    $\GTh = \etc{\ofa{\formal{p}_j}{\sch{b}_j}}$ and $\etc{\xi_k} \neq \emp$,
  \item $\coftype{\lst{P}}{\dsubst{\GG}{r}{z}}$,
  \item $\ceqtm[\Psi']{\subst{\dsubst{\lst{J}}{r}{z}}{\lst{P}}{\Gg}}{\lst{I}}{\dsubst{\td{\GD}{\psi_z}}{r}{z}}$,
  \item 
    $\Tm{\tyatrel{\subst{\dsubst{\GTh}{r}{z}}{\lst{P}}{\Gg}}{\dsubst{\td{\Gs}{\psi_z}}{r}{z}}}(\etc{N_j})$,
  \end{enumerate}
  then, abbreviating
  \begin{align*}
    \K_1^s &\eqdef \dsubst{\td{\K}{\psi_z}}{s}{z} \\
    \lst{P}^s &\eqdef \mcoe{z.\GG}{r}{s}{\lst{P}} \\
    (\forall j)\; N_j^s &\eqdef \coe{z.\tyatty{\subst{\sch{b}_j}{\lst{P}^z}{\Gg}}{\Gd.\td{\indcl{\K}{\Gd}}{\psi_z}}}{r}{s}{N_j} \\
    (\forall k)\; M_k^s &\eqdef \tcoe{z.\td{(\GD,\K)}{\psi_z}}{s}{r'}{\insttm{\Gth.\subst{\dsubst{\dsubst{\sch{m}_k}{\lst{r}}{\lst{x}}}{s}{z}}{\lst{P}^s}{\Gg}}{\K_1^s}{\etc{N_j^s}}}
  \end{align*}
  and
  \begin{align*}
    \intro* &\eqdef \intro[\K_1]{\lst{r}}{\lst{P}}{\etc{N_j}} \\
    O &\eqdef \fcom{z.\mcoe{z.\td{\GD}{\psi_z}}{z}{r'}{\subst{\lst{J}}{\lst{P}^z}{\Gg}}}{r'}{r}{\intro[\dsubst{\td{\K}{\psi_z}}{r'}{z}]{\lst{r}}{\lst{P}^{r'}}{\etc{N_j^{r'}}}}{\sys{\dsubst{\xi_k}{\lst{r}}{\lst{x}}}{y.M_k^y}}
  \end{align*}
  we have
  $\Tm{\ix{\opt{\Fcom}{\opt{\Intro{\K}{\ell}}{\Gs}}_{\dsubst{\psi_z}{r'}{z}}}{\mcoe{z.\td{\GD}{\psi_z}}{r}{r'}{\lst{I}}}}(\tcoe{z.\td{(\GD,\K)}{\psi_z}}{r}{r'}{\intro*},O)$.
\end{lemma}
\begin{proof}
  In addition to facts 1-3 from the previous proof, we also have the following.
  \begin{enumerate}
    \addtocounter{enumi}{3}
  \item By 1-3, the typing assumptions on $\etc{\sch{m}_k}$, \cref{lem:insttm-typing} (using the
    assumption that $\Gs$ supports $\K_{<\ell}$), and the fact that $\sigma \subseteq \TcoeI{\sigma}$, we have
    \begin{enumerate}
    \item
      $\Tm{\ix{\Gs_{\dsubst{\psi_z}{r'}{z};z}}{\mcoe{z.\td{\GD}{\psi}}{z}{r'}{\subst{\lst{J}}{\lst{P}^{z}}{\Gg}}}}_{\id \mid \dsubst{\xi_k}{\lst{r}}{\lst{x}},\dsubst{\xi_l}{\lst{r}}{\lst{x}}}(M_k^z,M_l^z)$
      for all $k,l$,
    \item
      $\Tm{\ix{\Gs_{\dsubst{\psi_z}{r'}{z}}}{\mcoe{z.\td{\GD}{\psi}}{r}{r'}{\lst{I}}}}_{\id \mid \dsubst{\xi_k}{\lst{r}}{\lst{x}}}(M_k^r,\tcoe{z.\td{(\GD,\K)}{\psi_z}}{r}{r'}{\insttm{\Gth.\subst{\dsubst{\dsubst{\sch{m}_k}{\lst{r}}{\lst{x}}}{r}{z}}{\lst{P}}{\Gg}}{\K_1}{\etc{N_j}}})$
      for all $k$,
    \item
      $\Tm{\ix{\Gs_{\dsubst{\psi_z}{r'}{z}}}{\subst{\dsubst{\lst{J}}{r'}{z}}{\lst{P}^{r'}}{\Gg}}}_{\id \mid \dsubst{\xi_k}{\lst{r}}{\lst{x}}}(M_k^{r'},\insttm{\Gth.\subst{\dsubst{\dsubst{\sch{m}_k}{\lst{r}}{\lst{x}}}{r'}{z}}{\lst{P}^{r'}}{\Gg}}{\K_1^{r'}}{\etc{N_j^{r'}}})$
      for all $k$.
    \end{enumerate}
  \item Applying \cref{lem:supports-I-intro} with 1-2 and then \cref{lem:supports-I-fcom} with 3-4, we obtain
    \begin{enumerate}
    \item $\Tm{\opt{\Fcom}{\opt{\Intro{\K}{\ell}}{\sigma}}_{\dsubst{\psi_z}{r'}{z}}}_{\id\mid\dsubst{\xi_k}{\lst{r}}{\lst{x}}}(O,M^r_k)$ for all $k$,
    \item $\Tm{\opt{\Fcom}{\opt{\Intro{\K}{\ell}}{\sigma}}_{\dsubst{\psi_z}{r'}{z}}}(O)$.
    \end{enumerate}
  \end{enumerate}
  We now proceed by \cref{lem:expansion}. Let $\psitd[']$ be given; we have two cases.
  \begin{itemize}
  \item There exists a least $k$ such that
    $\models \td{\dsubst{\xi_k}{\lst{r}}{\lst{x}}}{\psi'}$.

    Then $\td{\tcoe{z.\td{(\GD,\K)}{\psi_z}}{r}{r'}{\intro*}}{\psi'} \steps
    \td{\tcoe{z.\td{(\GD,\K)}{\psi_z}}{r}{r'}{\insttm{\Gth.\subst{\dsubst{\dsubst{\sch{m}_k}{\lst{r}}{\lst{x}}}{r}{z}}{\lst{P}}{\Gg}}{\K_1}{\etc{N_j}}}}{\psi'}$. Observing
    that the reduct is $M^r_k$, we apply 5(a).
  \item There is no such $k$.

    Then $\td{\tcoe{z.\td{(\GD,\K)}{\psi_z}}{r}{r'}{\intro*}}{\psi'} \steps \td{O}{\psi'}$. We apply 5(b). \qedhere
  \end{itemize}
\end{proof}

\begin{theorem}
  \label{thm:tcoe-sigma-supports}
  $\Gs$ supports $\K$.
\end{theorem}
\begin{proof}
  We prove that $\Gs$ supports every prefix $\K_1 \constrspre \K$ by induction on the form of $\K_1$.
  \begin{enumerate}
  \item $\K_1 = \nilconstrs$.

    Then we have to show $\Fhcom{\Gs} \cup \Fcoe{\Gs} \subseteq \Gs$. This holds by
    \cref{cor:tcoe-sigma-supports-fhcom,cor:tcoe-sigma-supports-fcoe}.

  \item $\K_1 = \snocconstrs{\K_2}{\ell : \ldots}$.

    Then we have to show that $\F_{\K_2}(\Gs) \cup \Intro{\K}{\ell}{\Gs} \subseteq \Gs$.  By
    induction hypothesis we know $\F_{\K_2}(\Gs) \subseteq \Gs$, so it remains to show
    $\Intro{\K}{\ell}{\Gs} \subseteq \Gs$. We prove the stronger statement that
    $\multi{\Fcom}{\opt{\Intro{\K}{\ell}}{\Gs}} \subseteq \Gs$.  By universal property of $\Gs$,
    it suffices to show
    $\multi{\Fcom}{\opt{\Intro{\K}{\ell}}{\Gs}} \subseteq
    \TcoeI{\multi{\Fcom}{\opt{\Intro{\K}{\ell}}{\Gs}}}$.

    By the universal property of $\multi{\Fcom}$ and definition of
    $\opt{\Intro{\K}{\ell}}$, it is then enough to show that
    \begin{enumerate}
    \item $\Gs \subseteq \TcoeI{\multi{\Fcom}{\opt{\Intro{\K}{\ell}}{\Gs}}}$,
    \item $\Intro{\K}{\ell}{\Gs} \subseteq \TcoeI{\multi{\Fcom}{\opt{\Intro{\K}{\ell}}{\Gs}}}$,
    \item $\Fcom{\TcoeI{\multi{\Fcom}{\opt{\Intro{\K}{\ell}}{\Gs}}}} \subseteq \TcoeI{\multi{\Fcom}{\opt{\Intro{\K}{\ell}}{\Gs}}}$.
    \end{enumerate}
    We prove these in turn.
    \begin{enumerate}
    \item This holds because $\Gs = \TcoeI{\Gs} \subseteq \TcoeI{\multi{\Fcom}{\opt{\Intro{\K}{\ell}}{\Gs}}}$.
    \item Suppose $\ix{\Intro{\K}{\ell}{\Gs}_\psi}{\lst{I}}(V,V')$ holds for some $\psitd$ and
      $\coftype[\Psi']{\lst{I}}{\td{\GD}{\psi}}$. To show
      $\ix{\TcoeI{\multi{\Fcom}{\opt{\Intro{\K}{\ell}}{\Gs}}}_\psi}{\lst{I}}(V,V')$, we need to show that
      $\ix{\indrel{\K}_\psi}{\lst{I}}(V,V')$ holds and that for every $\dimj[\Psi']{r,r'}$,
      $\tds{(\Psi',z)}{\psi_z}{\Psi}$ with $\dsubst{\psi_z}{r}{z} = \psi$, and $W,W' \in \{V,V'\}$, we have
      \begin{enumerate}[label=\arabic*.]
        \setcounter{enumiii}{3}
      \item
        $\Tm{\ix{\multi{\Fcom}{\opt{\Intro{\K}{\ell}}{\Gs}}_{\dsubst{\psi_z}{r'}{z}}}{\mcoe{z.\td{\GD}{\psi_z}}{r}{r'}{\lst{I}}}}(\tcoe{z.\td{(\GD,\K)}{\psi_z}}{r}{r'}{W},\tcoe{z.\td{(\GD',\K')}{\psi_z}}{r}{r'}{W'})$,
      \item
        $\Tm{\ix{\multi{\Fcom}{\opt{\Intro{\K}{\ell}}{\Gs}}_{\dsubst{\psi_z}{r}{z}}}{\lst{I}}}(\tcoe{z.\td{(\GD,\K)}{\psi_z}}{r}{r}{W},W)$.
      \end{enumerate}
      We know that $\ix{\indrel{\K}_\psi}{\lst{I}}(V,V')$ holds because $\Gs \subseteq \indrel{\K}$ and
      $\indrel{\K}$ supports $\K$.  We prove
      the other two statements as follows.
      \begin{enumerate}[label=\arabic*.]
        \setcounter{enumiii}{3}
      \item By either \cref{lem:tcoe-beta-intro0} or \cref{lem:tcoe-beta-intron}, depending on whether
        $\K[\ell]$ has a boundary, we have that
        \begin{itemize}
        \item $\Tm{\ix{\multi{\Fcom}{\opt{\Intro{\K}{\ell}}{\Gs}}_{\dsubst{\psi_z}{r'}{z}}}{\mcoe{z.\td{(\GD,\K)}{\psi_z}}{r}{r'}{\lst{I}}}}(\tcoe{z.\td{(\GD,\K)}{\psi_z}}{r}{r'}{W},O)$,
        \item $\Tm{\ix{\multi{\Fcom}{\opt{\Intro{\K}{\ell}}{\Gs}}_{\dsubst{\psi_z}{r'}{z}}}{\mcoe{z.\td{(\GD,\K)}{\psi_z}}{r}{r'}{\lst{I}}}}(\tcoe{z.\td{(\GD',\K')}{\psi_z}}{r}{r'}{W'},O')$
        \end{itemize}
        hold, where $O$ and $O'$ are as defined in the appropriate lemma. The right-hand sides of these
        equations are themselves equal in $\Tm{\multi{\Fcom}{\opt{\Intro{\K}{\ell}}{\Gs}}}$ by
        \cref{lem:tcoe-sigma-coe-kan,lem:supports-I-intro,lem:supports-I-fcom,cor:tcoe-sigma-supports-fcom},
        where we use the induction hypothesis that $\Gs$ supports $\K_2$ in order to apply
        \cref{lem:supports-I-intro}.
      \item Again, we have
        $\Tm{\ix{\multi{\Fcom}{\opt{\Intro{\K}{\ell}}{\Gs}}_{\dsubst{\psi_z}{r}{z}}}{\lst{I}}}(\tcoe{z.\td{(\GD,\K)}{\psi_z}}{r}{r}{W},O)$
        (this time with $r'$ replaced with $r$ in $O$), and the right-hand side is equal to $W$ in
        $\Tm{\multi{\fhcom}{\opt{\Intro{\K}{\ell}}{\Gs}}}$ by
        \cref{lem:tcoe-sigma-coe-kan,lem:supports-I-intro,lem:supports-I-fcom,cor:tcoe-sigma-supports-fcom}.
      \end{enumerate}

    \item We have
      $\Fcom{\TcoeI{\multi{\Fcom}{\opt{\Intro{\K}{\ell}}{\Gs}}}} \subseteq
      \TcoeI{\opt{\Fcom}{\multi{\Fcom}{\opt{\Intro{\K}{\ell}}{\Gs}}}}$ by
      \cref{cor:fcom-coerceable}, and the result follows because
      $\opt{\Fcom} \circ \multi{\Fcom} = \multi{\Fcom}$. \qedhere
    \end{enumerate}
  \end{enumerate}
\end{proof}

\begin{corollary}
  \label{lem:ind-coe-kan}
  $\ceqtypep{\indcl[\td{\GD}{\psi}]{\td{\K}{\psi}}{\lst{I}}}{\indcl[\td{\GD'}{\psi}]{\td{\K'}{\psi}}{\lst{I}'}}$
  are equally $\coe$-Kan for all $\psitd$ and $\ceqtm[\Psi']{\lst{I}}{\lst{I}'}{\td{\GD}{\psi}}$.
\end{corollary}
\begin{proof}
  By \cref{thm:tcoe-sigma-supports}, we have $\Gs = \indrel{\K}$. The result thus follows by
  \cref{lem:tcoe-sigma-coe-kan}.
\end{proof}


\subsection{Elimination}

For elimination, we separate our presentation into two parts. First, we specify the data which is provided to
the eliminator for $\indcl{\K}{-}$. Second, we prove the typing rules for said eliminator.

\subsubsection{Elimination data}

\begin{definition}
  The grammar of \emph{elimination lists} is given by
  \[
    \begin{array}{rcl}
      \E &::=& \nilelim \mid \snocelim{\E}{\ell : \lst{x}.\Gg.\Gh.\Gr.R}\; \text{(where $|\Gh| = |\Gr|$)}
    \end{array}
  \]
  As with constructor lists, we write $\E[\ell]$ for the entry at label $\ell$, $\E_{<\ell}$ for the prefix
  preceding $\ell$, and $\E \elimpre \E'$ to mean that $\E$ is a prefix of $\E'$.  We say that an elimination
  list $\E$ \emph{matches} a constructor list $\K$ if for every $\ell$ with $\E[\ell] = \lst{x}.\Gg.\Gh.\Gr.R$
  and $\K[\ell] = \constr{\lst{x}}{\GG}{\Gg.\lst{I}}{\Gg.\GTh}{\sys{\xi_k}{\Gg.\Gth.\sch{m}_k}}$, we have
  $|\Gth| = |\Gh| = |\Gr|$.
\end{definition}

A case $\lst{x}.\Gg.\Gh.\Gr.R$ in an an elimination list gives access to the dimension parameters $\lst{x}$,
non-recursive arguments $\Gg$, and recursive arguments $\Gh$ to the constructor, along with the results $\Gr$
of the recursive calls on arguments in $\Gh$. To state the types of the recursive call results, we define a
dependent interpretation function for argument types.

\begin{definition}[Syntactic dependent type interpretation]
  Let an argument type $\sch{b}$ and terms $\open{\Gd,h}{\tmj{D}}$, and $\tmj{N}$ be given. We define a term
  $\tyatty*{\sch{b}}{\Gd.h.D}{N}$ as follows.
  \[
    \begin{array}{rcl}
      \tyatty*{\argvar{\lst{I}}}{\Gd.h.D}{N} &\eqdef& \subst{\subst{D}{\lst{I}}{\Gd}}{N}{h} \\
      \tyatty*{\argpi{b}{B}{\sch{c}}}{\Gd.h.D}{N} &\eqdef& \picl{b}{B}{\tyatty*{\sch{c}}{\Gd.h.D}{\app{N}{b}}}
    \end{array}
  \]
  For a context $\GTh = \etc{\sch{b}_j}$ and terms $\etc{N_j}$, we will write $\tyatty*{\GTh}{\Gd.h.D}{\etc{N_j}}$ for the list
  $\etc{\tyatty*{\sch{b}_j}{\Gd.h.D}{N_j}}$.
\end{definition}

\begin{definition}[Semantic dependent type interpretation]
  Let $\cwfargtype{\GD}{\sch{b}}$ and a value $(\Gd{:}\GD,\indcl{\K}{\Gd})$-indexed $\Psi$-relation $\Ga$ be
  given. We define a value $\tyatrel{\sch{b}}{\Gd.\indcl{\K}{\Gd}}$-indexed $\Psi$-relation
  $\tyatrel*{\sch{b}}{\Ga}$ by recursion on the structure of $\sch{b}$:
  \[
    \begin{array}{rcl}
      \ix{\tyatrel*{\argvar{\lst{I}}}{\Ga}_\psi}{N} &\eqdef& \ix{\Ga_\psi}{\td{\lst{I}}{\psi},N} \\
      \ix{\tyatrel*{\argpi{b}{B}{\sch{c}}}{\Ga}_\psi}{N} &\eqdef& \Pirel{\td{B}{\psi}}{(\psi',M) \mapsto \ix{\tyatrel*{\subst{\td{\sch{c}}{\psi\psi'}}{M}{b}}{\td{\Ga}{\psi\psi'}}}{\app{\td{N}{\psi'}}{M}}}
    \end{array}
  \]
\end{definition}

\begin{proposition}
  If
  \begin{enumerate}
  \item $\ceqargtype{\GD}{\sch{b}}{\sch{b}'}$,
  \item $\eqtypek{\ofc{\Gd}{\GD},\oft{h}{\indcl{\K}{\Gd}}}{D}{D'}$,
  \end{enumerate}
  then
  \begin{enumerate}
  \item $\eqtypek{\oft{b}{\tyatty{\sch{b}}{\Gd.\indcl{\K}{\Gd}}}}{\tyatty*{\sch{b}}{\Gd.h.D}{b}}{\tyatty*{\sch{b}'}{\Gd.h.D'}{b}}$, and
  \item $\vper{\tyatty*{\sch{b}}{\Gd.h.D}{N}} = \ix{\tyatrel*{\sch{b}}{(\psi,\lst{I},M) \mapsto \vper{\subst{\subst{\td{D}{\psi}}{\lst{I}}{\Gd}}{M}{h}}}}{N}$
    for any $\coftype{N}{\tyatty{\sch{b}}{\Gd.\indcl{\K}{\Gd}}}$.
  \end{enumerate}
\end{proposition}

To state the coherence conditions required of the eliminator cases, which ensure that the boundary of a
constructor's case lines up with the cases for that constructor's boundary, we also define dependent
instantiation of boundary terms.

\begin{definition}[Dependent boundary interpretation]
  Let a constructor list $\K$, matching elimination list $\E$, and terms $\open{h}{\tmj{D}}$ and
  $\tmj{\lst{I},\lst{N},\lst{S}}$ with $|\lst{N}| = |\lst{S}|$ be given. For any $\open{\Gth}{\bndj{\sch{m}}}$
  with $\fb{\sch{m}} \subseteq \Gth$, $\fl{\sch{m}} \subseteq \dom{\K} \cap \dom{\E}$ and
  $|\Gth| = |\lst{N}|$, we define a term $\insttm*{\Gth.\sch{m}}{\K}{\E}{\Gd.h.D}{\lst{N}}{\lst{S}}$ by
  \[
    \begin{array}{rcl}
      \insttm*{\Gth.\Gth[j]}{\K}{\E}{\Gd.h.D}{\lst{N}}{\lst{S}} &\eqdef& \lst{S}[j] \\[.5em]
      \insttm*{\Gth.\bndintro{\lst{r}}{\lst{P}}{\lst{\sch{n}}}}{\K}{\E}{\Gd.h.D}{\lst{N}}{\lst{S}} &\eqdef& \subst{\subst{\subst{\dsubst{R}{\lst{r}}{\lst{x}}}{\lst{P}}{\Gg}}{\insttm{\Gth.\lst{\sch{n}}}{\K}{\lst{N}}}{\Gh}}{\insttm*{\Gth.\lst{\sch{n}}}{\K}{\E}{\Gd.h.D}{\lst{N}}{\lst{S}}}{\Gr} \\
                                                     && \text{where $\E[\ell] = \lst{x}.\Gg.\Gh.\Gr.R$} \\[.5em]
      \insttm*{\Gth.\bndfhcom*{\lst{I}}{\xi_i}}{\K}{\E}{\Gd.h.D}{\lst{N}}{\lst{S}} &\eqdef& \com{y.\subst{\subst{D}{\lst{I}}{\Gd}}{F^y}{h}}{r}{r'}{\insttm*{\Gth.\sch{m}}{\K}{\E}{\Gd.h.D}{\lst{N}}{\lst{S}}}{\sys{\xi_i}{y.\insttm*{\Gth.\sch{n}_i}{\K}{\E}{\Gd.h.D}{\lst{N}}{\lst{S}}}} \\[.5em]
                                                                &&\text{where $F^y = \insttm{\Gth.\bndfhcom{\lst{I}}{r}{y}{\sch{m}}{\sys{\xi_i}{\sch{n}_i}}}{\K}{\lst{N}}$} \\[.5em]
      \insttm*{\Gth.\bndfcoe{z.\lst{I}}{r}{r'}{\sch{m}}}{\K}{\E}{\Gd.h.D}{\lst{N}}{\lst{S}} &\eqdef& \coe{z.\subst{\subst{D}{\lst{I}}{\Gd}}{F^z}{h}}{r}{r'}{\insttm*{\Gth.\sch{m}}{\K}{\E}{\Gd.h.D}{\lst{N}}{\lst{S}}} \\[.5em]
      &&\text{where $F^z = \insttm{\Gth.\bndfcoe{z.\lst{I}}{r}{z}{\sch{m}}}{\K}{\lst{N}}$} \\[.5em]
      \insttm*{\Gth.\bndlam{a}{\sch{n}}}{\K}{\E}{\Gd.h.D}{\lst{N}}{\lst{S}} &\eqdef& \lam{a}{(\insttm*{\Gth.\sch{n}}{\K}{\E}{\Gd.h.D}{\lst{N}}{\lst{S}})} \\[.5em]
      \insttm*{\Gth.\bndapp{\sch{n}}{M}}{\K}{\E}{\Gd.h.D}{\lst{N}}{\lst{S}} &\eqdef& \app{\insttm*{\Gth.\sch{n}}{\K}{\E}{\Gd.h.D}{\lst{N}}{\lst{S}}}{M}
    \end{array}
  \]
  Here and henceforth, we write $\insttm*{\Gth.\etc{\sch{m}_n}}{\K}{\E}{\Gd.h.D}{\lst{N}}{\lst{S}}$ for a list
  of terms $\open{\Gth}{\etc{\sch{m}_n}}$ to mean
  $\etc{\insttm*{\Gth.\sch{m}_n}{\K}{\E}{\Gd.h.D}{\lst{N}}{\lst{S}}}$.

\end{definition}

We will establish a typing rule for dependent boundary interpretation in \cref{lem:insttm*-typing} after
defining a typing judgment for elimination lists.

\begin{proposition}[Basic facts on boundary interpretation]~
  \label{prop:insttm*-facts}
  \begin{enumerate}
  \item $\subst{\insttm*{\Gth.\sch{m}}{\K}{\E}{\Gd.h.D}{\lst{N}}{\lst{S}}}{P}{a} = \insttm*{\Gth.\subst{\sch{m}}{P}{a}}{\subst{\K}{P}{a}}{\subst{\E}{P}{a}}{\Gd.h.\subst{D}{P}{a}}{\subst{\lst{N}}{P}{a}}{\subst{\lst{S}}{P}{a}}$.
  \item If $\open{\Gf}{\bndj{\sch{m}}}$ and $\open{\Gth}{\bndj{\lst{\sch{n}}}}$ with $|\Gf| = |\lst{\sch{n}}|$, then
    \[
      \insttm*{\Gth.\subst{\sch{m}}{\lst{\sch{n}}}{\Gf}}{\K}{\E}{\Gd.h.D}{\lst{N}}{\lst{S}} = \insttm*{\Gf.\sch{m}}{\K}{\E}{\Gd.h.D}{\insttm{\Gth.\lst{\sch{n}}}{\K}{\lst{N}}}{\insttm*{\Gth.\lst{\sch{n}}}{\K}{\E}{\Gd.h.D}{\lst{N}}{\lst{S}}}.
    \]
  \item If $\K \constrspre \K'$ and $\E \constrspre \E'$, then
    $\insttm*{\Gth.\sch{m}}{\K}{\E}{\Gd.h.D}{\lst{N}}{\lst{S}} = \insttm*{\Gth.\sch{m}}{\K'}{\E'}{\Gd.h.D}{\lst{N}}{\lst{S}}$.
  \end{enumerate}
\end{proposition}

\begin{definition}[Elimination list typing]
  \label{def:elim-list}
  Presupposing $\cwfconstrs{\GD}{\K}$ and $\wftypek{\oft{\Gd}{\GD},\oft{h}{\indcl{\K}{\Gd}}}{D}$, the judgment
  $\ceqcasespart{\GD}{\E}{\E'}{\K}{\Gd.h.D}$ is inductively defined by the following rules.
  \begin{mathpar}
    \Infer
    { }
    {\ceqcasespart{\GD}{\nilelim}{\nilelim}{\K}{\Gd.h.D}}
    \and
    \Infer
    {\ceqcasespart{\GD}{\E}{\E'}{\K}{\Gd.h.D} \\
      \height{\K}{\ell} = |\E| \\
      \K[\ell] = \constr{\lst{x}}{\GG}{\Gg.\lst{I}}{\Gg.\GTh}{\sys{\xi_k}{\Gg.\Gth.\sch{m}_k}} \\
      \eqtm[\Psi,\lst{x}]{\ofc{\Gg}{\GG},\ofc{\Gh}{\tyatty{\GTh}{\indcl{\K}{\lst{I}}}},\ofc{\Gr}{\tyatty*{\GTh}{\Gd.h.D}{\Gh}}}{R}{R'}{\subst{D}{\intro{\lst{x}}{\Gg}{\Gh}}{h}} \\
      (\forall k)\;\eqtm[\Psi,\lst{x}\mid\xi_k]{\ofc{\Gg}{\GG},\ofc{\Gh}{\tyatty{\GTh}{\indcl{\K}{\lst{I}}}},\ofc{\Gr}{\tyatty*{\GTh}{\Gd.h.D}{\Gh}}}{R}{\insttm*{\Gth.\sch{m}_k}{\K}{\E}{\Gd.h.D}{\Gh}{\Gr}}{\subst{D}{\intro{\lst{x}}{\Gg}{\Gh}}{h}}
    }
    {\ceqcasespart{\GD}{\snocelim{\E}{\ell:\lst{x}.\Gg.\Gh.\Gr.R}}{\snocelim{\E'}{\ell:\lst{x}.\Gg.\Gh.\Gr.R'}}{\K}{\Gd.h.D}}
  \end{mathpar}
  We say that $\ceqcases{\GD}{\E}{\E'}{\K}{h.D}$ when $\ceqcasespart{\GD}{\E}{\E'}{\K}{h.D}$ and
  $|\K| = |\E|$.
\end{definition}

\begin{lemma}[Dependent boundary interpretation typing]
  \label{lem:insttm*-typing}
  If
  \begin{enumerate}
  \item $\ceqconstrs{\GD}{\K}{\K'}$,
  \item $\eqtypek{\ofc{\Gd}{\GD},\oft{h}{\indcl{\K}{\Gd}}}{D}{D'}$,
  \item $\ceqcases{\GD}{\E}{\E'}{\K}{\Gd.h.D}$,
  \item $\ceqargtm{\GD}{\ofc{\Gth}{\GTh}}{\sch{m}}{\sch{m}'}{\sch{b}}$,
  \item $\ceqtm{\lst{N}}{\lst{N}'}{\tyatty{\GTh}{\Gd.\indcl{\K}{\Gd}}}$,
  \item $\ceqtm{\lst{S}}{\lst{S}'}{\tyatty*{\GTh}{\Gd.h.D}{\lst{N}}}$,
  \end{enumerate}
  then
  $\ceqtm{\insttm*{\Gth.\sch{m}}{\K}{\E}{\Gd.h.D}{\lst{N}}{\lst{S}}}{\insttm*{\Gth.\sch{m}'}{\K'}{\E'}{\Gd.h.D'}{\lst{N}}{\lst{S}}}{\tyatty*{\sch{b}}{\Gd.h.D}{\insttm{\Gth.\sch{m}}{\K}{\lst{N}}}}$.
\end{lemma}
\begin{proof}
  By strong induction on $\height{\K}{\sch{m},\sch{m}'}$ and an inner induction on the rules defining
  $\ceqargtm{\GD}{\ofc{\Gth}{\GTh}}{\sch{m}}{\sch{m}'}{\sch{b}}$. We prove a few representative cases.
  \begin{description}
  \item{\rulename{Hyp}} Then $\sch{m} = \sch{m}' = \Gth[j]$ and $\sch{b} = \GTh[j]$ for some $j$, so
    $\insttm*{\Gth.\sch{m}}{\K}{\E}{\Gd.h.D}{\lst{N}}{\lst{S}} = \lst{S}[j]$,
    $\insttm*{\Gth.\sch{m}'}{\K'}{\E'}{\Gd.h.D}{\lst{N'}}{\lst{S'}} = \lst{S}'[j]$, and
    $\tyatty*{\sch{b}}{\Gd.h.D}{\insttm{\Gth.\sch{m}}{\K}{\lst{N}}} = \tyatty*{\GTh[j]}{\Gd.h.D}{\lst{N}}$.
    We therefore want to show $\ceqtm{\lst{S}[j]}{\lst{S}'[j]}{\tyatty*{\GTh[j]}{\Gd.h.D}{\lst{N}[j]}}$. This
    holds by assumption.
  \item{\rulename{$\bndintro$-I}} Then
    $\sch{m} = \bndintro{\lst{r}}{\lst{P}}{\lst{\sch{n}}}$,
    $\sch{m}' = \bndintro{\lst{r}}{\lst{P}'}{\lst{\sch{n}}'}$, and $\sch{b} = \argvar{\lst{I}}$, so
    \begin{itemize}
    \item $\insttm*{\Gth.\sch{m}}{\K}{\E}{\Gd.h.D}{\lst{N}}{\lst{S}} = \subst{\subst{\subst{\dsubst{R}{\lst{r}}{\lst{x}}}{\lst{P}}{\Gg}}{\insttm{\Gth.\lst{\sch{n}}}{\K}{\lst{N}}}{\Gh}}{\insttm*{\Gth.\lst{\sch{n}}}{\K}{\E}{\Gd.h.D}{\lst{N}}{\lst{S}}}{\Gr}$,
    \item $\insttm*{\Gth.\sch{m}'}{\K'}{\E'}{\Gd.h.D}{\lst{N}'}{\lst{S}'} = \subst{\subst{\subst{\dsubst{R'}{\lst{r}}{\lst{x}}}{\lst{P}'}{\Gg}}{\insttm{\Gth.\lst{\sch{n}}'}{\K'}{\lst{N}'}}{\Gh}}{\insttm*{\Gth.\lst{\sch{n}}'}{\K'}{\E'}{\Gd.h.D}{\lst{N}'}{\lst{S}'}}{\Gr}$,
    \item $\tyatty*{\sch{b}}{\Gd.h.D}{\insttm{\Gth.\sch{m}}{\K}{\lst{N}}} = \subst{\subst{D}{\subst{\lst{I}}{\lst{P}}{\Gg}}{\Gd}}{\intro{\lst{r}}{\lst{P}}{\insttm{\Gth.\lst{\sch{n}}}{\K}{\lst{N}}}}{h}$,
    \end{itemize}
    where $\K[\ell] = \constr{\lst{x}}{\GG}{\Gg.\lst{I}}{\Gg.\GF}{\sys{\xi_k}{\Gg.\Gf.\sch{m}_k}}$,
    $\E[\ell] = \lst{x}.\Gg.\Gh.\Gr.R$, and $\E'[\ell] = \lst{x}.\Gg.\Gh.\Gr.R'$. We know that
    \begin{itemize}
    \item $\eqtm[\Psi,\lst{x}]{\ofc{\Gg}{\GG},\ofc{\Gh}{\tyatty{\GF}{\indcl{\K}{\lst{I}}}},\ofc{\Gr}{\tyatty*{\GF}{\Gd.h.D}{\Gh}}}{R}{R'}{\subst{D}{\intro{\lst{x}}{\Gg}{\Gh}}{h}}$,
    \item $\ceqtm{\lst{P}}{\lst{P}'}{\GG}$ from the premises of \rulename{$\bndintro$-I},
    \item
      $\ceqtm{\insttm{\Gth.\lst{\sch{n}}}{\K}{\lst{N}}}{\insttm{\Gth.\lst{\sch{n}}'}{\K'}{\lst{N}'}}{\tyatty{\subst{\GF}{\lst{P}}{\Gg}}{\Gd.\indcl{\K}{\Gd}}}$
      by the premises of \rulename{$\bndintro$-I} and \cref{lem:insttm-typing},
    \item
      $\ceqtm{\insttm*{\Gth.\lst{\sch{n}}}{\K}{\E}{\Gd.h.D}{\lst{N}}{\lst{S}}}{\insttm*{\Gth.\lst{\sch{n}}'}{\K'}{\E'}{\Gd.h.D'}{\lst{N}'}{\lst{S}'}}{\tyatty*{\subst{\GF}{\lst{P}}{\Gg}}{\Gd.h.D}{\insttm{\Gth.\lst{\sch{n}}}{\K}{\lst{N}}}}$
      by induction hypothesis.
    \end{itemize}
    Plugging $\lst{r}$ and these equations in for $\Gg,\Gh,\Gr$ in $R,R'$, we get the equality of
    $\insttm*{\Gth.\sch{m}}{\K}{\E}{\Gd.h.D}{\lst{N}}{\lst{S}}$ and
    $\insttm*{\Gth.\sch{m}'}{\K'}{\E'}{\Gd.h.D'}{\lst{N}'}{\lst{S}'}$ in
    $\subst{\subst{D}{\subst{\lst{I}}{\lst{P}}{\Gg}}{h}}{\intro{\lst{r}}{\lst{P}}{\insttm{\Gth.\lst{\sch{n}}}{\K}{\lst{N}}}}{h}$.
    
  \item{\rulename{$\bndintro$-B}}
    Then $\sch{m} = \bndintro{\lst{r}}{\lst{P}}{\lst{\sch{n}}}$, $\sch{m}' = \subst{\subst{\dsubst{\sch{m}_k}{\lst{r}}{\lst{x}}}{\lst{P}}{\Gg}}{\lst{\sch{n}}}{\Gf}$, and $\sch{b} = \argvar{\lst{I}}$, so
    \begin{itemize}
    \item $\insttm*{\Gth.\sch{m}}{\K}{\E}{\Gd.h.D}{\lst{N}}{\lst{S}} = \subst{\subst{\subst{\dsubst{R}{\lst{r}}{\lst{x}}}{\lst{P}}{\Gg}}{\insttm{\Gth.\lst{\sch{n}}}{\K}{\lst{N}}}{\Gh}}{\insttm*{\Gth.\lst{\sch{n}}}{\K}{\E}{\Gd.h.D}{\lst{N}}{\lst{S}}}{\Gr}$,
    \item
      $\insttm*{\Gth.\sch{m}'}{\K'}{\E'}{\Gd.h.D'}{\lst{N}'}{\lst{S}'} =
      \insttm*{\Gth.\subst{\subst{\dsubst{\sch{m}_k}{\lst{r}}{\lst{x}}}{\lst{P}}{\Gg}}{\lst{\sch{n}}}{\Gf}}{\K'}{\E'}{\Gd.h.D'}{\lst{N}'}{\lst{S}'}$,
      which by \cref{prop:insttm*-facts}(b) is equal to
      $\insttm*{\Gf.\subst{\dsubst{\sch{m}_k}{\lst{r}}{\lst{x}}}{\lst{P}}{\Gg}}{\K'}{\E'}{\Gd.h.D'}{\insttm{\Gth.\lst{\sch{n}}}{\K}{\lst{N}}}{\insttm*{\Gth.\lst{\sch{n}}}{\K'}{\E'}{\Gd.h.D'}{\lst{N}}{\lst{S}}}$,
    \item $\tyatty*{\sch{b}}{\Gd.h.D}{\insttm{\Gth.\sch{m}}{\K}{\lst{N}}} = \subst{\subst{D}{\subst{\lst{I}}{\lst{P}}{\Gg}}{\Gd}}{\intro{\lst{r}}{\lst{P}}{\insttm{\Gth.\lst{\sch{n}}}{\K}{\lst{N}}}}{h}$,
    \end{itemize}
    where
    $\K[\ell] = \constr{\lst{x}}{\GG}{\Gg.\lst{I}}{\Gg.\GF}{\sys{\xi_k}{\Gg.\Gf.\sch{m}_k}}$,
    $\E[\ell] = \lst{x}.\Gg.\Gh.\Gr.R$, and $\E'[\ell] =
    \lst{x}.\Gg.\Gh.\Gr.R'$. By $\ceqcases{\GD}{\E}{\E'}{\K}{\Gd.h.D}$, we have
    \[
      \eqtm[\Psi,\lst{x}\mid\xi_k]{\ofc{\Gg}{\GG},\ofc{\Gh}{\tyatty{\GF}{\indcl{\K}{\lst{I}}}},\ofc{\Gr}{\tyatty*{\GF}{\Gd.h.D}{\Gh}}}{R}{\insttm*{\Gf.\sch{m}_k}{\K}{\E}{\Gd.h.D}{\Gh}{\Gr}}{\subst{D}{\intro{\lst{x}}{\Gg}{\Gh}}{h}}.
    \]
    
    By induction hypothesis, using the fact that $\height{\K}{\sch{m}_k} < \height{\K}{\sch{m},\sch{m}'}$, we
    know that $\insttm*{\Gf.\sch{m}_k}{\K}{\E}{\Gd.h.D}{\Gh}{\Gr}$ is equal at this type to
    $\insttm*{\Gf.\sch{m}_k}{\K'}{\E'}{\Gd.h.D'}{\Gh}{\Gr}$. We also have:
    \begin{itemize}
    \item $\coftype{\lst{P}}{\GG}$ from the premises of \rulename{$\bndintro$-B},
    \item
      $\ceqtm{\insttm{\Gth.\lst{\sch{n}}}{\K}{\lst{N}}}{\insttm{\Gth.\lst{\sch{n}}}{\K'}{\lst{N}'}}{\tyatty{\subst{\GF}{\lst{P}}{\Gg}}{\Gd.\indcl{\K}{\Gd}}}$
      by the premises of \rulename{$\bndintro$-B} and \cref{lem:insttm-typing},
    \item
      $\ceqtm{\insttm*{\Gth.\lst{\sch{n}}}{\K}{\E}{\Gd.h.D}{\lst{N}}{\lst{S}}}{\insttm*{\Gth.\lst{\sch{n}}'}{\K'}{\E'}{\Gd.h.D'}{\lst{N}'}{\lst{S}'}}{\tyatty*{\subst{\GF}{\lst{P}}{\Gg}}{\Gd.h.D}{\insttm{\Gth.\lst{\sch{n}}}{\K}{\lst{N}}}}$
      by induction hypothesis.
    \end{itemize}
    Plugging these into the above equation, we get the equality of
    $\insttm*{\Gth.\sch{m}}{\K}{\E}{\Gd.h.D}{\lst{N}}{\lst{S}}$ and
    $\insttm*{\Gth.\sch{m}'}{\K'}{\E'}{\Gd.h.D'}{\lst{N}'}{\lst{S}'}$ in
    $\subst{\subst{D}{\subst{\lst{I}}{\lst{P}}{\Gg}}{\Gd}}{\intro{\lst{r}}{\lst{P}}{\insttm{\Gth.\lst{\sch{n}}}{\K}{\lst{N}}}}{h}$. \qedhere
  \end{description}
\end{proof}

\subsubsection{Elimination rules}

\begin{figure}
  \begin{mdframed}
    \input{opsem/elim}
  \end{mdframed}
  \caption{Operational semantics of $\elim$}
  \label{fig:ind-opsem-elim}
\end{figure}

For the remainder of this section, we fix $\ceqctxk{\GD}{\GD'}$, families
$\eqtypek{\ofc{\Gd}{\GD},\oft{h}{\indcl{\K}{\Gd}}}{D}{D'}$, and $\ceqcases{\GD}{\E}{\E'}{\K}{\Gd.h.D}$.

Recursive calls to the eliminator are mediated by an operator $\func$, which gives the action of an argument
type $\sch{b}$ on a map $\Gd.h.R$ out of the family $\Gd.\indcl{\K}{\Gd}$. The operator, defined in
\cref{fig:ind-opsem-elim}, satisfies the following typing rule.

\begin{lemma}[Action of argument types]
  \label{lem:func-typing}
  If
  \begin{enumerate}
  \item $\ceqargtype{\GD}{\sch{b}}{\sch{b}'}$,
  \item $\Ga \subseteq \indrel{\K}$ is a value $\GD$-indexed $\Psi$-PER,
  \item for every $\psitd$, $\ceqtm[\Psi']{\lst{I}}{\lst{I}'}{\td{\GD}{\psi}}$, and
    $\Tm{\ix{\Ga_\psi}{\lst{I}}}(M,M')$, we have \\
    $\ceqtm[\Psi']{\subst{\subst{\td{F}{\psi}}{\lst{I}}{\Gd}}{M}{a}}{\subst{\subst{\td{F'}{\psi}}{\lst{I}'}{\Gd}}{M'}{a}}{\subst{\subst{\td{D}{\psi}}{\lst{I}}{\Gd}}{M}{a}}$,
  \end{enumerate}
  then for every $\psitd$ and $\Tm{\tyatrel*{\sch{b}}{\Ga}}_\psi(N,N')$, we have \\
  $\ceqtm[\Psi']{\func{\td{\sch{b}}{\psi}}{\Gd.h.\td{R}{\psi}}{N}}{\func{\td{\sch{b}'}{\psi}}{\Gd.h.\td{R'}{\psi}}{N'}}{\tyatty*{\sch{b}}{\Gd.h.D}{N}}$.
\end{lemma}
\begin{proof}
  By induction on the derivation of $\ceqargtype{\GD}{\sch{b}}{\sch{b}'}$. 
\end{proof}

Given $\GTh = \etc{\formal{p}_j : \sch{b}_j}$ and terms $\etc{N_j}$, we will write
$\func{\GTh}{\Gd.h.R}{\etc{N_j}}$ for the list $\etc{\func{\sch{b}_j}{\Gd.h.R}{N_j}}$.

As with $\coe$, we will prove the elimination typing rule by first defining a
subrelation $\Gs \subseteq \indrel{\K}$ on which the eliminator is
well-behaved, then showing that the eliminator satisfies $\fhcom$-$\beta$ and
$\intro*$-$\beta$ rules, then using these $\beta$-rules to show that $\Gs$ is
closed under $\Fhcom$ and $\Intro{\K}{\ell}$. The $\elim$ proof is in some ways
conceptually simpler than the $\coe$ proof, because we are no longer mapping
back into the inductive type.

\begin{definition}
  We define a value $\GD$-indexed $\Psi$-PER $\Gs \subseteq \indrel{\K}$ by taking
  $\ix{\Gs_\psi}{\lst{I}}(V,V')$ to hold when for every $\ceqtm[\Psi']{\lst{J}}{\lst{J}'}{\td{\GD}{\psi}}$
  with $\ceqtm[\Psi']{\lst{I}}{\lst{J}}{\td{\GD}{\psi}}$ and every $W,W' \in \{V,V'\}$, we have
  $\ceqtm[\Psi']{\elim{\Gd.h.\td{D}{\psi}}{\lst{J}}{W}{\td{\E}{\psi}}}{\elim{\Gd.h.\td{D'}{\psi}}{\lst{J'}}{W'}{\td{\E'}{\psi}}}{\subst{\subst{\td{D}{\psi}}{\lst{I}}{\Gd}}{V}{h}}$.
\end{definition}

\begin{lemma}[Extension to terms]
  \label{lem:elimi-term-extension}
  For every $\psitd$, $\ceqtm[\Psi']{\lst{I}}{\lst{I}'}{\td{\GD}{\psi}}$, and
  $\ix{\Gs_\psi}{\lst{I}}(M,M')$, we have
  $\ceqtm[\Psi']{\elim{\Gd.h.\td{D}{\psi}}{\lst{I}}{M}{\td{\E}{\psi}}}{\elim{\Gd.h.\td{D'}{\psi}}{\lst{I}'}{M'}{\td{\E'}{\psi}}}{\subst{\subst{\td{D}{\psi}}{\lst{I}}{\Gd}}{M}{h}}$.
\end{lemma}
\begin{proof}
  By \cref{lem:elimination}, as the eliminator is eager and $\indrel{\K}$ is
  value-coherent.
\end{proof}

\begin{lemma}[$\fhcom$-$\beta$]
  \label{lem:elim-beta-fhcom}
  For any $\psitd$, $\coftype[\Psi']{\lst{I}}{\td{\GD}{\psi}}$, $\dimj[\Psi']{r,r'}$, and $\etc{x_i}$ valid,
  if
  \begin{enumerate}
  \item $\Tm{\ix{\Gs_\psi}{\lst{I}}}(M)$,
  \item $\Tm{\ix{\Gs_\psi}{\lst{I}}}_{\id\mid \xi_i,\xi_j}(N_i,N_j)$ for all $i,j$,
  \item $\Tm{\ix{\Gs_\psi}{\lst{I}}}_{\id\mid \xi_i}(M,\dsubst{N_i}{r}{y})$ for all $i$,
  \end{enumerate}  
  then, abbreviating $F^s := \fhcom{r}{s}{M}{\sys{\xi_i}{y.N_i}}$, we have
  \begin{align*}
    \elim{\Gd.h.\td{D}{\psi}}{\lst{I}}{F^{r'}}{\td{\E}{\psi}} 
    &\eq 
    \com{y.\subst{\subst{\td{D}{\psi}}{\lst{J}}{\Gd}}{F^y}{h}}{r}{r'}{\elim{\Gd.h.\td{D}{\psi}}{\lst{J}}{M}{\td{\E}{\psi}}}{\sys{\xi_i}{y.\elim{\Gd.h.\td{D}{\psi}}{\lst{J}}{N_i}{\td{\E}{\psi}}}}
  \end{align*}
  in $\subst{\subst{\td{D}{\psi}}{\lst{I}}{\Gd}}{F^{r'}}{h}$.
\end{lemma}
\begin{proof}
  By \cref{lem:expansion}. Let $\psitd[']$ be given. We have three cases.
  \begin{enumerate}
  \item There exists a least $i$ such that $\models \td{\xi_i}{\psi'}$.

    Then
    $\td{\elim{\Gd.h.\td{D}{\psi}}{\lst{I}}{F^{r'}}{\td{\E}{\psi}}}{\psi'} \steps
    \td{\elim{\Gd.h.\td{D}{\psi}}{\lst{I}}{\dsubst{N_i}{r'}{y}}{\td{\E}{\psi}}}{\psi'}$. By
    \cref{lem:elimi-term-extension,lem:supports-I-fhcom,prop:com}, the reduct is equal to
    $\td{\com{y.\subst{\subst{\td{D}{\psi}}{\lst{J}}{\Gd}}{F^y}{h}}{r}{r'}{\elim{\Gd.h.\td{D}{\psi}}{\lst{J}}{M}{\td{\E}{\psi}}}{\sys{\xi_i}{y.\elim{\Gd.h.\td{D}{\psi}}{\lst{J}}{N_i}{\td{\E}{\psi}}}}}{\psi'}$
    in $\td{\subst{\subst{\td{D}{\psi}}{\lst{I}}{\Gd}}{F^{r'}}{h}}{\psi'}$.
  \item $\td{r}{\psi'} = \td{r'}{\psi'}$ and $\not\models \td{\xi_i}{\psi'}$ for all $i$.
    
    Then
    $\td{\elim{\Gd.h.\td{D}{\psi}}{\lst{I}}{F^{r'}}{\td{\E}{\psi}}}{\psi'} \steps
    \td{\elim{\Gd.h.\td{D}{\psi}}{\lst{I}}{M}{\td{\E}{\psi}}}{\psi'}$. By
    \cref{lem:elimi-term-extension,lem:supports-I-fhcom,prop:com}, the reduct is equal to
    $\td{\com{y.\subst{\subst{\td{D}{\psi}}{\lst{J}}{\Gd}}{F^y}{h}}{r}{r'}{\elim{\Gd.h.\td{D}{\psi}}{\lst{J}}{M}{\td{\E}{\psi}}}{\sys{\xi_i}{y.\elim{\Gd.h.\td{D}{\psi}}{\lst{J}}{N_i}{\td{\E}{\psi}}}}}{\psi'}$
    in $\td{\subst{\subst{\td{D}{\psi}}{\lst{I}}{\Gd}}{F^{r'}}{h}}{\psi'}$.

  \item $\td{r}{\psi'} = \td{r'}{\psi'}$ and $\not\models \td{\xi_i}{\psi'}$ for all $i$.

    Then
    $\td{\elim{\Gd.h.\td{D}{\psi}}{\lst{I}}{F^{r'}}{\td{\E}{\psi}}}{\psi'} \steps
    \td{\com{y.\subst{\subst{\td{D}{\psi}}{\lst{J}}{\Gd}}{F^y}{h}}{r}{r'}{\elim{\Gd.h.\td{D}{\psi}}{\lst{J}}{M}{\td{\E}{\psi}}}{\sys{\xi_i}{y.\elim{\Gd.h.\td{D}{\psi}}{\lst{J}}{N_i}{\td{\E}{\psi}}}}}{\psi'}$.
    By \cref{lem:elimi-term-extension,lem:supports-I-fhcom,prop:com}, the reduct is in
    $\td{\subst{\subst{\td{D}{\psi}}{\lst{I}}{\Gd}}{F^{r'}}{h}}{\psi'}$. \qedhere
  \end{enumerate}
\end{proof}

\begin{lemma}[$\fcoe$-$\beta$]
  \label{lem:elim-beta-fcoe}
  For any $\psitd$, $\coftype[\Psi']{\lst{I}}{\td{\GD}{\psi}}$, and $\tmj[\Psi']{r,r'}$, if
  \begin{enumerate}
  \item $\coftype[\Psi',z]{\lst{J}}{\td{\GD}{\psi}}$ with $\ceqtm[\Psi']{\dsubst{\lst{J}}{r'}{z}}{\lst{I}}{\td{\GD}{\psi}}$,
  \item $\Tm{\ix{\Gs_\psi}{\dsubst{\lst{J}}{r}{z}}}(M)$,
  \end{enumerate}
  then
  \begin{align*}
    \elim{\Gd.h.\td{D}{\psi}}{\lst{I}}{\fcoe{z.\lst{J}}{r}{r'}{M}}{\td{\E}{\psi}}
    &\eq
    \coe{z.\subst{\subst{\td{D}{\psi}}{\lst{J}}{\Gd}}{\fcoe{z.\lst{J}}{r}{z}{M}}{h}}{r}{r'}{\elim{\Gd.h.\td{D}{\psi}}{\dsubst{\lst{J}}{r}{z}}{M}{\td{\E}{\psi}}}
  \end{align*}
  in $\subst{\subst{\td{D}{\psi}}{\lst{I}}{\Gd}}{\fcoe{z.\lst{J}}{r}{r'}{M}}{h}$.
\end{lemma}

\begin{proof}
  By \cref{lem:expansion}. Let $\psitd[']$ be given. We have two cases.
  \begin{enumerate}
  \item $\td{r}{\psi'} = \td{r'}{\psi'}$.

    Then
    $\td{\elim{Gd.h.\td{D}{\psi}}{\lst{I}}{\fcoe{z.\lst{J}}{r}{r'}{M}}{\td{\E}{\psi}}}{\psi'} \steps
    \td{\elim{\Gd.h.\td{D}{\psi}}{\lst{I}}{M}{\td{\E}{\psi}}}{\psi'}$. By
    \cref{lem:elimi-term-extension,lem:supports-I-fcoe} and the assumption that $D$ is Kan, the reduct is
    equal to
    $\coe{z.\subst{\subst{\td{D}{\psi}}{\lst{J}}{\Gd}}{\fcoe{z.\lst{J}}{r}{z}{M}}{h}}{r}{r'}{\elim{\Gd.h.\td{D}{\psi}}{\dsubst{\lst{J}}{r}{z}}{M}{\td{\E}{\psi}}}$
    in $\subst{\subst{\td{D}{\psi}}{\lst{I}}{\Gd}}{\fcoe{z.\lst{J}}{r}{r'}{M}}{h}$.
  \item $\td{r}{\psi'} \neq \td{r'}{\psi'}$.

    Then
    $\td{\elim{\Gd.h.\td{D}{\psi}}{\lst{I}}{\fcoe{z.\lst{J}}{r}{r'}{M}}{\td{\E}{\psi}}}{\psi'} \steps
    \td{\coe{z.\subst{\subst{\td{D}{\psi}}{\lst{J}}{\Gd}}{\fcoe{z.\lst{J}}{r}{z}{M}}{h}}{r}{r'}{\elim{\Gd.h.\td{D}{\psi}}{\dsubst{\lst{J}}{r}{z}}{M}{\td{\E}{\psi}}}}{\psi'}$. By
    \cref{lem:elimi-term-extension,lem:supports-I-fcoe}, the reduct is in
    $\subst{\subst{\td{D}{\psi}}{\lst{I}}{\Gd}}{\fcoe{z.\lst{J}}{r}{r'}{M}}{h}$. \qedhere
  \end{enumerate}
\end{proof}

As with the interleaved proofs of \cref{lem:supports-I-intro,lem:insttm-typing},
we will extract a sub-lemma of the $\intro*$-$\beta$ rule establishing a
property of the term interpretation functions. In this case, the property is a
sort of $\beta$-rule for the eliminator applied to boundary terms.

\begin{definition}
  We say that the property $\ElimBnd{n}$ holds for some $n \le |\K|$ when for all $\psitd$ and
  \begin{enumerate}
  \item $\cwfargtm[\Psi']{\td{\GD}{\psi}}[\td{\K}{\psi}]{\ofc{\Gth}{\GTh}}{\sch{m}}{\sch{b}}$ with $\height{\td{\K}{\psi}}{\sch{m}} = n$,
  \item $\Tm{\tyatrel{\GTh}{\Gs}}(\lst{N})$,
  \end{enumerate}
  we have
  \begin{align*}
    \func{\sch{b}}{\Gd.h.\elim{h.\Gd.\td{D}{\psi}}{\Gd}{h}{\td{\E}{\psi}}}{\insttm{\Gth.\sch{m}}{\td{\K}{\psi}}{\lst{N}}}
    &\eq
    \insttm*{\Gth.\sch{m}}{\td{\K}{\psi}}{\td{\E}{\psi}}{\Gd.h.D}{\lst{N}}{\func{\GTh}{h.\elim{\Gd.h.\td{D}{\psi}}{\Gd}{h}{\td{\E}{\psi}}}{\lst{N}}}
  \end{align*}
  in $\tyatty*{\sch{b}}{\Gd.h.\td{D}{\psi}}{\insttm{\Gth.\sch{m}}{\td{\K}{\psi}}{\lst{N}}}$.
\end{definition}

\begin{lemma}[$\intro*$-$\beta$]
  \label{lem:elim-beta-intro}
  Let $\ell \in \K$ such that $\ElimBnd{n}$ holds for all $n < \height{\K}{\ell}$. For any $\psitd$ and
  $\coftype[\Psi']{\lst{I}}{\td{\GD}{\psi}}$, if
  \begin{enumerate}
  \item $\td{\K}{\psi}[\ell] = \constr{\lst{x}}{\GG}{\Gg.\lst{J}}{\Gg.\GTh}{\sys{\xi_k}{\Gg.\Gth.\sch{m}_k}}$
    and $\td{\E}{\psi}[\ell] = \lst{x}.\Gg.\Gh.\Gr.R$,
  \item $\ceqconstrs[\Psi']{\GD}{\td{\K}{\psi}}{\K'}$,
  \item $\coftype[\Psi']{\lst{P}}{\GG}$,
  \item $\ceqtm[\Psi']{\subst{\lst{J}}{\lst{P}}{\Gg}}{\lst{I}}{\Gd}$,
  \item $\tyatrel{\subst{\GTh}{\lst{P}}{\Gg}}{\td{\Gs}{\psi}}(\lst{N})$,
  \end{enumerate}
  then, abbreviating $\intro* \eqdef \intro[\K']{\lst{r}}{\lst{P}}{\lst{N}}$, we have
  \begin{align*}
    \elim{\Gd.h.\td{D}{\psi}}{\lst{I}}{\intro*}{\td{\E}{\psi}}
    &\eq
    \subst{\subst{\subst{\dsubst{R}{\lst{r}}{\lst{x}}}{\lst{P}}{\Gg}}{\lst{N}}{\Gh}}{\func{\subst{\GTh}{\lst{P}}{\Gg}}{\Gd.h.\elim{\Gd.h.\td{D}{\psi}}{\Gd}{h}{\td{\E}{\psi}}}{\lst{N}}}{\Gr}
  \end{align*}
  in $\subst{\subst{\td{D}{\psi}}{\lst{I}}{\Gd}}{\intro*}{h}$.
\end{lemma}
\begin{proof}
  By \cref{lem:expansion}. Let $\psitd[']$ be given. We have two cases.
  \begin{itemize}
  \item There exists a least $k$ such that
    $\models \td{\dsubst{\xi_k}{\lst{r}}{\lst{x}}}{\psi'}$.

    Then
    $\td{\elim{\Gd.h.\td{D}{\psi}}{\lst{I}}{\intro*}{\td{\E}{\psi}}}{\psi'} \steps
    \td{\elim{\Gd.h.\td{D}{\psi}}{\lst{I}}{\insttm{\Gth.\subst{\dsubst{\sch{m}_k}{\lst{r}}{\lst{x}}}{\lst{P}}{\Gg}}{\K'}{\lst{N}}}{\td{\E}{\psi}}}{\psi'}$. By
    $\ElimBnd{|\K_{<\ell}|}$, we have
    \begin{gather*}
      \td{\elim{\Gd.h.\td{D}{\psi}}{\lst{I}}{\insttm{\Gth.\subst{\dsubst{\sch{m}_k}{\lst{r}}{\lst{x}}}{\lst{P}}{\Gg}}{\K'}{\lst{N}}}{\td{\E}{\psi}}}{\psi'} \\
      \eq \\
      \td{\insttm*{\Gth.\subst{\dsubst{\sch{m}_k}{\lst{r}}{\lst{x}}}{\lst{P}}{\Gg}}{\td{\K}{\psi}}{\td{\E}{\psi}}{h.\td{D}{\psi}}{\lst{N}}{\func{\subst{\GTh}{\lst{P}}{\Gg}}{\Gd.h.\elim{\Gd.h.\td{D}{\psi}}{\Gd}{h}{\td{\E}{\psi}}}{\lst{N}}}}{\psi'}
    \end{gather*}
    in a type which is equal to $\td{\subst{\subst{\td{D}{\psi}}{\lst{I}}{\Gd}}{\intro*}{h}}{\psi'}$ by
    \cref{lem:supports-I-intro}(a).  The right-hand side of this equation is equal to $\td{O}{\psi'}$ in
    $\td{\subst{\subst{\td{D}{\psi}}{\lst{I}}{\Gd}}{\intro*}{h}}{\psi'}$ by the assumptions on $R$ in
    $\cwfcases{\GD}{\E}{\K}{\Gd.h.D}$ together with \cref{lem:func-typing,lem:elimi-term-extension}.
  \item $\not\models \td{\dsubst{\xi_k}{\lst{r}}{\lst{x}}}{\psi'}$ for all $k$.

    Then $\td{\elim{\Gd.h.\td{D}{\psi}}{\lst{I}}{\intro*}{\td{\E}{\psi}}}{\psi'} \steps \td{O}{\psi'}$, and the reduct
    is in
    $\td{\subst{\subst{\td{D}{\psi}}{\lst{I}}{\Gd}}{\intro*}{h}}{\psi'}$ by the assumptions on $R$ in
    $\cwfcases{\GD}{\E}{\K}{\Gd.h.D}$ together with
    \cref{lem:func-typing,lem:elimi-term-extension}. \qedhere
  \end{itemize}
\end{proof}

\begin{lemma}
  \label{lem:boundary-naturality}
  $\ElimBnd{n}$ holds for all $n \in \N$.
\end{lemma}
\begin{proof}
  By induction on $n$ and the derivation of
  $\cwfargtm[\Psi']{\td{\GD}{\psi}}[\td{\K}{\psi}]{\ofc{\Gth}{\GTh}}{\sch{m}}{\sch{b}}$. Assume
  $\ElimBnd{m}$ holds for all $m < n$. We will prove a few representative cases.
  We abbreviate
  $\lst{S} :=
  \func{\GTh}{\Gd.h.\elim{\Gd.h.\td{D}{\psi}}{\Gd}{h}{\td{\E}{\psi}}}{\lst{N}}$
  to save space.

  \begin{description}
  \item{\rulename{Hyp}} Then $\sch{m} = \Gth[j]$ and $\sch{b} = \GTh[j]$ for
    some $j$, so
    \begin{enumerate}
    \item $\insttm{\Gth.\sch{m}}{\td{\K}{\psi}}{\lst{N}} = \lst{N}[j]$,
    \item $\insttm*{\Gth.\sch{m}}{\td{\K}{\psi}}{\td{\E}{\psi}}{\Gd.h.\td{D}{\psi}}{\lst{N}}{\lst{S}} = \func{\GTh[j]}{\Gd.h.\elim{\Gd.h.\td{D}{\psi}}{\Gd}{h}{\td{\E}{\psi}}}{\lst{N}[j]}$,
    \end{enumerate}
    and we want to show
    $\ceqtm[\Psi']{\func{\GTh[j]}{\Gd.h.\elim{\Gd.h.\td{D}{\psi}}{\Gd}{h}{\td{\E}{\psi}}}{\lst{N}[j]}}{\func{\GTh[j]}{\Gd.h.\elim{\Gd.h.\td{D}{\psi}}{\Gd}{h}{\td{\E}{\psi}}}{\lst{N}[j]}}{\tyatty*{\GTh[j]}{\Gd.h.\td{D}{\psi}}{\lst{N}[j]}}$. This
    holds by \cref{lem:func-typing,lem:elimi-term-extension} and the assumption
    $\Tm{\tyatrel{\GTh}{\td{\Gs}{\psi}}}(\lst{N})$.
  \item{\rulename{$\bndintro$-I}} Then
    $\sch{m} = \bndintro{\lst{r}}{\lst{P}}{\lst{\sch{n}}}$ and $\sch{b} = \argvar{\lst{I}}$, so
    \begin{itemize}
    \item $\insttm{\Gth.\sch{m}}{\td{\K}{\psi}}{\lst{N}} = \intro[\td{\K}{\psi}]{\lst{r}}{\lst{P}}{\insttm{\Gth.\lst{\sch{n}}}{\td{\K}{\psi}}{\lst{N}}}$,
    \item $\insttm*{\Gth.\sch{m}}{\td{\K}{\psi}}{\td{\E}{\psi}}{\Gd.h.\td{D}{\psi}}{\lst{N}}{\lst{S}} = \subst{\subst{\subst{\dsubst{R}{\lst{r}}{\lst{x}}}{\lst{P}}{\Gg}}{\etc{\insttm{\Gth.\lst{\sch{n}}}{\td{\K}{\psi}}{\lst{N}}}}{\Gd}}{\insttm*{\Gth.\lst{\sch{n}}}{\td{\K}{\psi}}{\td{\E}{\psi}}{\Gd.h.\td{D}{\psi}}{\lst{N}}{\lst{S}}}{\Gr}$,
    \end{itemize}
    where $\ell \in \td{\K}{\psi}$,
    $\td{\K}{\psi}[\ell] =
    \constr{\lst{x}}{\GG}{\Gg.\lst{J}}{\Gg.\GF}{\sys{\xi_k}{\Gg.\Gf.\sch{m}_k}}$ and
    $\td{\E}{\psi}[\ell] = \lst{x}.\Gg.\Gh.\Gr.R$. In this case, we want to show
    that
    \begin{gather*}
      \elim{\Gd.h.\td{D}{\psi}}{\lst{I}}{\intro[\td{\K}{\psi}]{\lst{r}}{\lst{P}}{\insttm{\Gth.\lst{\sch{n}}}{\td{\K}{\psi}}{\lst{N}}}}{\td{\E}{\psi}}
      \eq
      \subst{\subst{\subst{\dsubst{R}{\lst{r}}{\lst{x}}}{\lst{P}}{\Gg}}{\insttm{\Gth.\lst{\sch{n}}}{\td{\K}{\psi}}{\lst{N}}}{\Gh}}{\insttm*{\Gth.\lst{\sch{n}}}{\td{\K}{\psi}}{\td{\E}{\psi}}{\Gd.h.\td{D}{\psi}}{\lst{N}}{\lst{S}}}{\Gr}
    \end{gather*}
    in
    $\subst{\subst{\td{D}{\psi}}{\lst{I}}{\Gd}}{\intro[\td{\K}{\psi}]{\lst{r}}{\lst{P}}{\insttm{\Gth.\lst{\sch{n}}}{\td{\K}{\psi}}{\lst{N}}}}{h}$.
    By \cref{lem:elim-beta-intro}, which we can apply since $\height{\K}{\ell} \le n$, the left-hand side is
    equal to
    \begin{gather*}
      \tag{$*$}
      \subst{\subst{\subst{\dsubst{R}{\lst{r}}{\lst{x}}}{\lst{P}}{\Gg}}{\insttm{\Gth.\lst{\sch{n}}}{\td{\K}{\psi}}{\lst{N}}}{\Gh}}{\func{\subst{\GF}{\lst{P}}{\Gg}}{\Gd.h.\elim{\Gd.h.\td{D}{\psi}}{\Gd}{h}{\td{\E}{\psi}}}{\insttm{\Gth.\lst{\sch{n}}}{\td{\K}{\psi}}{\lst{N}}}}{\Gr}
    \end{gather*}
    in
    $\subst{\subst{\td{D}{\psi}}{\lst{I}}{\Gd}}{\intro[\td{\K}{\psi}]{\lst{r}}{\lst{P}}{\insttm{\Gth.\lst{\sch{n}}}{\td{\K}{\psi}}{\lst{N}}}}{h}$.
    Finally, we can apply the inner induction hypothesis to the terms
    $\cwfargtm[\Psi']{\td{\GD}{\psi}}[\td{\K}{\psi}]{\GTh}{\lst{\sch{n}}}{\subst{\GF}{\lst{P}}{\Gg}}$
    to get
    \begin{gather*}
      \func{\subst{\GF}{\lst{P}}{\Gg}}{\Gd.h.\elim{\Gd.h.\td{D}{\psi}}{\Gd}{h}{\td{\E}{\psi}}}{\insttm{\Gth.\lst{\sch{n}}}{\td{\K}{\psi}}{\lst{N}}}
      \eq
      \insttm*{\Gth.\lst{\sch{n}}}{\td{\K}{\psi}}{\td{\E}{\psi}}{\Gd.h.\td{D}{\psi}}{\lst{N}}{\lst{S}}
    \end{gather*}
    in
    $\tyatty*{\subst{\GF}{\lst{P}}{\Gg}}{\Gd.h.\td{D}{\psi}}{\insttm{\Gth.\lst{\sch{n}}}{\td{\K}{\psi}}{\lst{N}}}$. Replacing
    the left-hand side of this equation by the right in the $\Gr$ position in ($*$) brings us to our
    destination. \qedhere
  \end{description}
\end{proof}

The proof of the final theorem is essentially mechanical: we prove the
eliminator is well-behaved on each possible input by referring to the
appropriate $\beta$ rule.

\begin{theorem}
  \label{thm:elimi-supports-K}
  $\Gs$ supports $\K$.
\end{theorem}
\begin{proof}
  We need to show that $\Fhcom{\Gs} \subseteq \Gs$, $\Fcoe{\Gs} \subseteq \Gs$, and
  $\Intro{\K}{\ell}{\Gs} \subseteq \Gs$ for each $\ell \in \K$.
  \begin{enumerate}
  \item $\Fhcom{\Gs} \subseteq \Gs$.

    Suppose we have $\ix{\Fhcom{\Gs}_\psi}{\lst{I}}(V,V'))$. For any
    $W,W' \in \{V,V'\}$, each of the terms $\elim{\Gd.h.\td{D}{\psi}}{\lst{I}}{W}{\td{\E}{\psi}}$ and
    $\elim{\Gd.h.\td{D'}{\psi}}{\lst{I}}{W'}{\td{\E'}{\psi}}$ reduces per \cref{lem:elim-beta-fhcom}, and the
    reducts are equal by \cref{lem:elimi-term-extension} and \cref{prop:com}.
  \item $\Fcoe{\Gs} \subseteq \Gs$.

    As with the previous proof, but using \cref{lem:elim-beta-fcoe} and the $\coe$-Kan conditions for $D$.

  \item $\Intro{\K}{\ell}{\Gs} \subseteq \Gs$.

    As with the previous proofs, but using \cref{lem:elim-beta-intro} and
    $\ceqcases{\GD}{\E}{\E'}{\K}{\Gd.h.D}.$. \qedhere
  \end{enumerate}
\end{proof}

\begin{corollary}
  \label{cor:elim}
  $\eqtm{\ofc{\Gd}{\GD},\oft{h}{\indcl{\K}{\Gd}}}{\elim{\Gd.h.D}{\Gd}{h}{\E}}{\elim{\Gd.h.D'}{\Gd}{h}{\E'}}{D}$.
\end{corollary}
\begin{proof}
  By \cref{thm:elimi-supports-K}, we have $\indrel{\K} = \Gs$. Apply
  \cref{lem:elimi-term-extension}.
\end{proof}



\section{Examples}
\label{sec:example}

In this section, we show how to encode various inductive types in our schema. We
will also discuss opportunities for optimizations and alternative constructions
in special cases.

For the sake of readability, we use $\emp$ rather than $\cdot$ to denote empty lists and omit these where
unambiguous, for example writing $\indcl{\K}$ rather than $\indcl[\emp]{\K}{\emp}$. We will write constructor
operators simply as $\formal{\ell}$ and $\ell$ rather than $\bndintro$ and $\intro$ and leave the reader to
infer the annotations.

\subsection{$\W$-types}

Let $\cwftypek{A}$, $\wftypek{\oft{a}{A}}{B}$ be given. We can define their
\emph{$\W$-type} \cite{martin-lof82} as
$\Wcl{A}{a.B} \eqdef \indcl{\K_{\Wcl{A}{a.B}}}$ where
\begin{align*}
  \K_{\Wcl{A}{a.B}} &\eqdef \listconstrs{
      \wsup : \constr{\emp}{A}{a.\emp}{a.\argarr{B}{\argvar}}{\emp}
  }
\end{align*}
and derive a typing rule for the eliminator:
\begin{align*}
    \welim{h.D}{M}{a.g.r.R} &\eqdef \elim{h.D}{\emp}{M}{\listelim{
                              \wsup : \emp.a.g.r.R
                              }}
\end{align*}
\begin{mathpar}
  \Infer
  {\wftypek{\oft{h}{\Wcl{A}{a.B}}}{D} \\
    \coftype{M}{\Wcl{A}{a.B}} \\\\
    \oftype{\oft{a}{A},\oft{g}{\arr{B}{\Wcl{A}{a.B}}},\oft{r}{\picl{b}{B}{\subst{D}{\app{g}{b}}{h}}}}{R}{\subst{D}{\wsup{a}{g}}{h}}}
  {\coftype{\welim{h.D}{M}{a.g.r.R}}{\subst{D}{M}{h}}}
\end{mathpar}

\subsection{Torus}
\label{sec:example:torus}

The most natural way to define the torus in the cubical setting is \`a la
\citet[\S IV.E]{licata15}. We set $\Torus \eqdef \indcl{\K_\Torus}$ where
\begin{align*}
  \K_\Torus &\eqdef \listconstrs{
    \begin{array}{lcl}
      \tbase &:& \constr{\emp}{\emp}{\emp.\emp}{\emp.\emp}{\emp} \\
      \tlpa &:& \constr*{x}{\emp}{\emp.\emp}{\emp.\emp}{
                \begin{array}{l}
                  \tube{x=0}{\tbase*}, \\
                  \tube{x=1}{\tbase*}
                \end{array}
      } \\
      \tlpb &:& \constr*{y}{\emp}{\emp.\emp}{\emp.\emp}{
                \begin{array}{l}
                  \tube{y=0}{\tbase*}, \\
                  \tube{y=1}{\tbase*}
                \end{array}
      } \\
      \tsurf &:& \constr*{x,y}{\emp}{\emp.\emp}{\emp.\emp}{
                 \begin{array}{l}
                   \tube{x=0}{\tlpb*{y}},
                   \tube{y=0}{\tlpa*{x}}, \\
                   \tube{x=1}{\tlpb*{y}},
                   \tube{y=1}{\tlpa*{x}}
                 \end{array}}
    \end{array}
  }  
\end{align*}
The eliminator is then given by
\begin{align*}
  \telim{h.D}{M}{R_{\tbase}}{x.R_{\tlpa}}{y.R_{\tlpb}}{x.y.R_{\tsurf}} &\eqdef
    \elim*{h.D}{\emp}{M}{\listelim{
      \begin{array}{lcl}
        \tbase &:& \emp.\emp.\emp.R_{\tbase}, \\
        \tlpa &:& x.\emp.\emp.R_{\tlpa}, \\
        \tlpb &:& y.\emp.\emp.R_{\tlpb}, \\
        \tsurf &:& (x,y).\emp.\emp.R_{\tsurf}
      \end{array}
  }}
\end{align*}
and satisfies the typing rule
\begin{mathpar}
  \Infer
  {\coftype{R_{\tbase}}{\subst{D}{\tbase}{h}} \\
    \coftype[\Psi,x]{R_{\tlpa}}{\subst{D}{\tlpa{x}}{h}} \\
    (\forall \Ge)\;\ceqtm[\Psi]{\dsubst{R_{\tlpa}}{\Ge}{x}}{R_{\tbase}}{\subst{D}{\tlpa{\Ge}}{h}} \\
    \coftype[\Psi,y]{R_{\tlpb}}{\subst{D}{\tlpb{y}}{h}} \\
    (\forall \Ge)\;\ceqtm[\Psi]{\dsubst{R_{\tlpb}}{\Ge}{y}}{R_{\tbase}}{\subst{D}{\tlpb{\Ge}}{h}} \\
    \coftype[\Psi,x,y]{R_{\tsurf}}{\subst{D}{\tsurf{x}{y}}{h}} \\
    (\forall \Ge)\;\ceqtm[\Psi,y]{\dsubst{R_{\tsurf}}{\Ge}{x}}{R_{\tlpb}}{\subst{D}{\tsurf{\Ge}{y}}{h}} \\
    (\forall \Ge)\;\ceqtm[\Psi,x]{\dsubst{R_{\tsurf}}{\Ge}{y}}{R_{\tlpa}}{\subst{D}{\tsurf{x}{\Ge}}{h}} \\    
  }
  {\coftype{\telim{h.D}{M}{R_{\tbase}}{x.R_{\tlpa}}{y.R_{\tlpb}}{x.y.R_{\tsurf}}}{\subst{D}{M}{h}}}
\end{mathpar}
We can also define the torus in a ``globular'' style more reminiscent of the
HoTT Book's definition \cite[\S6.6]{hott-book}. We take the same specifications
for $\tbase$,$\tlpa$,$\tlpb$, but change $\tsurf$ to
\begin{align*}
  \tsurf &: \constr*{x,y}{\emp}{\emp.\emp}{\emp.\emp}{
             \begin{array}{l}
               \tube{x=0}{\tbase*}, \\
               \tube{x=1}{\tbase*}, \\
               \tube{y=0}{\bigbndfhcom{}{0}{1}{\tlpa*{x}}{
               \begin{array}{l}
                 \tube{x=0}{z.\tbase*}, \\
                 \tube{x=1}{z.\tlpb*{z}}
               \end{array}}}, \\
               \tube{y=1}{\bigbndfhcom{}{0}{1}{\tlpb*{x}}{
               \begin{array}{l}
                 \tube{x=0}{z.\tbase*}, \\
                 \tube{x=1}{z.\tlpa*{z}}
               \end{array}}}
             \end{array}}
\end{align*}
With this definition, the coherence conditions on $\tsurf$ in the typing rule for the eliminator become
\begin{itemize}
\item $\ceqtm[\Psi,y]{\dsubst{R_{\tsurf}}{\Ge}{x}}{R_{\tbase}}{\subst{D}{\tsurf{\Ge}{y}}{h}}$ for $\Ge = 0,1$,
\item $\ceqtm[\Psi,x]{\dsubst{R_{\tsurf}}{0}{y}}{\bigcom{z.\subst{D}{F}{h}}{0}{1}{R_{\tlpa}}{
      \begin{array}{l}
        \tube{x=0}{R_{\tbase}}, \\
        \tube{x=1}{R_{\tlpb}}
      \end{array}
    }}{\subst{D}{\tsurf{x}{0}}{h}}$ \\
  where $F = \fhcom{0}{z}{\tlpa{x}}{\tube{x=0}{z.\tbase},\tube{x=1}{z.\tlpb{z}}}$,
\item $\ceqtm[\Psi,x]{\dsubst{R_{\tsurf}}{1}{y}}{\bigcom{z.\subst{D}{F}{h}}{0}{1}{\dsubst{R_{\tlpb}}{x}{y}}{
      \begin{array}{l}
        \tube{x=0}{R_{\tbase}}, \\
        \tube{x=1}{\dsubst{R_{\tlpa}}{y}{x}}
      \end{array}
    }}{\subst{D}{\tsurf{x}{1}}{h}}$ \\
  where $F = \fhcom{0}{z}{\tlpb{x}}{\tube{x=0}{z.\tbase},\tube{x=1}{z.\tlpa{z}}}$.
\end{itemize}

The torus is a \emph{closed} inductive type: it has no parameters and no free
dimension variables. For such types, we can optimize by making coercion trivial:
\[
  \coe{z.\Torus}{r}{r'}{M} \steps M.
\]
For zero-dimensional closed inductive types, such as $\bool$ or $\nat$, we can go even further and make
composition trivial as well:
\[
  \hcom*{\nat}{\xi_i} \steps M.
\]

\subsection{$\W$-Quotients}

$\W$-quotients \cite[\S3.2]{sojakova16} extend $\W$-types by adding a recursive
path constructor. Path constructor elements connect point constructor elements
as specified by two provided functions. Let $\cwftypek{A}$,
$\wftypek{\oft{a}{A}}{B}$, $\cwftypek{C}$, and $\coftype{F_0,F_1}{\arr{C}{A}}$
be given. We define $\WQcl{A}{a.B}{C}{F_0}{F_1} \eqdef \indcl{\K_{\WQ}}$ where
\begin{align*}
  \K_\WQ &\eqdef \listconstrs{
    \begin{array}{lcl}
      \wqsup &:& \constr{\emp}{A}{a.\emp}{a.\argarr{B}{\argvar}}{\emp}, \\
      \wqcell &:& \constr*{x}{C}{c.\emp}{c.(\argarr{\subst{B}{F_0(c)}{a}}{\argvar},\argarr{\subst{B}{F_1(c)}{a}}{\argvar})}{
                  \begin{array}{l}
                    \tube{x=0}{c.(\formal{g_0},\formal{g_1}).\wqsup*{\app{F_0}{c}}{\formal{g_0}}}, \\
                    \tube{x=1}{c.(\formal{g_0},\formal{g_1}).\wqsup*{\app{F_1}{c}}{\formal{g_1}}}
                  \end{array}}
    \end{array}
  }
\end{align*}
The eliminator is given by
\begin{align*}
  \wqelim{h.D}{M}{a.g.r.R_{\wqsup}}{x.c.g_0.g_1.r_0.r_1.R_{\wqcell}} &\eqdef
  \elim{h.D}{\emp}{M}{\listelim{
    \begin{array}{lcl}
      \wqsup &:& \emp.a.g.r.R_{\wqsup}, \\
      \wqcell &:& x.c.(g_0,g_1).(r_0,r_1).R_{\wqcell}
    \end{array}
  }}
\end{align*}
Abbreviating $\WQcl{A}{a.B}{C}{F_0}{F_1}$ as $\WQ$, the eliminator satisfies the typing rule
\begin{mathpar}
  \Infer
  {\wftypek{\oft{h}{\WQ}}{D} \\
    \coftype{M}{\WQ} \\\\
    \oftype{\oft{a}{A},\oft{g}{\arr{B}{\WQ}},\oft{r}{\picl{b}{B}{\subst{D}{\app{g}{b}}{h}}}}{R_{\wqsup}}{\subst{D}{\wqsup{a}{g}}{h}} \\
    \GG_{\wqcell} = (\oft{c}{C},\etc{\oft{g_\Ge}{\arr{\subst{B}{\app{F_\Ge}{c}}{a}}{\WQ}}},\etc{\oft{r_\Ge}{\picl{b}{\subst{B}{\app{F_\Ge}{c}}{a}}{\subst{D}{\app{g_\Ge}{b}}{h}}}}) \\
    \oftype[\Psi,x]{\GG_{\wqcell}}{R_{\wqcell}}{\subst{D}{\wqcell{x}{c}{g_0}{g_1}}{h}} \\
    \eqtm{\GG_{\wqcell}}{\dsubst{R_{\wqcell}}{0}{x}}{\subst{R_{\wqsup}}{\app{F_0}{c},g_0,r_0}{a,g,r}}{\subst{D}{\wqcell{0}{c}{g_0}{g_1}}{h}} \\
    \eqtm{\GG_{\wqcell}}{\dsubst{R_{\wqcell}}{1}{x}}{\subst{R_{\wqsup}}{\app{F_1}{c},g_1,r_1}{a,g,r}}{\subst{D}{\wqcell{1}{c}{g_0}{g_1}}{h}}    
  }
  {\coftype{\wqelim{h.D}{M}{a.h.r.R_{\wqsup}}{x.c.h_0.h_1.r_0.r_1.R_{\wqcell}}}{\subst{D}{M}{h}}}
\end{mathpar}

$\W$-quotients carve out a space of higher inductive types which are in a
certain sense recursive only at the level of points. Although the $\wqcell$
constructor does take recursive arguments, the recursive arguments of a
$\wqcell$ term are fully determined by the 0-dimensional $\wqsup$ elements at
its boundary. The form of the 1-dimensional constructor can therefore vary only
in the type $C$ of its non-recursive parameter and the functions $F_0,F_1$ which
form the non-recursive part of the boundary term.

\subsection{Higher truncations}

Encoding the higher truncations \cite[\S7.3]{hott-book} in our schema requires
some indirection. One option is to use a \emph{hub-and-spokes} construction as
in the HoTT Book. Assuming we have already defined the $n$-spheres, we could
then define the $n$-truncation as
$\Trunc[n]{A} \eqdef \indcl{\K_{\Trunc[n]{A}}}$ where
\begin{align*}
  \K_{\Trunc[n]{A}} &\eqdef \listconstrs{
    \begin{array}{lcl}
      \trpt &:& \constr{\emp}{A}{a.\emp}{a.\emp}{\emp} \\
      \trhub &:& \constr{\emp}{\emp}{\emp.\emp}{\emp.\argarr{\C[n+1]}{\argvar}}{\emp} \\
      \trspoke &:& \constr*{x}{\C[n+1]}{s.\emp}{s.\argarr{\C[n+1]}{\argvar}}{
                   \begin{array}{l}
                     \tube{x=0}{s.\formal{f}.\trhub*{\formal{f}}} \\
                     \tube{x=1}{s.\formal{f}.\bndapp{\formal{f}}{s}}
                   \end{array}
      }
    \end{array}
  }
\end{align*}
The idea of this definition is to construct $\Trunc[n]{A}$ by recursively
contracting every $(n+1)$-sphere to a $\trhub$ point. We can define the
eliminator as
\begin{align*}
  \trhselim{h.D}{M}{a.R_{\trpt}}{f.r_f.R_{\trhub}}{s.f.r_f.R_{\trspoke}} &\eqdef
  \elim*{h.D}{\emp}{M}{\listelim{
    \begin{array}{lcl}
      \trpt &:& \emp.a.\emp.\emp.R_{\trpt}, \\
      \trhub &:& \emp.\emp.f.r_f.R_{\trhub}, \\
      \trspoke &:& x.s.f.r_f.R_{\trspoke}
    \end{array}
  }}
\end{align*}
which satisfies the typing rule
\begin{mathpar}
  \Infer
  {\wftypek{\oft{h}{\Trunc[n]{A}}}{D} \\
    \coftype{M}{\Trunc[n]{A}} \\
    \oftype{\oft{a}{A}}{R_{\trpt}}{\subst{D}{\trpt{a}}{h}} \\
    \oftype{\oft{f}{\arr{\C[n+1]}{\Trunc[n]{A}}},\oft{r_f}{\picl{s}{\C[n+1]}{\subst{D}{\app{f}{s}}{h}}}}{R_{\trhub}}{\subst{D}{\trhub{f}}{h}} \\
    \GG_{\trspoke} = (\oft{s}{\C[n+1]},\oft{f}{\arr{\C[n+1]}{\Trunc[n]{A}}},\oft{r_f}{\picl{s}{\C[n+1]}{\subst{D}{\app{f}{s}}{h}}}) \\
    \oftype[\Psi,x]{\GG_{\trspoke}}{R_{\trspoke}}{\subst{D}{\trspoke{x}{s}{f}}{h}} \\
    \eqtm{\GG_{\trspoke}}{\dsubst{R_{\trspoke}}{0}{x}}{R_{\trhub}}{\subst{D}{\trspoke{0}{s}{f}}{h}} \\
    \eqtm{\GG_{\trspoke}}{\dsubst{R_{\trspoke}}{1}{x}}{\app{r_f}{s}}{\subst{D}{\trspoke{1}{s}{f}}{h}}}
  {\coftype{\trhselim{h.D}{M}{a.R_{\trpt}}{f.r.R_{\trhub}}{s.f.r_f.R_{\trspoke}}}{\subst{D}{M}{h}}}
\end{mathpar}

\subsection{Localization }
\label{sec:localization}

Given a family of maps $\oftype{\oft{i}{I}}{F_i}{\arr{S_i}{T_i}}$, a type
$\cwftypek{A}$ is \emph{$F$-local} if precomposition by $F_i$ gives an
equivalence between $\arr{T_i}{A}$ and $\arr{S_i}{A}$ for all $i$. The
\emph{localization} $\Loc{F}{A}$ of an type $\cwftypek{A}$ at $F$ is the
universal $F$-local type with a map $\arr{A}{\Loc{F}{A}}$. \citet{shulman11}
constructs localization as a higher inductive type which we can encode in our schema.

\begin{align*}
  \K_{\Loc} &\eqdef \listconstrs{
    \begin{array}{lcl}
      \loc &:& \constr{\emp}{A}{\_.\emp}{\_.\emp}{\emp}, \\
      \ext &:& \constr{\emp}{(\oft{i}{I},\oft{t}{T_i})}{(i,\_).\emp}{(i,\_).\argarr{S_i}{\argvar}}{\emp}, \\
      \ext['] &:& \constr{\emp}{(\oft{i}{I},\oft{t}{T_i})}{(i,\_).\emp}{(i,\_).\argarr{S_i}{\argvar}}{\emp}, \\
      \rtr &:& \constr*{x}{(\oft{i}{I},\oft{s}{S_i})}{(i,\_).\emp}{(i,\_).\argarr{S_i}{\argvar}}{
               \begin{array}{l}
                 \tube{x=0}{(i,s).\formal{g}.\bndapp{\formal{g}}{s}},\\
                 \tube{x=1}{(i,s).\formal{g}.\ext*{i}{\app{F_i}{s}}{\formal{g}}}
               \end{array}}, \\
      \rtr['] &:& \constr*{x}{(\oft{i}{I},\oft{t}{T_i})}{(i,\_).\emp}{(i,\_).\argarr{T_i}{\argvar}}{
                \begin{array}{l}
                 \tube{x=0}{(i,t).\formal{h}.\bndapp{\formal{h}}{t}},\\
                 \tube{x=1}{(i,t).\formal{h}.\ext*[']{i}{t}{\bndlam{s}{\bndapp{\formal{h}}{\app{F_i}{s}}}}}
               \end{array}}
    \end{array}
  }
\end{align*}

The constructor $\loc$ includes $A$ in $\Loc{F}{A}$. The constructors $\ext$ and
$\rtr$ give a right inverse to precomposition by $F_i$ for each $i$, while the
constructors $\ext[']$ and $\rtr[']$ give a left inverse. Per
\cite[\S4.3]{hott-book}, this data makes $- \circ F_i$ an equivalence for each
$i$. We will not write out the eliminator for this inductive type, but it is not
hard to see that any function from $A$ into an $F$-local type factors through
$\Loc{F}{A}$.

\subsection{Identity types}
\label{sec:example:identity}

Given a type $\cwftypek{A}$, we define its identity family by
$\Id{A}{M}{M'} \eqdef \indcl[(A,A)]{\K_{\Id}}{(M,M')}$ where
\begin{align*}
  \K_{\Id} \eqdef \listconstrs{
    \begin{array}{lcl}
      \refl &:& \constr{\emp}{A}{a.(a,a)}{a.\emp}{\emp}
    \end{array}
  }
\end{align*}
As eliminator, we obtain exactly the $\fibelim$ eliminator for the Martin-L\"of identity type.
\[
  \Idelim{a.b.p.C}{a.R}{M}{N}{P} \eqdef \elim{(a,b).p.C}{(M,N)}{\left[\emp.a.\emp.\emp.R\right]}{P}
\]
\begin{mathpar}
  \Infer
  {\wftypek{\oft{a,b}{A},\oft{p}{\Id{A}{a}{b}}}{C} \\
    \oftype{\oft{a}{A}}{R}{\subst{\subst{C}{a}{b}}{\refl{a}}{p}} \\
    \coftype{M,N}{A} \\
    \coftype{P}{\Id{A}{M}{N}}}
  {\coftype{\Idelim{a.b.p.C}{a.R}{M}{N}{P}}{\subst{\subst{\subst{C}{M}{a}}{N}{b}}{P}{p}}}
\end{mathpar}
The eliminator satisfies a $\beta$-rule for $\refl$ up to exact equality.
\[
  \ceqtm{\Idelim{a.b.p.C}{a.R}{M}{M}{\refl{M}}}{\subst{R}{M}{a}}{\subst{\subst{\subst{C}{M}{a}}{M}{b}}{\refl{M}}{p}}
\]
On the other hand, we have a type $\Path{A}{M}{N}$ of paths between $M$ and $N$. The elements of the latter
are terms varying in a bound dimension variable, as shown below (see \cite[\S5.3]{chtt-iii}).
\begin{mathpar}
  \Infer
  {\coftype[\Psi,x]{P}{A} \\ \ceqtm{\dsubst{P}{0}{x}}{M}{A} \\ \ceqtm{\dsubst{P}{1}{x}}{N}{A}}
  {\coftype{\dlam{x}{P}}{\Path{A}{M}{N}}}
  \and
  \Infer
  {\coftype{Q}{\Path{A}{M}{N}}}
  {\coftype{\dapp{Q}{r}}{A}}
  \and
  \Infer
  {\coftype{Q}{\Path{A}{M}{N}}}
  {\ceqtm{\dapp{Q}{0}}{M}{A}}
  \and
  \Infer
  {\coftype{Q}{\Path{A}{M}{N}}}
  {\ceqtm{\dapp{Q}{1}}{N}{A}}
  \and
  \Infer
  {\coftype[\Psi,x]{P}{A} \\ \ceqtm{\dsubst{P}{0}{x}}{M}{A} \\ \ceqtm{\dsubst{P}{1}{x}}{N}{A}}
  {\ceqtm{\dapp{(\dlam{x}{P})}{r}}{\dsubst{P}{r}{x}}{A}}
\end{mathpar}
We can construct an equivalence between $\Path{A}{M}{N}$ and $\Id{A}{M}{N}$.
\begin{align*}
  \lam{p}{\ \coe{x.\Id{A}{M}{\dapp{p}{x}}}{0}{1}{\refl{M}}} &\in \Path{A}{M}{N} \to \Id{A}{M}{N} \\
  \lam{i}{\ \Idelim{a.b.\Path{A}{a}{b}}{a.(\dlam{\_}{a})}{M}{N}{i}} &\in \Id{A}{M}{N} \to \Path{A}{M}{N}
\end{align*}
(We leave it to the reader to prove that these are mutual inverses.) While $\Path{A}{M}{N}$ is then an
identity type ``up to equivalence,'' it does not appear to be an identity type up to exact equality, in the
sense that there are no known terms $\refl'$ and $\Idelim'$ which validate the validate the rules an identity
type should satisfy for $\Path{A}{M}{N}$. Specifically, while we can set $\refl'(M) \eqdef \dlam{\_}{M}$ and
define a $\Idelim'$ which has the right type, it does not appear possible to do this in a way which satisfies
the $\beta$ rule up to exact equality. This is related to the failure of \emph{regularity} in CHiTT
\cite{coquand-regularity}: it is not generally the case that $\coe{\_.A}{r}{r'}{M} \eq M \in A$. If this were
the case, then the forward map of the equivalence above would take $\dlam{\_}{M}$ to $\refl{M}$, making it
possible to transport the identity type structure from $\Id{A}{M}{N}$ to $\Path{A}{M}{N}$ while preserving the
$\beta$-rule for $\Idelim$. This issue exists in all known univalent cubical type theories; see
\cref{sec:related} for more details.

\subsection{Making an example of a non-example}

\citet[\S9]{lumsdaine17} give an example of a higher inductive type $\Blass$ which is not modeled in ZF and
therefore cannot be encoded using only pushouts and natural numbers, adapting a result of
\citet[\S9]{blass83}. This type also cannot be encoded in our schema, as it requires the definition of a
boundary term by natural number recursion. However, we can encode it if we extend the specification language
with a natural number recursor. Assume we are defining a cubical type system which contains the strict natural
numbers type $\nat$ defined in Part III (or a weak natural numbers type defined using our schema) and
$(-1)$-truncations. We extend the boundary term language with
\[
  \begin{array}{rcl}
    \sch{m} &::=& \cdots \mid \bndnatrec{M}{\sch{m}}{a.\formal{p}.\sch{m}}
  \end{array}
\]
and add the following rules to the formal type system.
\begin{mathpar}
  \Infer
  {\ceqtm{M}{M'}{\nat} \\
    \ceqargtm{\GD}{\GTh}{\sch{n}}{\sch{n}'}{\sch{a}} \\
    \eqargtm{\oft{a}{\nat}}{\GD}{\GTh,\ofa{\formal{p}}{\sch{a}}}{\sch{q}}{\sch{q}'}{\sch{a}}}
  {\ceqargtm{\GD}{\GTh}{\bndnatrec{M}{\sch{n}}{a.\formal{p}.\sch{q}}}{\bndnatrec{M'}{\sch{n}'}{a.\formal{p}.\sch{q}'}}{\sch{a}}}
  \and
  \Infer
  {\cwfargtm{\GD}{\GTh}{\sch{n}}{\sch{a}} \\
   \wfargtm{\oft{a}{\nat}}{\GD}{\GTh,\ofa{\formal{p}}{\sch{a}}}{\sch{q}}{\sch{a}}}
  {\ceqargtm{\GD}{\GTh}{\bndnatrec{\z}{\sch{n}}{a.\formal{p}.\sch{q}}}{\sch{n}}{\sch{a}}}
  \and
  \Infer
  {\coftype{M}{\nat} \\
    \cwfargtm{\GD}{\GTh}{\sch{n}}{\sch{a}} \\
    \wfargtm{\oft{a}{\nat}}{\GD}{\GTh,\ofa{\formal{p}}{\sch{a}}}{\sch{q}}{\sch{a}}}
  {\ceqargtm{\GD}{\GTh}{\bndnatrec{\suc{M}}{\sch{n}}{a.\formal{p}.\sch{q}}}{\subst{\subst{\sch{q}}{M}{a}}{\bndnatrec{M}{\sch{n}}{a.\formal{p}.\sch{q}}}{\formal{p}}}{\sch{a}}}
\end{mathpar}
We extend the interpretation functions by
\[
  \insttm{\Gth.\bndnatrec{M}{\sch{n}}{a.\formal{p}.\sch{q}}}{\K}{\lst{N}} \eqdef \natrec{M}{\insttm{\Gth.\sch{n}}{\K}{\lst{N}}}{a.r.\insttm{\Gth,\formal{p}.\sch{q}}{\K}{\lst{N},r}}
\]
and
\begin{gather*}
  \insttm*{\Gth.\bndnatrec{M}{\sch{n}}{a.\formal{p}.\sch{q}}}{\K}{\E}{\Gd.h.D}{\lst{N}}{\lst{S}} \eqdef \\
  \natrec{M}{\insttm*{\Gth.\sch{n}}{\K}{\E}{\Gd.h.D}{\lst{N}}{\lst{S}}}{a.r.\insttm*{\Gth,\formal{p}.\sch{q}}{\K}{\E}{\Gd.h.D}{(\lst{N},\insttm{\Gth.\bndnatrec{a}{\sch{n}}{a.\formal{p}.\sch{q}}}{\K}{\lst{N}})}{(\lst{S},r)}}.
\end{gather*}
We will not list the many constructors of the inductive type $\Blass$ here. Suffice to say that the addition
of $\natrec$ is needed to encode the constructor (4) in \cite[\S9]{lumsdaine17}, specifically in the
definition of the functions $\formal{h}_k : \argarr{\nat}{\argvar}$ by recursion on $k$.

This blind spot in our schema is of course not limited to natural numbers, but arises whenever one wishes to
define a boundary term by recursion on some element of a positive type. In this section, we have seen that our
language nonetheless suffices to define the majority of constructs with established uses in homotopy type
theory; it remains to be seen whether this will change as new applications come to light.


\section{Relationship with HoTT and HITs}

We speak of cubical inductive types as a specific realization of the general and vague concept of higher
inductive type. In the same way that cubical type theory is a higher type theory, i.e., a type theory for
reasoning explicitly about higher-dimensional objects, cubical inductive types are higher inductive types:
types generated by explicit higher-dimensional constructors.

Cubical inductive types occupy a specific niche on two axes. First, as the name suggests, their constructors
are $n$-cubes; this is not the only option. The higher inductive types of the HoTT Book could be called
\emph{globular inductive types} (GITs), types generated by constructors which map into either the type itself
or into its iterated path/identity types. Although there is an obvious correspondence between 1-dimensional
CITs and 1-dimensional GITs, the situation is murkier at higher dimensions. As an example, consider the two
presentations of the torus from \cref{sec:example:torus}. The first defines the torus by
point constructor $\tbase$, two path constructors $\tlpa{r}$ and $\tlpb{s}$, and a 2-dimensional square
constructor $\tsurf{r}{s}$, with boundaries specified like so:

\[
  \begin{tikzpicture}
    \draw (0 , 2) [->] to node [above] {\small $x$} (0.5 , 2) ;
    \draw (0 , 2) [->] to node [left] {\small $y$} (0 , 1.5) ;
    \node (tl) at (1.5 , 2) {$\tbase$} ;
    \node (tr) at (5.5 , 2) {$\tbase$} ;
    \node (bl) at (1.5 , 0) {$\tbase$} ;
    \node (br) at (5.5 , 0) {$\tbase$} ;
    \draw (tl) [->] to node [above] {$\tlpa{x}$} (tr) ;
    \draw (tl) [->] to node [left] {$\tlpb{y}$} (bl) ;
    \draw (tr) [->] to node [right] {$\tlpb{y}$} (br) ;
    \draw (bl) [->] to node [below] {$\tlpa{x}$} (br) ;
    \node at (3.5 , 1) {$\tsurf{x}{y}$} ;
  \end{tikzpicture}
\]

With GITs, we need a different way of specifying the square constructor $\tsurf$. We might define the torus
$\Torus$ as generated by a constructor $\base$ in $\Torus$, constructors $\tlpa$,$\tlpb$ in
$\Path{\Torus}{\base}{\base}$, and a constructor $\tsurf$ in
$\Path{\Path{\Torus}{\base}{\base}}{\tlpa \cdot \tlpb}{\tlpb \cdot \tlpa}$, but it is not immediately clear
that this is equivalent to our CIT definition. Converting between GITs and CITs becomes increasingly difficult
with increasing constructor dimensionality. For GITs in particular, stating an eliminator becomes more complex
as constructor dimensionality increases.

The second distinction we want to make is between what we will call $\Path$-HITs and $\Id$-HITs. As described
in \cref{sec:example:identity}, the path type $\Path{A}{M}{N}$ is meaningfully distinct from the identity type
$\Id{A}{M}{N}$, which we define in \cref{sec:example:identity} as the indexed inductive type generated by
reflexivity.  $\Path$-HITs and $\Id$-HITs, then, are higher inductive types specified in terms of
$\Path$ and $\Id$ respectively. Our cubical inductive types are
$\Path$-HITs, as their higher constructors produce elements of path types. In contrast, the higher inductive types of the HoTT Book are
$\Id$-HITs, as their constructors produce elements of identity types. This distinction seems to be orthogonal
to the shape distinction; for example, \citet{licata15} define an indexed inductive type of squares and
specify the torus as a cubical inductive type in terms of that type. 

Of course, any $\Path$-HIT is homotopy equivalent to the corresponding $\Id$-HIT, but again there may be
differences at the level of exact equality. The eliminator for an $\Id$-HIT as specified in the HoTT Book
satisfies an exact computation rule on point constructors, but the corresponding ``computation'' rules for
higher constructors hold only up to identification \cite[\S6.2]{hott-book}. For $\Id$-HITs, it is not clear
whether it is sensible to ask for exact computation rules for higher constructors, for reasons described in
the HoTT Book. In contrast, our
$\Path$-HITs satisfy exact computation rules for constructors of any dimensionality. By way of the
$\Path \simeq \Id$ equivalence, a 1-dimensional $\Path$-HIT is an $\Id$-HIT in the sense of the HoTT Book, but
the converse implication fails because we do not obtain these exact equations.


\section{Related Work}
\label{sec:related}

The concept of inductive types with higher-dimensional constructors originated at the 2011 Oberwolfach
meeting, in discussions between Andrej Bauer, Peter Lumsdaine, Mike Shulman, and Michael Warren (see \cite[\S6
Notes]{hott-book}).  Since then, there has been an abundance of work seeking to make the concept precise, none
of which has been the final word. (This paper makes no pretensions to that throne.) The HoTT Book presents
many examples of higher inductive types, including types with recursive constructors, indexed inductive types,
and inductive-inductive types, but only sketches a general schema \cite[\S6.13]{hott-book}.

In the non-higher setting, schemata for inductive types in dependent type theory
\cite{martin-lof82,constable85,coquand88,dybjer94} typically provide inductive types which are fixed-points of
\emph{strictly positive operators}, that is, the syntactic class of type operators in which the type variable
never occurs in the domain of a function type. As we move to the higher setting, we can ask whether this class
of argument types remains sufficient for general use. Moreover, we have the additional dimension of boundary
terms: when we give a path constructor, what can its endpoints be?

In pursuit of a general notion of higher inductive type, Sojakova \cite{sojakova14,sojakova15,sojakova16}
introduced the class of $\W$-quotients (also called $\W$-suspensions) and showed that they could be
characterized as homotopy-initial algebras, building on work on ordinary inductive types in HoTT
\cite{awodey12,awodey17}. A $\W$-quotient is generated by a recursive 0-constructor (\`a la $\W$-types
\cite{martin-lof82}) and a recursive 1-constructor which connects instances of the point
constructor. $\W$-quotients suffice to define types such as pushouts and modular arithmetic types, but cannot
directly be used to define types like truncations with constructors whose boundaries are not point
constructors. (However, since pushouts can be written as $\W$-quotients, some of these can be encoded
indirectly; see below.)

More recently, \citet{basold17,dybjer17,kaposi18} have introduced schemata for HITs in formal Martin-L\"of
type theory, using a syntactic grammar of argument types and boundary terms. These schemata allow for
recursive constructors, so can be used to define HITs like the $(-1)$-truncation directly. Our work can be
viewed a cubical counterpart to this line of work. It is much simpler to handle constructors of arbitrary
dimensionality in the cubical setting; while \cite{kaposi18} also handle the general case, each new dimension
creates additional complexity. More importantly, the cubical setting allows us to give an operational
semantics and canonicity result for instances of our schema.

Another line of work seeks to reduce more complex HITs to simpler ones. Van Doorn \cite{van-doorn16} and
\citet{kraus16} gave two different constructions of the $(-1)$-truncation from non-recursive HITs, each
obtaining the truncation as the sequential homotopy colimit of an $\omega$-indexed sequence of
types. (Homotopy colimits indexed by $\omega$ can be defined using pushouts and a natural numbers type.)
\citet{rijke17} later gave a construction of $n$-truncations in general, using a definition of the image of a
term which is again constructed via pushouts and a sequential colimit. We expect that many HITs can be defined
in this way. On the other hand, the complexity of these definitions makes them unwieldy for computational
purposes, and they generally support ``computation'' rules for path constructors only up to a path. Moreover,
\citet[\S9]{lumsdaine17} give an example of a HIT which cannot be constructed from pushouts and the natural
numbers (and indeed is not constructible in ZF).

On the semantic side, \citet{lumsdaine17} developed the notion of \emph{cell monad with parameters}, a
semantic notion of specification for a higher inductive type, and gave a class of model categories for which
all such higher inductive types exist. This class does not obviously correspond to a particular syntactic
schema, but includes, in some form, all of the examples we present in \cref{sec:example}. However, their work
does not allow boundary terms to use the fibrant structure of the type being defined; we allow homogeneous
composition in boundary terms, though not coercion. Also, for reasons related to fibrant replacement, their
approach suffers from size issues when dealing with parameterized HITs. For example, a pushout type may not
lie in the same universe as its constituent parts. In our setting, we can be more careful about the free
fibrant structure we add; it is not clear to us whether this is possible at their level of generality.

One of the central motivations for investigating cubical type theory, particularly in the work of Brunerie and
Licata \cite{brunerie14,licata15}, was as a convenient language for specifying and proving theorems about
higher inductive types. Even in traditional homotopy type theory, cubes proved to be a useful organizing
principle. It also seemed that a primitive cubical type theory would allow for eliminators with exact
computation rules on path constructors, which was believed to be problematic in standard homotopy type theory
\cite[\S6.2]{hott-book}. \Citet{bch} gave the first constructive model of type theory in cubical sets, but
this model is believed to be incompatible with HITs due to a lack of diagonals in the cube
category. \citet{cchm} dodged this issue by adding diagonals (and reversals and connections) to the cube
category, defining a univalent formal type theory with a circle and $(-1)$-truncation type. \citet{huber16b}
then proved a canonicity result for this type theory. More recently, \citet{coquand18} have defined additional
examples of cubical inductive types, sketched a schema, and proven consistency with a model in cubical
sets. The previous parts of this series \cite{chtt-i,chtt-ii,chtt-iii} include a circle type, which satisfies
a canonicity theorem by definition.  \citet{cctt} define a formal Cartesian cubical type theory with a
suspension type, and their formalized model generalizes this to a pushout type. \citet{isaev14} has proposed a
type theory with an interval type supporting a general class of \emph{data types with conditions} (and dual
\emph{records with conditions}), which is quite similar in spirit to our schema. For a broader overview of the
development of cubical type theory, we refer the reader to \cite{cctt}.

A primary motivation for defining indexed inductive types in cubical type theory is to obtain an identity
type, the indexed inductive family $\Id{A}{-}{-}$ generated by the reflexive identification. As mentioned
above, it appears that the native $\Path{A}{-}{-}$ family cannot be used as an identity type in the known
univalent cubical type theories. \citet[\S9]{swan14} gives an algebraic weak factorization system for (a
category equivalent to) the cubical sets of \citet{bch}, and shows how to use this to define an identity type
with an exact computation rule for the eliminator applied to reflexivity. The idea is to define the identity
type as a subset of the factorization of the diagonal, the restriction to a subset being necessary to ensure
stability under substitution. This is quite conceptually similar to our definition, but the specifics of
Swan's construction less obviously generalize beyond identity types.\footnote{See notes at
  \url{http://www.cs.cmu.edu/~ecavallo/notes/muri17.pdf} for a more detailed comparison of the two
  constructions.} Following Swan's ideas, \citet[\S9.1]{cchm} defines an identity type for cubical sets with
diagonals and connections, with elements of the identity type being elements of the path type paired with an
element of the face lattice on which they are degenerate. Finally, \citet{bezem17} give yet another definition
based on a cofibration-trivial fibration factorization. Rather than using a construction tailored to the
identity type, we obtain it as a particular instance of our schema.


\appendix
\part*{Appendix}
\section{Lemmas}
\label{app:lemmas}

\begin{definition}
  For a $\Psi$-relation $\Ga$, define a $\Psi$-relation $\lift{\Ga}$ by
  \[
    \lift{\Ga}_\psi(M,M') \iffdef M \evals V \land M' \evals V' \land \Ga_\psi(V,V').
  \]
\end{definition}

\begin{lemma}[Introduction]
  \label{lem:introduction}
  Let $\Ga$ be a value $\Psi$-PER. If for all $\psitd$, either
  $\Ga_\psi(\td{M}{\psi},\td{M'}{\psi})$ or
  $\Tm{\Ga}_\psi(\td{M}{\psi},\td{M'}{\psi})$, then $\Tm{\Ga}(M,M')$.
\end{lemma}
\begin{proof}
  Let $\tdss{_1}{_1}{}$ and $\tdss{_2}{_2}{_1}$ be given. We divide into three cases.
  \begin{description}
  \item{(aa)} $\Ga_{\psi_1}(\td{M}{\psi_1},\td{M'}{\psi_1})$ and $\Ga_{\psi_1\psi_2}(\td{M}{\psi_1\psi_2},\td{M'}{\psi_1\psi_2})$.

    Then $\td{M}{\psi_1} \evals \td{M}{\psi_1}$ and
    $\td{M'}{\psi_1} \evals \td{M'}{\psi_1}$ with
    $\Ga_{\psi_1\psi_2}(\td{M}{\psi_1\psi_2},\td{M'}{\psi_1\psi_2})$, so
    $\lift{\Ga}_{\psi_1\psi_2}(\td{M}{\psi_1\psi_2},\td{M'}{\psi_1\psi_2})$.
  \item{(ab)}
    $\Ga_{\psi_1}(\td{M}{\psi_1},\td{M'}{\psi_1})$ and $\Tm{\Ga}_{\psi_1\psi_2}(\td{M}{\psi_1\psi_2},\td{M'}{\psi_1\psi_2})$.

    Then $\td{M}{\psi_1} \evals \td{M}{\psi_1}$ and
    $\td{M'}{\psi_1} \evals \td{M'}{\psi_1}$ with
    $\Tm{\Ga}_{\psi_1\psi_2}(\td{M}{\psi_1\psi_2},\td{M'}{\psi_1\psi_2})$, so
    $\lift{\Ga}_{\psi_1\psi_2}(\td{M}{\psi_1\psi_2},\td{M'}{\psi_1\psi_2})$.
  \item{(b$*$)} $\Tm{\Ga}_{\psi_1}(\td{M}{\psi_1},\td{M'}{\psi_1})$.

    By $\Tm{\Ga}_{\psi_1}(\td{M}{\psi_1},\td{M'}{\psi_1})$, we have
    $\td{M}{\psi_1} \evals M_1$ and $\td{M'}{\psi_1} \evals M_1'$ with
    $\lift{\Ga}_{\psi_1\psi_2}(\td{M_1}{\psi_2},\td{M}{\psi_1\psi_2},\td{M'_1}{\psi_2},\td{M'}{\psi_1\psi_2})$. \qedhere
  \end{description}
\end{proof}

\begin{lemma}[Coherent expansion]
  \label{lem:expansion}
  Let $\Ga$ be a value $\Psi$-PER and let $\tmj{M,M'}$. If for all $\psitd$,
  there exists $M''$ such that $\td{M}{\psi} \msteps M''$ and
  $\Tm{\Ga}_\psi(M'',\td{M'}{\psi})$, then $\Tm{\Ga}(M,M')$.
\end{lemma}
\begin{proof}
  (Or see \cite[Lemma 41]{chtt-iii}.) Let $\tdss{_1}{_1}{}$ and
  $\tdss{_2}{_2}{_1}$ be given. By assumption, there exists $M_1$ such that
  $\td{M}{\psi_1} \msteps M_1$ and $\Tm{\Ga}_{\psi_1}(M_1,\td{M'}{\psi_1})$. By
  $\Tm{\Ga}_{\psi_1}(M_1,\td{M'}{\psi_1})$, we know that $M_1 \evals V_1$,
  $\td{M'}{\psi_1} \evals V_1'$, $\td{V_1}{\psi_2} \evals V_2$,
  $\td{M_1'}{\psi_2} \evals V_2'$, and $\td{M'}{\psi_1\psi_2} \evals V_{12}'$
  with $\Ga_{\psi_1\psi_2}(V_2,V_2',V_{12}')$. We also have some $M_{12}$ such
  that $\td{M}{\psi_1\psi_2} \msteps M_{12}$ and
  $\Tm{\Ga}_{\psi_1\psi_2}(M_{12},\td{M'}{\psi_1\psi_2})$. By
  $\Tm{\Ga}_{\psi_1\psi_2}(M_{12},\td{M'}{\psi_1\psi_2})$, we have
  $M_{12} \evals W_{12}$ and $\td{M'}{\psi_1\psi_2} \evals W_{12}'$ with
  $\Ga_{\psi_1\psi_2}(W_{12},W_{12}')$. Note that $V_{12}' = W_{12}'$.

  Examining this data, we have $\td{M}{\psi_1} \evals V_1$,
  $\td{M'}{\psi_1} \evals V_1'$, $\td{V_1}{\psi_2} \evals V_2$,
  $\td{V_1'}{\psi_2} \evals V_2'$, $\td{M}{\psi_1\psi_2} \evals W_{12}$, and
  $\td{M'}{\psi_1\psi_2} \evals W_{12}'$ with
  $\Ga_{\psi_1\psi_2}(V_2,V_2',W_{12},W_{12}')$.
\end{proof}

\begin{corollary}[Restricted expansion]
  \label{cor:restricted-expansion}
  Let $\Ga$ be a value $\Psi$-PER and $\Xi$ be a constraint context.  Let $\Ga$ be
  a value $\Psi$-PER and let $\tmj{M,M'}$. If for all $\psitd$ with
  $\models \td{\Xi}{\psi}$, there exists $M''$ such that
  $\td{M}{\psi} \msteps M''$ and $\Tm{\Ga}_\psi(M'',\td{M'}{\psi})$, then
  $(\Tm{\Ga} \mid \Xi)(M,M')$.
\end{corollary}

\begin{lemma}[Value-coherent evaluation]
  \label{lem:value-coherent-evals}
  Let $\Ga$ be a value-coherent $\Psi$-PER. For any $\Tm{\Ga}_\psi(M)$, there is $V$ such that $M \evals V$
  and $\Tm{\Ga}_\psi(M,V)$.
\end{lemma}
\begin{proof}
  (Or see \cite[Lemma 38]{chtt-iii}.) By definition of $\Tm{\Ga}_\psi(M)$, there is $V$ such that $M \evals V$
  and $\Ga_\psi(V)$. By value-coherence, this implies $\Tm{\Ga}_\psi(V)$. To see that $\Tm{\Ga}_\psi(M,V)$
  holds, let $\tdss{_1}{_1}{'}$ and $\tdss{_2}{_2}{'}$ be given. By definition of $\Tm{\Ga}_\psi(M)$ applied
  with substitutions $\id$ and $\psi_1$, we have $\td{M}{\psi_1} \evals M_1$ and $\td{V}{\psi_1} \evals V_1$
  with $\Ga_{\psi\psi_1}(M_1,V_1)$. By value-coherence, this implies $\Tm{\Ga}_{\psi\psi_1}(M_1,V_1)$, which
  in turn implies that $\td{M_1}{\psi_2} \evals M_2$ and $\td{V_1}{\psi_2} \evals V_2$ with
  $\Ga_{\psi\psi_1\psi_2}(M_2,V_2)$. By definition of $\Tm{\Ga}_\psi(M)$ applied with substitutions $\id$ and
  $\psi_1\psi_2$, we have $\td{M}{\psi_1\psi_2} \evals M_{12}$ and $\td{V}{\psi_1\psi_2} \evals V_{12}$ with
  $\Ga_{\psi\psi_1\psi_2}(M_{12},V_{12})$. Finally, $\Tm{\Ga}_\psi(M)$ implies that
  $\Ga_{\psi\psi_1\psi_2}(M_2,M_{12})$, so by transitivity we have
  $\Ga_{\psi\psi_1\psi_2}(M_2,V_2,M_{12},V_{12})$ as desired.
\end{proof}

\begin{definition}
  \label{def:eager}
  We say that $\open{a}{\tmj{N}}$ is \emph{eager} if for all $\psitd$ and
  $\tmj[\Psi']{M}$, we have $\subst{\td{N}{\psi}}{M}{a} \evals W$ iff there
  exists $\tmj[\Psi']{V}$ such that $M \evals V$ and
  $\subst{\td{N}{\psi}}{V}{a} \evals W$.
\end{definition}

\begin{lemma}[Elimination]
  \label{lem:elimination}
  Let $\Ga$ be a value-coherent $\Psi$-PER and $\Gb$ be a value $\Psi$-PER over
  $\Tm{\Ga}$. Suppose $\relseq{a:\Gg}{\ix{\Tm{\Gb}}{a}(N,N')}$ for some
  $\Gg \subseteq \Ga$. If $\open{a}{\tmj{N,N'}}$ are eager, then
  $\relseq{a:\Tm{\Gg}}{\ix{\Tm{\Gb}}{a}(N,N')}$.
\end{lemma}
\begin{proof}
  Let $\psitd$ and $\tmj[\Psi']{M,M'}$ be given with $\Tm{\Gg}_\psi(M,M')$. We
  want to show that
  $\Tm{\ix{\Gb_\psi}{M}}(\subst{\td{N}{\psi}}{M}{a},\subst{\td{N'}{\psi}}{M'}{a})$ holds,
  so let $\tdss{_1}{_1}{'}$ and $\tdss{_2}{_2}{_1}$ be given.

  By $\Tm{\Gg}_\psi(M,M')$, we know there exist $\td{M}{\psi_1} \evals M_1$ and
  $\td{M'}{\psi_1} \evals M'_1$ such that $\Gg_{\psi\psi_1}(M_1,M_1')$ holds and
  $\lift{\Gg}_{\psi\psi_1\psi_2}(\td{M_1}{\psi_2},\td{M}{\psi_1\psi_2},\td{M'_1}{\psi_2},\td{M'}{\psi_1\psi_2})$
  holds. By assumption,
  $\Tm{\ix{\Gb_{\psi\psi_1}}{M_1}}(\subst{\td{N}{\psi\psi_1}}{M_1}{a},\subst{\td{N'}{\psi\psi_1}}{M'_1}{a})$
  holds. Because $\Gg_{\psi\psi_1}(M_1,M_1')$ implies
  $\Ga_{\psi\psi_1'}(M_1,M_1')$ and $\Ga$ is value-coherent, we have
  $\Tm{\Ga}_{\psi\psi_1}(\td{M}{\psi_1},M_1)$, so we can adjust the index for
  $\Tm{\ix{\Gb_{\psi}}{M}}_{\psi_1}(\subst{\td{N}{\psi\psi_1}}{M_1}{a},\subst{\td{N'}{\psi\psi_1}}{M'_1}{a})$. From
  this, we have $\subst{\td{N}{\psi\psi_1}}{M_1}{a} \evals N_1$ and
  $\subst{\td{N'}{\psi\psi_1}}{M'_1}{a} \evals N'_1$ with
  $\lift{\ix{\Gb_{\psi}}{M}}_{\psi_1\psi_2}(\td{N_1}{\psi_2},\subst{\td{N}{\psi\psi_1\psi_2}}{\td{M_1}{\psi_2}}{a},\td{N'_1}{\psi_2},\subst{\td{N'}{\psi\psi_1\psi_2}}{\td{M'_1}{\psi_2}}{a})$.

  Now, from
  $\lift{\Gg}_{\psi_1\psi_2}(\td{M_1}{\psi_2},\td{M}{\psi_1\psi_2},\td{M'_1}{\psi_2},\td{M'}{\psi_1\psi_2})$,
  we know that $\td{M_1}{\psi_2} \evals M_2$ and
  $\td{M}{\psi_1\psi_2} \evals M_{12}$ with $\Gg_{\psi_1\psi_2}(M_2,M_{12})$. By
  assumption, this implies
  $\Tm{\ix{\Gb_{\psi\psi_1\psi_2}}{M_2}}(\subst{\td{N}{\psi\psi_1\psi_2}}{M_2}{a},\subst{\td{N}{\psi\psi_1\psi_2}}{M_{12}}{a})$. Again
  because $\Ga$ is value-coherent, we can obtain
  $\Tm{\ix{\Gb_{\psi}}{M}}_{\psi_1\psi_2}(\subst{\td{N}{\psi\psi_1\psi_2}}{M_2}{a},\subst{\td{N}{\psi\psi_1\psi_2}}{M_{12}}{a})$
  by adjusting the index. In particular,
  $\lift{\ix{\Gb_{\psi}}{M}}_{\psi_1\psi_2}(\subst{\td{N}{\psi\psi_1\psi_2}}{M_2}{a},\subst{\td{N}{\psi\psi_1\psi_2}}{M_{12}}{a})$
  holds. As $\open{a}{\tmj{N}}$ is eager, we know that
  $\subst{\td{N}{\psi\psi_1\psi_2}}{M_2}{a}$ and
  $\subst{\td{N}{\psi\psi_1\psi_2}}{\td{M_1}{\psi_2}}{a}$ converge to the same
  value, as do $\subst{\td{N}{\psi\psi_1\psi_2}}{M_{12}}{a}$ and
  $\subst{\td{N}{\psi\psi_1\psi_2}}{\td{M}{\psi_1\psi_2}}{a}$. Thus
  $\lift{\ix{\Gb_{\psi}}{M}}_{\psi_1\psi_2}(\subst{\td{N}{\psi\psi_1\psi_2}}{\td{M_1}{\psi_2}}{a},\subst{\td{N}{\psi\psi_1\psi_2}}{\td{M}{\psi_1\psi_2}}{a})$
  holds.

  Similarly, we can show that $\lift{\ix{\Gb_{\psi}}{M}}_{\psi_1\psi_2}(\subst{\td{N'}{\psi\psi_1\psi_2}}{\td{M_1'}{\psi_2}}{a},\subst{\td{N'}{\psi\psi_1\psi_2}}{M'_{12}}{a})$ holds. Finally, we use transitivity of $\lift{\ix{\Gb_{\psi}}{M}}$ to find that
  $\lift{\ix{\Gb_{\psi}}{M}}_{\psi_1\psi_2}(\td{N_1}{\psi_2},\subst{\td{N}{\psi\psi_1\psi_2}}{\td{M}{\psi_1\psi_2}}{a},\td{N'_1}{\psi_2},\subst{\td{N'}{\psi\psi_1\psi_2}}{\td{M'}{\psi_1\psi_2}}{a})$ holds.
\end{proof}


\section{Operational semantics}
\label{sec:opsem}

\subsection{Formation}
\begin{mathpar}
  \Infer
  { }
  {\isval{\indcl{\K}{\lst{I}}}}
\end{mathpar}


\subsection{Introduction}

\paragraph{Constructor}
\begin{mathpar}
  \Infer
  {\K[\ell] = \constr{\lst{x}}{\GG}{\Gg.\lst{I}}{\Gg.\GTh}{\sys{\xi_k}{\Gg.\Gth.\sch{m}_k}} \\
    (\forall k)\;\not\models \dsubst{\xi_k}{\lst{r}}{\lst{x}}}
  {\isval{\intro{\lst{r}}{\lst{P}}{\lst{N}}}}
  \and
  \Infer
  {\K[\ell] = \constr{\lst{x}}{\GG}{\Gg.\lst{I}}{\Gg.\GTh}{\sys{\xi_k}{\Gg.\Gth.\sch{m}_k}} \\
    \models \dsubst{\xi_k}{\lst{r}}{\lst{x}} \\
    (\forall l < k)\, \not\models \dsubst{\xi_l}{\lst{r}}{\lst{x}}}
  {\intro{\lst{r}}{\lst{P}}{\lst{N}} \steps \insttm{\Gth.\subst{\dsubst{\sch{m}_k}{\lst{r}}{\lst{x}}}{\lst{P}}{\Gg}}{\K}{\lst{N}}}
\end{mathpar}


\paragraph{Formal homogeneous composition}
\begin{mathpar}
  \Infer
  {(\forall i)\; \not\models \xi_i \\ r \neq r'}
  {\isval{\fhcom*{\xi_i}}}
  \and
  \Infer
  {(\forall i)\; \not\models \xi_i \\ r = r'}
  {\fhcom*{\xi_i} \steps M}
  \and
  \Infer
  {\models \xi_i \\ (\forall j < i)\; \not\models \xi_j}
  {\fhcom*{\xi_i} \steps \dsubst{N_i}{r'}{y}}
\end{mathpar}


\paragraph{Formal coercion}
\begin{mathpar}
  \Infer
  {\lst{I} \neq \emp \\ r \neq r'}
  {\isval{\fcoe{z.\lst{I}}{r}{r'}{M}}}
  \and
  \Infer
  {\lst{I} \neq \emp}
  {\fcoe{z.\lst{I}}{r}{r}{M} \steps M}
  \and
  \Infer
  { }
  {\fcoe{z.\emp}{r}{r'}{M} \steps M}
\end{mathpar}


\paragraph{Formal heterogeneous composition}
\begin{mathpar}
  \Infer
  { }
  {\fcom{z.\lst{I}}{r}{r'}{M}{\sys{\xi_i}{y.N_i}}
    \steps
    \fhcom{r}{r'}{\fcoe{z.\lst{I}}{r}{r'}{M}}{\sys{\xi_i}{y.\fcoe{z.\lst{I}}{y}{r'}{N_i}}}}
\end{mathpar}


\subsection{Composition}
\begin{mathpar}
  \Infer
  { }
  {\hcom*{\indcl{\K}{\lst{I}}}{\xi_i} \steps \fhcom*{\xi_i}}
\end{mathpar}


\subsection{Coercion}
\paragraph{Multi-coercion ($\mcoe$)}
\begin{align*}
  \mcoe{z.\emp}{r}{r'}{{\emp}} &\eqdef {\emp} \\
  \mcoe{z.(\ofc{\Gg}{\GG},\oft{a}{A})}{r}{r'}{(\lst{M},M)} &\eqdef (\mcoe{z.\GG}{r}{r'}{\lst{M}},\coe{z.\subst{A}{\mcoe{z.\GG}{r}{z}{\lst{M}}}{\Gg}}{r}{r'}{M})
\end{align*}

\paragraph{Total space coercion ($\tcoe$)}
\begin{mathpar}
  \Infer
  {M \steps M'}
  {\tcoe{z.(\GD,\K)}{r}{r'}{M} \steps \tcoe{z.(\GD,\K)}{r}{r'}{M'}}
  \and
  \Infer
  {(\forall i) \not\models \xi_i \\
    s \neq s'}
  {\tcoe{z.(\GD,\K)}{r}{r'}{\fhcom{s}{s'}{M}{\sys{\xi_i}{y.N_i}}} \steps
    \fhcom{s}{s'}{\tcoe{z.(\GD,\K)}{r}{r'}{M}}{\sys{\xi_i}{y.\tcoe{z.(\GD,\K)}{r}{r'}{N_i}}}}
  \and
  \Infer
  {\lst{I} \neq \emp \\ s \neq s'}
  {\tcoe{z.(\GD,\K)}{r}{r'}{\fcoe{y.\lst{I}}{s}{s'}{M}} \steps
    \fcoe{y.\coe{z.\GD}{r}{r'}{\lst{I}}}{s}{s'}{\tcoe{z.(\GD,\K)}{r}{r'}{M}}}
  \and
  \Infer
  {\K[\ell] = \constr{\lst{x}}{\GG}{\Gg.\lst{I}}{\Gg.\GTh}{\emp} \\
    \GTh = \etc{\oft{\formal{p}_j}{\sch{b}_j}} \\\\
    (\forall s)\; \lst{P}^s = \mcoe{z.\GG}{r}{s}{\lst{P}} \\
    (\forall s,j)\; N_j^s = \coe{z.\tyatty{\subst{\sch{b}_j}{\lst{P}^z}{\Gg}}{\Gd.\indcl{\K}{\Gd}}}{r}{s}{N_j}}
  {\tcoe{z.(\GD,\K)}{r}{r'}{\intro[\K']{\lst{r}}{\lst{P}}{\etc{N_j}}} \steps
    \fcoe{z.\mcoe{z.\GD}{z}{r'}{\subst{\lst{I}}{\lst{P}^z}{\Gg}}}{r'}{r}{\intro[\dsubst{\K}{r'}{z}]{\lst{r}}{\lst{P}^{r'}}{\etc{N_j^{r'}}}}}
  \and
  \and
  \Infer
  {\K[\ell] = \constr{\lst{x}}{\GG}{\Gg.\lst{I}}{\Gg.\GTh}{\sys{\xi_k}{\Gg.\Gth.\sch{m}_k}} \\
    \etc{\xi_i} \neq \emp \\
    \GTh = \etc{\oft{\formal{p}_j}{\sch{b}_j}} \\
    (\forall k)\; \not\models \dsubst{\xi_k}{\lst{r}}{\lst{x}} \\\\
    (\forall s)\; \lst{P}^s = \mcoe{z.\GG}{r}{s}{\lst{P}} \\
    (\forall s,j)\; N_j^s = \coe{z.\tyatty{\subst{\sch{b}_j}{\lst{P}^z}{\Gg}}{\Gg.\indcl{\K}{\Gg}}}{r}{s}{N_j} \\\\
    (\forall s,k)\; M_k^s = \tcoe{z.(\GD,\K)}{s}{r'}{\insttm{\Gth.\subst{\dsubst{\dsubst{\sch{m}_k}{\lst{r}}{\lst{x}}}{s}{z}}{\lst{P}^s}{\Gg}}{\dsubst{\K}{s}{z}}{\etc{N_j^s}}}
  }
  {\tcoe{z.(\GD,\K)}{r}{r'}{\intro[\K']{\lst{r}}{\lst{P}}{\etc{N_j}}} \\\\
    \steps \\\\
    \fcom{z.\mcoe{z.\GD}{z}{r'}{\subst{\lst{I}}{\lst{P}^z}{\Gg}}}{r'}{r}{\intro[\dsubst{\K}{r'}{z}]{\lst{r}}{\lst{P}^{r'}}{\etc{N_j^{r'}}}}{\sys{\dsubst{\xi_k}{\lst{r}}{\lst{x}}}{z.M_k^z}}}
\end{mathpar}

\paragraph{Coercion ($\coe$)}
\begin{mathpar}
  \Infer
  { }
  {\coe{z.\indcl{\K}{\lst{I}}}{r}{r'}{M} \steps \fcoe{z.\mcoe{z.\GD}{z}{r'}{\lst{I}}}{r}{r'}{\tcoe{z.(\GD,\K)}{r}{r'}{M}}}
\end{mathpar}


\subsection{Elimination}
\paragraph{Action of argument types}
\begin{align*}
  \func{\argvar{\lst{I}}}{\Gd.h.R}{M} &\eqdef \subst{\subst{R}{\lst{I}}{\Gd}}{M}{h} \\
  \func{\argpi{b}{B}{\sch{c}}}{\Gd.h.R}{M} &\eqdef \lam{b}{\func{\sch{c}}{\Gd.h.R}{\app{M}{b}}}
\end{align*}
\paragraph{Elimination}
\begin{mathpar}
  \Infer
  {M \steps M'}
  {\elim{\Gd.h.D}{\lst{I}}{M}{\E} \steps
    \elim{\Gd.h.D}{\lst{I}}{M'}{\E}}
  \and
  \Infer
  {(\forall i)\;\not\models \xi_i \\
    r \neq r'} 
  {\elim{\Gd.h.D}{\lst{I}}{\fhcom{r}{r'}{M}{\sys{\xi_i}{N_i}}}{\E} \steps \\
    \com{y.\subst{\subst{D}{\lst{J}}{\Gd}}{\fhcom{r}{y}{M}{\sys{\xi_i}{N_i}}}{h}}{r}{r'}{\elim{\Gd.h.D}{\lst{I}}{M}{\E}}{\sys{\xi_i}{y.\elim{\Gd.h.D}{\lst{I}}{N_i}{\E}}}}
  \and
  \Infer
  {r \neq r'}
  {\elim{\Gd.h.D}{\lst{I}}{\fcoe{z.\lst{J}}{r}{r'}{M}}{\E} \steps
    \coe{z.\subst{\subst{D}{\lst{J}}{\Gd}}{\fcoe{z.\lst{J}}{r}{z}{M}}{h}}{r}{r'}{\elim{\Gd.h.D}{\dsubst{\lst{J}}{r}{z}}{M}{\E}}}
  \and
  \Infer
  {\K[\ell] = \constr{\lst{x}}{\GG}{\Gg.\lst{J}}{\Gg.\GTh}{\sys{\xi_k}{\Gg.\Gth.\sch{m}_k}} \\
    \E[\ell] = \lst{x}.\Gg.\Gh.\Gr.R \\
    \GTh = \etc{\ofa{p_j}{\sch{b}_j}} \\
    (\forall k)\; \not\models \dsubst{\xi_k}{\lst{r}}{\lst{x}}}
  {\elim{\Gd.h.D}{\lst{I}}{\intro{\lst{r}}{\lst{P}}{\etc{N_j}}}{\E}
    \steps
    \subst{\subst{\subst{\dsubst{R}{\lst{r}}{\lst{x}}}{\lst{P}}{\Gg}}{\etc{N_j}}{\Gh}}{\etc{\func{\subst{\sch{b}_j}{\lst{P}}{\Gg}}{\Gd.h.\elim{\Gd.h.D}{\Gd}{h}{\E}}{N_j}}}{\Gr}}
\end{mathpar}
  


\section{Selected proof theory rules}
\label{app:proof-theory}

\subsection{Formation}
\begin{mathpar}
  \Infer[(\ref{def:schema})]
  {\cwftypek{\GD}}
  {\ceqconstrs{\GD}{\nilconstrs}{\nilconstrs}}
  \and
  \Infer
  {\ceqconstrs{\GD}{\K}{\K'} \\
    \ceqtypek{\GG}{\GG'} \\
    \eqtm{\ofc{\Gg}{\GG}}{\lst{I}}{\lst{I}'}{\GD} \\
    \ceqargtype{\GD}{\GTh}{\GTh'} \\
    \text{$\etc{\xi_k}$ valid or empty} \\
    \fd{\etc{\xi_k}} \subseteq \lst{x} \\
    (\forall k,l)\; \eqargtm[\Psi,\lst{x}]<\xi_k,\xi_l>{\ofc{\Gg}{\GG}}{\GD}{\ofc{\Gth}{\GTh}}{\sch{m}_k}{\sch{m}_l}{\argvar{\lst{I}}}
  }
  {\ceqconstrs{\GD}{\snocconstrs{\K}{\constr{\lst{x}}{\GG}{\Gg.\lst{I}}{\Gg.\GTh}{\sys{\xi_k}{\Gg.\Gth.\sch{m}_k}}}}{\snocconstrs{\K'}{\constr{\lst{x}}{\GG'}{\Gg.\lst{I}'}{\Gg.\GTh'}{\sys{\xi_k}{\Gg.\Gth.\sch{m}'_k}}}}}
  \and
  \Infer[(\ref{lem:formation})]
  {\ceqtypek{\GD}{\GD'} \\ \ceqconstrs{\GD}{\K}{\K'} \\ \ceqtm{\lst{I}}{\lst{I}'}{\GD}}
  {\ceqtypek{\indcl{\K}{\lst{I}}}{\indcl[\GD']{\K'}{\lst{I}'}}}
\end{mathpar}

\subsection{Introduction}

\paragraph{Constructor}
\begin{mathpar}
  \Infer[(\ref{lem:supports-I-intro})]
  {\ceqconstrs{\GD}{\K}{\K'} \\
    \K[\ell] = \constr{\lst{x}}{\GG}{\Gg.\lst{I}}{\Gg.\GTh}{\sys{\xi_k}{\Gg.\Gth.\sch{m}_k}} \\
    \ceqtm{\lst{P}}{\lst{P}'}{\GG} \\
    \ceqtm{\lst{N}}{\lst{N}'}{\tyatty{\subst{\GTh}{\lst{P}}{\Gg}}{\Gd.\indcl{\K}{\Gd}}}}
  {\ceqtm{\intro{\lst{r}}{\lst{P}}{\lst{N}}}{%
      \left\{
        \begin{array}{ll}
          \intro[\K']{\lst{r}}{\lst{P}'}{\lst{N}'}, \\
          \insttm{\Gth.\subst{\dsubst{\sch{m}_k}{\lst{r}}{\lst{x}}}{\lst{P}}{\Gg}}{\K}{\lst{N}}, &\text{if $\models \dsubst{\xi_k}{\etc{r}}{\etc{x}}$}
        \end{array}
      \right\}
    }{\indcl{\K}{\subst{\lst{I}}{\lst{P}}{\Gg}}}}
\end{mathpar}

\paragraph{Homogeneous composition}
\begin{mathpar}
  \Infer[(\ref{lem:supports-I-fhcom})]
  {\coftype{\lst{I}}{\GD} \\
    \ceqtm{M}{M'}{\indcl{\K}{\lst{I}}} \\
    (\forall i,j)\; \ceqtm[\Psi,y]<\xi_i,\xi_j>{N_i}{N'_j}{\indcl{\K}{\lst{I}}} \\
    (\forall i)\; \ceqtm<\xi_i>{\dsubst{N_i}{r}{y}}{M}{\indcl{\K}{\lst{I}}}}
  {\ceqtm{\fhcom{r}{r'}{M}{\sys{\xi_i}{N_i}}}{%
      \left\{
        \begin{array}{ll}
          \fhcom{r}{r'}{M'}{\sys{\xi_i}{N'_i}}, \\
          \dsubst{N_i}{r'}{y}, &\text{if $\models \xi_i$} \\
          M, &\text{if $r = r'$}
        \end{array}
      \right\}
    }{\indcl{\K}{\lst{I}}}}
\end{mathpar}

\paragraph{Coercion}
\begin{mathpar}
  \Infer[(\ref{lem:supports-I-fcoe})]
  {\ceqtm[\Psi,z]{\lst{I}}{\lst{I}'}{\GD} \\
    \ceqtm{M}{M'}{\indcl{\K}{\dsubst{\lst{I}}{r}{z}}}}
  {\ceqtm{\fcoe{z.\lst{I}}{r}{r'}{M}}{%
      \left\{
        \begin{array}{ll}
          \fcoe{z.\lst{I}'}{r}{r'}{M'}, \\
          M, &\text{if $r = r'$ or $\lst{I} = \emp$}
        \end{array}
      \right\}
    }{\indcl{\K}{\dsubst{\lst{I}}{r'}{z}}}}
\end{mathpar}

\paragraph{Heterogeneous composition}
\begin{mathpar}
  \Infer[(\ref{lem:supports-I-fcom})]
  {\ceqtm[\Psi,z]{\lst{I}}{\lst{I}'}{\GD} \\
    \ceqtm{M}{M'}{\indcl{\K}{\dsubst{\lst{I}}{r}{y}}} \\
    (\forall i,j)\; \ceqtm[\Psi,y]<\xi_i,\xi_j>{N_i}{N'_j}{\indcl{\K}{\lst{I}}} \\
    (\forall i)\; \ceqtm<\xi_i>{\dsubst{N_i}{r}{y}}{M}{\indcl{\K}{\dsubst{\lst{I}}{r}{y}}}}
  {\ceqtm{\fcom{y.\lst{I}}{r}{r'}{M}{\sys{\xi_i}{N_i}}}{\fcom{y.\lst{I}'}{r}{r'}{M'}{\sys{\xi_i}{N'_i}}}{\indcl{\K}{\dsubst{\lst{I}}{r'}{y}}}}
\end{mathpar}


\subsection{Coercion}

\paragraph{Total space coercion}
\begin{mathpar}
  \Infer[(\ref{thm:tcoe-sigma-supports})]
  {\ceqtypek[\Psi,z]{\GD}{\GD'} \\
    \ceqconstrs[\Psi,z]{\GD}{\K}{\K'} \\
    \ceqtm{\lst{I}}{\lst{I}'}{\dsubst{\GD}{r}{z}} \\
    \ceqtm{M}{M'}{\indcl[\dsubst{\GD}{r}{z}]{\dsubst{\K}{r}{z}}{\lst{I}}}}
  {\ceqtm{\tcoe{z.(\GD,\K)}{r}{r'}{M}}{%
      \left\{
        \begin{array}{ll}
          \tcoe{z.(\GD',\K')}{r}{r'}{M'}, \\
          M, &\text{if $r = r'$}
        \end{array}
      \right\}
    }{\indcl[\dsubst{\GD}{r'}{z}]{\dsubst{\K}{r'}{z}}{\mcoe{z.\GD}{r}{r'}{\lst{I}}}}}
\end{mathpar}


\subsection{Elimination}

\begin{mathpar}
  \Infer[(\ref{def:elim-list})]
  { }
  {\ceqcasespart{\GD}{\nilelim}{\nilelim}{\K}{\Gd.h.D}}
  \and
  \Infer[(\ref{def:elim-list})]
  {\ceqcasespart{\GD}{\E}{\E'}{\K}{\Gd.h.D} \\
    \height{\K}{\ell} = |\E| \\
    \K[\ell] = \constr{\lst{x}}{\GG}{\Gg.\lst{I}}{\Gg.\GTh}{\sys{\xi_k}{\Gg.\Gth.\sch{m}_k}} \\
    \eqtm[\Psi,\lst{x}]{\ofc{\Gg}{\GG},\ofc{\Gh}{\tyatty{\GTh}{\indcl{\K}{\lst{I}}}},\ofc{\Gr}{\tyatty*{\GTh}{\Gd.h.D}{\Gh}}}{R}{R'}{\subst{D}{\intro{\lst{x}}{\Gg}{\Gh}}{h}} \\
    (\forall k)\;\eqtm[\Psi,\lst{x}\mid\xi_k]{\ofc{\Gg}{\GG},\ofc{\Gh}{\tyatty{\GTh}{\indcl{\K}{\lst{I}}}},\ofc{\Gr}{\tyatty*{\GTh}{\Gd.h.D}{\Gh}}}{R}{\insttm*{\Gth.\sch{m}_k}{\K}{\E}{\Gd.h.D}{\Gh}{\Gr}}{\subst{D}{\intro{\lst{x}}{\Gg}{\Gh}}{h}}
  }
  {\ceqcasespart{\GD}{\snocelim{\E}{\ell:\lst{x}.\Gg.\Gh.\Gr.R}}{\snocelim{\E'}{\ell:\lst{x}.\Gg.\Gh.\Gr.R'}}{\K}{\Gd.h.D}}
  \and
  \Infer[(\ref{def:elim-list})]
  {\ceqcasespart{\GD}{\E}{\E'}{\K}{\Gd.h.D} \\ |\K| = |\E|}
  {\ceqcases{\GD}{\E}{\E'}{\K}{\Gd.h.D}}
\end{mathpar}

\begin{mathpar}
  \Infer[(\ref{cor:elim})]
  {\ceqconstrs{\GD}{\K}{\K'} \\
    \eqtypek{\ofc{\Gd}{\GD},\oft{h}{\indcl{\K}}{\Gd}}{D}{D'} \\
    \ceqtm{\lst{I}}{\lst{I}'}{\GD} \\
    \ceqtm{M}{M'}{\indcl{\K}{\lst{I}}} \\
    \ceqcases{\GD}{\E}{\E'}{\K}{\Gd.h.D}}
 {\ceqtm{\elim{\Gd.h.D}{\lst{I}}{M}{\E}}{\elim{\Gd.h.D'}{\lst{I}'}{M'}{\E'}}{\subst{\subst{D}{\lst{I}}{\Gd}}{M}{h}}}
\end{mathpar}


\bibliographystyle{plainnat}
\bibliography{main}

\end{document}
